\documentclass[preprints,review,accept,moreauthors,pdftex]{mdpi} 

\UseRawInputEncoding

\firstpage{1} 
\pubvolume{1}
\issuenum{1}
\articlenumber{0}
\pubyear{2021}
\copyrightyear{2020}
\datereceived{} 
\dateaccepted{} 
\datepublished{} 
\hreflink{https://doi.org/} 

\usepackage{xcolor, ulem, soul, tabularx}
\setcitestyle{authoryear, round, comma, sort&compress, super, aysep={}}
\usepackage{multicol}
\usepackage{makecell}
\usepackage{deluxetable}
\usepackage{sidecap}



\newcommand{\red}[1]{{\textcolor{black}{#1}}}

\providecommand{\e}[1]{\ensuremath{\times 10^{#1}}}
\newcommand{\Msun}{\ensuremath{{M_{\odot}}}}
\newcommand{\Lsun}{\ensuremath{{L_{\odot}}}}

\newcommand{\myr}{${\rm M_{\odot}\,yr^{-1}}$}
\newcommand{\um}{$\mu$m}

\newcommand{\rxv}{$(r/r_{200})\times(\Delta v/\sigma_v)$}

\newcommand{\alphaUnits}{${(\rm M_{\odot}}\,\rm(K\,km\,s^{-1}\,pc^{-2})^{-1})$}

\renewcommand{\arraystretch}{1.5}
\renewcommand\sout[1]{}

\Title{From Clusters to Proto-clusters: the Infrared Perspective on Environmental Galaxy Evolution}

\Author{Stacey Alberts $^{1*}$\orcidA{
} and Allison Noble $^{2,3}$\orcidB}

\AuthorNames{Stacey Alberts and Allison Noble}

\address{%
$^{1}$ \quad Steward Observatory, University of Arizona, 933 N. Cherry Avenue, Tucson, AZ 85721 USA\\
$^{2}$ \quad School of Earth and Space Exploration, Arizona State University, Tempe, AZ 85287, USA \\
$^{3}$ \quad Beus Center for Cosmic Foundations, Arizona State University, Tempe, AZ 85287, USA}

\corres{Correspondence: salberts@arizona.edu}

\abstract{Environment is one of the primary drivers of galaxy evolution\red{; via multiple mechanisms, it can control the critical process}\sout{, capable} of transforming galaxies from star forming to quiescent\red{, commonly termed "quenching"}\sout{via multiple mechanisms}. Despite its importance, however, we still do not have a clear view of how \red{environmentally-driven}\sout{environmental} quenching proceeds even in the most extreme environments: galaxy clusters and their progenitor proto-clusters. Recent advances in infrared capabilities have enabled transformative progress not only in the identification of these structures but in detailed analyses of quiescence, obscured star formation, and molecular gas in (proto-)cluster galaxies across cosmic time. In this review, we will discuss the current state of the literature regarding the quenching of galaxies in (proto-)clusters from the observational, infrared perspective.  Our improved understanding of environmental galaxy evolution comes from unique observables across the distinct regimes of the near-, mid-, and far-infrared, crucial in the push to high redshift where \red{massive} galaxy growth is dominated by highly extincted, infrared-bright galaxies.}


\begin{document}

\section{Introduction}\label{sec:intro}

\begin{SCfigure}[0.8][!htb]
    \parbox{0.5\columnwidth}{\centering\includegraphics[width=0.5\columnwidth]{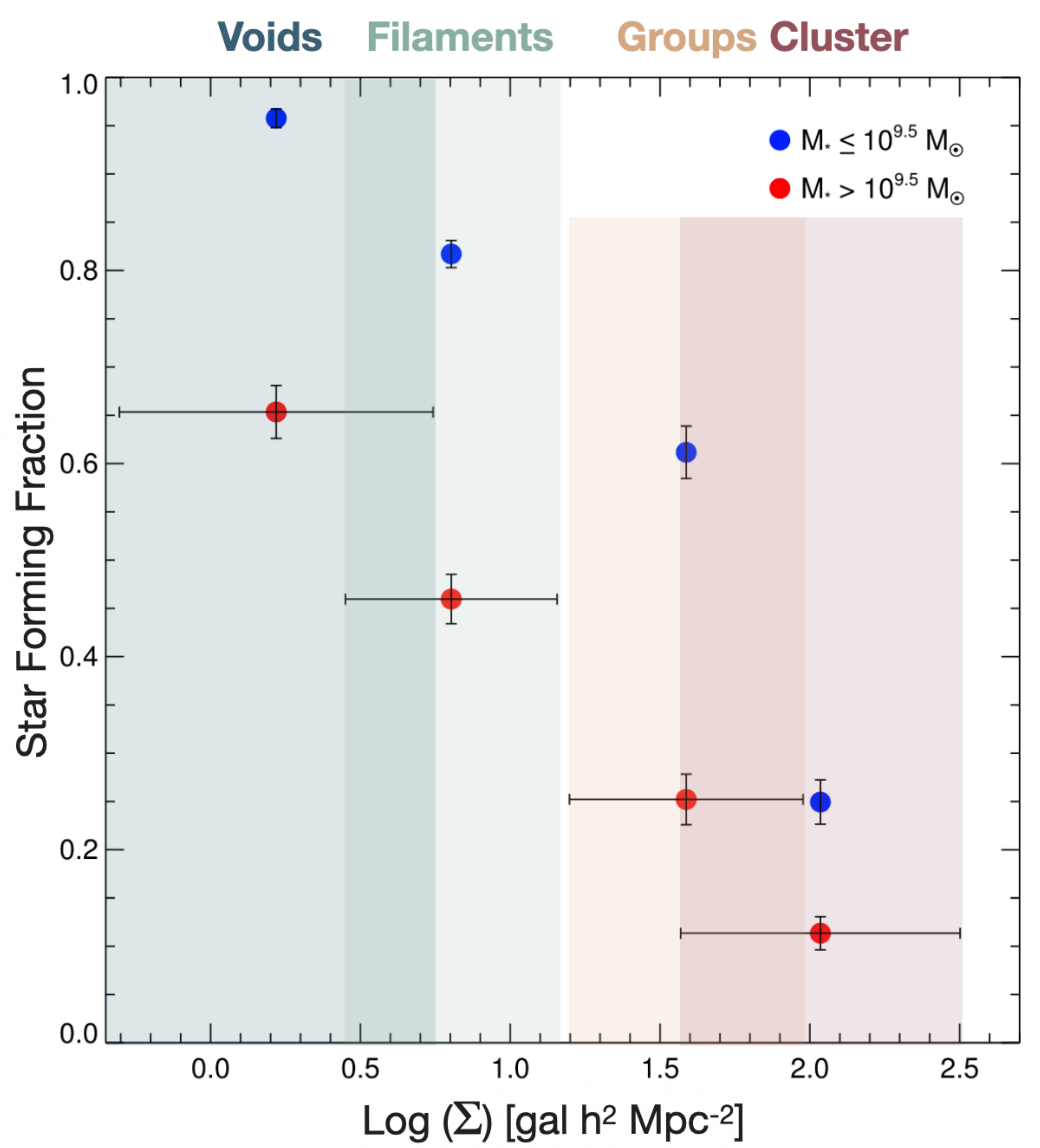}}
    \caption{ $-$ The star-forming fraction across four environments (voids, filaments, groups, and the cluster) in the Coma Supercluster region. Two stellar mass bins are shown: log $M_{\star}/\Msun\leq9.5$ (blue) and log $M_{\star}/\Msun>9.5$ (red).  At all masses, the star-forming fraction progressively decreases with increasing local galaxy density. Note that the fraction for low-mass galaxies in underdense void regions is nearly one. Horizontal error bars indicate the standard deviation of each local density bin (omitted from the lower mass bin for clarity).  Figure adapted from Figure~9 in \citealt{cybulski2014}. }
    \label{fig:coma}
\end{SCfigure}

In the four decades since \citealt{dressler1980} presented the morphological properties of galaxies in local galaxy clusters, a rich literature has emerged on the connection between galaxy evolution and environment.
We have now firmly established that high-density regions exhibit both a morphology-density relation  \citep[e.g.][]{dressler1980, postman1984, dressler1997} and a star formation rate (SFR)-density relation \citep[e.g.][]{dressler1984, lewis2002, peng2010}, with local clusters preferentially hosting \red{early-type}\sout{elliptical}, quiescent galaxies (QGs).  Star-forming galaxies (SFGs) in $z\sim0$ clusters typically contain low atomic \citep[e.g.][]{solanes2001, gavazzi2005, catinella2013, jaffe2016} and molecular \citep[e.g.][]{fumagalli2009, boselli2014} gas content. Correspondingly, as molecular gas is the fuel for forming stars, their star formation (SF) activity is also lower (see \citealt{boselli2006} for a review).  As demonstrated by the Coma Supercluster region, these systematic differences occur continuously with increasing galaxy density, from voids to filaments to groups to clusters, over a wide range in stellar mass (Figure~\ref{fig:coma}; \citealt{cybulski2014}, see also \citealp{gavazzi2010, mahajan2010}).  From this, we can infer progressively divergent pathways for the evolution of galaxies from star forming to quenched as a function of
environment.  

Underscoring the importance of understanding this environmentally-driven galaxy evolution, \red{a quarter to }half of all massive galaxies live in groups \red{or low-mass clusters} up to $z\sim1.5$ \red{\citep[e.g.][]{gerke2012, knobel2012, tempel2016, boselli2022}} and recent simulations suggest that galaxies living in proto-clusters (the early formation stage of galaxy clusters) at $z\sim10$ contributed up to half of the cosmic star formation rate density \citep[SFRD;][]{chiang2017}.
Massive galaxy clusters (log $M_{\rm halo}/\Msun\gtrsim14$) host a far smaller percentage of the overall massive galaxy population \citep[$25\%$ ($<5\%$) at $z=0$ ($z\sim1-1.5$) for log $M_{\star}/\Msun>9$;][]{boselli2022}; however, they provide the best
astrophysical laboratories for exploring environmental processes and establishing the
boundary conditions of environmentally-driven quenching and morphological
transformation. Indeed, detailed observational studies support a complex interplay of
multiple quenching mechanisms, internal and external, driving the evolution of local
galaxies over a wide range in galaxy density \citep[e.g.][and references therein]{fraser-mckelvie2018, coccato2020}.

The relationship between $z\sim0$ galaxies and environment has been extensively
studied and reviewed \citep[e.g.][]{dressler1984, boselli2006, delucia2007, blanton2009, boselli2014} and the availability of new datasets continues to push the
state of the art in the local Universe.  The infrared-to-submillimeter regime alone has produced a wealth of observations over low-redshift cluster galaxies, including: the \textit{Herschel} Virgo Cluster Survey \citep[HeViCS;][]{Davies2010}; the Virgo Environment Traced in CO  survey \citep[VERTICO;][]{brown2021, zabel2022}; the ALMA Fornax Cluster
Survey \citep[AlFoCS;][]{zabel2019}; the GAs Stripping Phenomena in galaxies survey \citep[GASP;][]{poggianti2017, moretti2020}; and references therein.  In
parallel, we have established that this relationship evolves with cosmic time $-$ for
example, the fraction of optically blue galaxies increases with increasing redshift
\citep[the Butcher-Oemler Effect;][]{butcher1984} $-$ and on through to the epoch
dominated by proto-clusters ($z\gtrsim2$; see \citealt{overzier2016} for a review). The
subsequent emerging picture contains many complexities, but can be broken roughly
into three epochs. From $z\sim0\rightarrow1$, galaxy populations in massive clusters
remain largely quenched, comparable to their local counterparts
\citep[e.g.][]{muzzin2012, balogh2016}, though with evidence for a diversity in
evolutionary pathways from galaxy-to-galaxy.  The epoch $z=1-2$ sees the end
of ubiquitous quiescence, with some clusters showing a breakdown or reversal in the
local SFR-density relation.  To establish the largely quenched populations by
$z\sim1$, environmentally-driven rapid quenching has to ramp up in these systems
during this epoch \citep{muzzin2012, matharu2021}. The last epoch at $z\gtrsim2$ is
largely populated by proto-clusters, highly-extended (10-30$^{\prime}$) structures in
the process of collapse, which host substantial star formation.

Infrared observations ($1\mu$m - 3mm)  are increasingly playing a critical role in our understanding of (proto-)cluster evolution due to the remarkable rise in observational
capabilities, starting with \textit{IRAS} in the 1980s through the recent successful launch of
the James Webb Space Telescope \citep[\textit{JWST};][]{gardner2006} in 2021. Four decades of space-based IR missions (e.g., \textit{IRAS}, \textit{ISO}, \textit{Spitzer}, \textit{WISE},
\textit{Herschel}, \red{\textit{AKARI}}, \textit{Planck}\footnote{Referenced facilities:  \textit{InfraRed Astronomy Satellite} \citep[\textit{IRAS};][]{neugebauer1984}; \textit{Infrared Space Observatory} \citep[\textit{ISO};][]{kessler1996}; \textit{Spitzer Space Telescope} \citep{werner2004}; \textit{Wide-field Infrared Survey Explorer} \citep[\textit{WISE};][]{wright2010}; \textit{Herschel Space Telescope} \citep{pilbratt2010}; \red{\textit{AKARI}} \citep{murakami2007}; \textit{Planck} \citep{planckcollaboration2011}} have provided a range of sensitivities, resolutions, and mapping
speeds with low background levels and at wavelengths that are not achievable on
earth due to low atmospheric transmission. From the ground, bolometer arrays (e.g., SCUBA, SCUBA-2, AzTEC, LABOCA\footnote{Referenced instruments: Submillimeter Common-User Bolometer Array \citep[SCUBA;][]{holland1999}, SCUBA-2 \citep{holland2013}, AzTEC \citep{wilson2008}, Large Apex BOlometer CAmera \citep[LABOCA;][]{siringo2009})} on
single-dish observatories e.g. JCMT, LMT, SPT, ACT\footnote{Referenced facilities: James Clerk Maxwell Telescope \citep[JCMT;][]{holland1999}, Large Millimeter Telescope/Gran Telescopio Milim\'{e}trico Alfonso Serrano \citep[LMT;][]{hughes2010}, South Pole Telescope \citep[SPT; ][]{carlstrom2011}, Atacama Cosmology Telescope \citep[ACT;][]{hincks2010}})  have delivered low-resolution
surveys over large areas, while interferometric facilities (e.g., NOEMA, PdBI, ALMA, JVLA\footnote{Referenced facilities: the NOrthern Extended Millimeter Array \citep[NOEMA;][]{chenu2016, krips2022}; the Plateau de Bure Interferometer \citep[PdBI;][]{guilloteau1992}; the Atacama Large Millimeter Array \citep[ALMA;][]{Wootten2009}, and the Karl G. Jansky Very Large Array \citep[JVLA;][]{Perley2011}}) have
enabled unparalleled sensitivity and resolution for detailed targeted studies.  The
extragalactic discovery space opened by these facilities includes the
characterization of significant, sometimes dominant, populations of moderate to
extremely dust-obscured galaxies.  For an expanded discussion of infrared facilities,
see \citealt{casey2014, farrah2019}. 

In this review, we present an overview of our
current understanding of galaxy populations in (proto-)clusters from the infrared
perspective, followed by pressing open questions and the future outlook.  Other relevant reviews in this Special Issue include ``The Role of AGN in Luminous Infrared Galaxies from the Multiwavelength Perspective'', \citealt{u2022}; ``Infrared Spectral Energy Distribution and Variability of Active Galactic Nuclei: Clues to the Structure of Circumnuclear Material'', \citealt{lyu2022}; ``The Past and Future of Mid-Infrared Studies of AGN'', \citealt{sajina2022}; ``Dust-obscured star formation: observational constraints from the last decade'', Zavala \& Casey 2022, in prep.; ``ALPINE: A Large Survey to Understand Teenage Galaxies", \citealt{faisst2022}. 

\subsection{Overview of Environment and Environmental Processes} \label{sec:environment}

\subsubsection{Defining Environment}

The definition of environment is not homogeneous across the literature
(see \citealt[][]{muldrew2012}).  One common approach is to focus on the ``global''
environment via large scale structures like galaxy clusters (or groups), while
another is to define the ``local'' environment in terms of the local galaxy density
field \citep[using a technique such as nearest neighbors, e.g.][]{kovac2010, scoville2013}.  Caution is warranted when comparing results using these different definitions as they may probe different populations and pathways. For example, centrals, the most massive galaxy in a given halo, and satellites are thought to follow different evolutionary tracks \citep[e.g.][]{hogg2003, vandenbosch2008, peng2010}. Global studies focus on satellite populations by design (minus the
group/cluster central known as the Brightest Group/Cluster Galaxy, BGG or BCG), while local environment studies may include both centrals and satellites and are sensitive to the scale used to measure the galaxy overdensity, which can complicate interpretation of the observed trends \citep{woo2013}. Additionally, environmental effects are usually identified
via comparison to a control or ``field'' sample, the nature of which can influence the results \citep{werner2022}.   Global studies typically adopt regions from surrounding ``blank'' fields, which will include voids, groups, and filaments, while local studies often compare to their lowest density bin, excluding intermediate densities.

In this review, we will focus mainly on studies using the global definition.  We define a galaxy cluster as a relaxed or unrelaxed (or merging) gravitationally-bound structure with a halo mass (at time of observation) of log $M_{200}/\Msun\gtrsim13.5$, acknowledging that many works refer to systems with log $M_{200}/\Msun\sim13-14$ as groups.  The definition of proto-cluster is even less clear in the literature; here we define a proto-cluster as an extended overdensity at $z\geq2$ which will collapse into a cluster by $z\sim0$. The $z\geq2$ redshift boundary is adopted for convenience as in reality the line is very blurred: the extended structure around clusters at $z<2$ may not have completely collapsed even by $z\sim0$ \citep{chiang2013} and proto-cluster cores at $z>2$ may have properties more typically associated with clusters \citep[i.e. quenched populations, X-ray emission;][but see \citealp{champagne2021}]{wang2018}. 

\subsubsection{Environment Quenching Mechanisms}

In a simplified schematic, galaxies are surrounded by a hot halo of ionized gas, the circumgalactic medium, which cools onto an extended reservoir of neutral hydrogen surrounding the disk.  This \textsc{Hi} gas subsequently condenses onto the disk of the galaxy, forming clouds of molecular gas (H$_2$), which are the nurseries for star formation.  Outflows in the form of stellar winds, jets, and/or galactic fountains \citep{bahe2015} can deliver material back out into the halo of the galaxy, forming a cosmic recycling of gas and material for the next generation of stars.  This baryon cycle in galaxies is often viewed as a ``bathtub" equilibrium model \citep[e.g.,][]{dekel2009, bouche2010, lilly2013}, where gas content is regulated by inflows and depleted through outflows and star formation consumption.  Importantly, in order to form stars on long timescales, galaxies require their disk gas reservoir to be replenished not only by gas recycling, but by external sources e.g., cold mode gas accretion \citep[e.g.][]{dekel2009} or gas-rich mergers. See \citealt{hatch2016a}, \citealt{tacconi2020}, and \citealt{saintonge2022} for recent reviews. 
  
In galaxy clusters, several environmental processes have been identified that may act on galaxies to enhance or shut down star formation.  They fall broadly into three categories. The first, hydrodynamical processes, depletes a galaxy's gas reservoirs though interaction with the hot ($10^7-10^8$ K), dense ($10^{-3}$ cm$^{-3}$) intracluster medium \citep[ICM;][]{sarazin1986}, often at high speeds ($\sim500-1,000$ km s$^{-1}$). Starvation \citep[also known as strangulation;][]{larson1980, balogh2000} heats or strips the diffuse hot halo of a galaxy upon entering the ICM, halting this replenishment of disk gas.  The effect of this process on a galaxy's SFR is generally thought to be slow, taking several Gyr \citep[e.g.][]{boselli2014, peng2015} and may start far from the cluster center \citep[up to five times the virial radius, $R_{\rm vir}$; e.g.][]{bahe2013}.  More aggressive hydrodynamical processes $-$ ram pressure stripping \citep[RPS;][]{gunn1972}, viscous stripping \citep{nulsen1982}, and thermal evaporation \citep{cowie1977} $-$ act directly to heat or strip the cold molecular gas tightly bound in a galaxy's disk.  Of these, we will focus on RPS\footnote{We note that starvation is likely a form of mild RPS that effects hot halo gas and not a fully distinct process.  However, for convenience, we refer to stripping of cold disk gas as RPS throughout this review.} in this review, which can range from mild to strong, gradually stripping the extended cold gas in a galaxy's outskirts to rapidly removing a significant portion of the central disk gas on short timescales.  The timescale to effect star formation similarly varies.  For extensive reviews on RPS, see \citealt{boselli2022, cortese2021}. {\bf Regardless of the specifics, quenching from hydrodynamical processes generally proceeds outside-in, affecting the galaxy outskirts first.}

The second category is gravitational mechanisms, either galaxy-to-galaxy interactions or perturbations induced by the halo, called tidal interactions.  Galaxy interactions can take place in the form of mergers (major or minor, gas-rich or gas-poor), interactions, or fly-bys, with the cumulative effect of multiple high-speed fly-bys called harassment \citep{moore1996, smith2010b, bialas2015}.  These processes are capable of inducing instabilities in the disk, which may drive gas inflows to the nucleus. In the local Universe, mergers are likely responsible for triggering central starbursts and Active Galactic Nuclei (AGN) that drive the quenching of extremely luminous galaxies \citep[e.g.][]{hopkins2008}.  The ability of mergers and interactions to trigger or quench starbursts or AGN at higher redshift, however, is still highly debated in the literature \citep[e.g.][]{kocevski2012, mechtley2016, shah2020}.  Potentially separate from quenching, gravitational processes may drive morphological transformation \citep[e.g.][]{springel2005, feldmann2011}, though the role of major mergers in forming ellipticals at high redshift is also highly uncertain \citep[e.g.][]{kaviraj2013, lofthouse2017}.  We will focus on the observational evidence for quenching by gravitational mechanisms; a full treatment of the role of interactions in morphological change in (proto-)cluster galaxies is beyond the scope of this review.

Finally, internal processes that operate in isolated galaxies occur and may even be enhanced or happen at earlier times in overdense environments.  This includes disk instabilities and stellar feedback as well as AGN feedback, which can heat and/or expel gas and is particularly relevant in quenching massive galaxies \citep[log $M_{\star}/\Msun>10.5$;][]{boselli2006, wylezalek2016, vogelsberger2014, vogelsberger2014a, henriques2015, henriques2019}. Luminous AGN may rapidly remove gas from galaxy centers e.g. \citep[e.g.][]{dimatteo2005, hopkins2008} while less luminous AGN activity may aid in starvation via modest outflows moving gas into the galaxy outskirts or hot halo \citep[e.g.][]{vogelsberger2014, vogelsberger2014a, schaye2015, zinger2020}. \red{Overdense environments can result in}\sout{Specific to overdense environments,} overconsumption \citep[e.g.,][]{mcgee2014, balogh2016} -- the depletion of gas via the combined effects of starvation, consumption by star formation, and modest feedback\red{. This} incorporates internal processes with environmentally-driven suppression of fresh gas accretion and recycling to effect quenching over a range of timescales, depending on SFR and feedback strength.  {\bf Unlike hydrodynamical processes, these mechanisms likely quench inside-out.}

How can we look for and separate these mechanism(s) in overdense environments? Resolved studies \citep[e.g.,][]{schaefer2017, fossati2018, poggianti2019, bluck2020} can look for signatures of quenching across the disk, separating outside-in from inside-out processes, as well as disturbed morphologies, ram-pressure stripped tails, and faint tidal features indicating interactions.  These studies, however, are still in their infancy, particularly at high redshift.  Aggregate studies of populations can instead look for trends in galaxy properties (stellar mass, SFR, AGN fraction, gas content) with environmental proxies (projected radius, local galaxy density, halo mass) over cosmic time. This allows for the quantification of properties such as: SFG and QG fractions; stellar mass functions; environmental quenching efficiencies (EQE); SFRs and deviations from the star-forming Main Sequence\footnote{The star-forming Main Sequence \citep[MS;][]{brinchmann2004, noeske2007, elbaz2007, daddi2007} is the observed correlation between the star formation and stellar mass of a galaxy\sout{at a given redshift}, which exhibits low scatter ($\sim0.3$ dex) \red{and a trend in the SFR per unit mass which increases with increasing redshift.  The latter results in the SFR for a MS galaxy rising by two orders of magnitude from $z\sim0$ to $z\sim1$}\sout{The amplitude of the MS increases with increasing redshift, which results in the typical galaxy star formation rate rising from $\sim1$ $M_{\odot}$ yr$^{-1}$ at $z\sim0$ to $\sim100$ $M_{\odot}$ yr$^{-1}$ at $z\sim1$} (see \citealt{schreiber2020} and references therein).
} (MS); and gas depletion timescales and gas fractions.  These will be discussed in Sections~\ref{sec:nir}-\ref{sec:submm} and
then placed in the context of the mechanisms described here in \S\,\ref{sec:wrap}.

\subsection{What do we learn from the infrared?}\label{sec:learn}

The infrared is traditionally broken into three regimes: the near-infrared (NIR), mid-infrared (MIR), and far-infrared (FIR), the latter of which includes the so-called submillimeter (submm) wavelengths.  This wavelength range contains a wealth of information, tracing stellar to star formation to gas properties.  Here we summarize the relevant observables and why they are important for (proto-)cluster studies.

\subsubsection{Near-Infrared}

In galaxies, the NIR is dominated by evolved stellar populations, the end product of galaxy growth via star formation and/or mergers. Continuum emission from low-mass stars produces a ubiquitous stellar bump feature peaked at $1.6\mu$m, seen in all galaxies with established stellar populations \citep[by $\sim10$ Myr after a young starburst;][]{sorba2010}, with the exception of luminous AGN hosts where the NIR is dominated by the hot dust continuum (see \citealt{lyu2022} for a review). NIR constraints near rest-frame $1\mu$m therefore provide a robust measure of the total stellar mass  -- with uncertainties largely driven by systematics \citep[$\sim0.3$ dex;][]{conroy2013} -- as well as sizes, morphologies, and tidal features\footnote{Sizes, morphologies, and disturbed features (indicating galaxy interactions) can be identified in the rest-NIR up to $z\sim1$ with the current capabilities of \textit{HST}. \red{These}\sout{Size} measurements\red{, however, are known to be sensitive to dust at the short wavelengths typically probed}\sout{at these short wavelengths are known to be adversely affected by dust} \citep[e.g.][]{popping2022}\red{; this uncertainty will be addressed by upcoming observations with \textit{JWST} (\S\,\ref{sec:future})}.  As such, we will not attempt a full overview of these measurements in this review, though they may be discussed in supporting contexts. }. Crucially for (proto-)cluster studies, the near-IR also provides a long wavelength anchor for color selections intended to separate SFG and QG populations.  The most widely used is the rest-frame UVJ color selection \citep{williams2009}, which uses rest J band \red{($\sim1\mu$m)}\sout{($\sim2\mu$m)} to break the degeneracy between stellar age and dust attenuation \citep[e.g.][]{whitaker2013, fumagalli2014, leja2019} with low (10-30$\%$) contamination in quiescent color space from SFGs \citep{belli2017, diaz-garcia2019, schreiber2018}.  From this, the relative SFG and QG fractions and environmental quenching efficiencies can be derived (see \S\,\ref{sec:nir}).

\subsubsection{Mid- to Far-Infrared}

Moving to longer wavelengths, the M/FIR regime is dominated by the reprocessed stellar light emitted in the infrared by small to large dust grains that are pervasive within (and between) (proto-)cluster galaxies (see \citealt{galliano2018} for a review). Within galaxies, dust is primarily heated by recent star formation, producing aromatic emission features in the MIR and a broad continuum peaked at $\sim70-100\mu$m in the FIR \citep{draine2003}. Luminous AGN can additionally generate a hot dust continuum in the N/MIR up to $\sim30\mu$m. Dust-obscured star formation is the dominant component in massive (log $M_{\star}/\Msun\gtrsim10$) galaxies at high redshift, with ultraviolet (UV) and optical emission (direct SF tracers) heavily attenuated \citep{whitaker2017}.  As a result, direct star formation tracers in the UV/optical can severely underestimate the true SFR and dust corrections can have large uncertainties.  In these obscured galaxies, the mid and far-IR near the dust peak provide robust SFR tracers (\citealt[][]{kennicutt2012}, and references therein).

In this review, we will refer to various classes of dusty SFGs (DSFGs): luminous
infrared galaxies (LIRGs, $10^{11}<L_{\rm IR}/\Lsun<10^{12}$), ultra-luminous infrared 
galaxies (ULIRGs; $L_{\rm IR}/\Lsun>10^{12}$) and submillimeter galaxies (SMGs;
$S_{250\mu \rm{m} - 2 \rm{mm}}>1$ mJy, see \citealt{casey2014} for  review). These populations are the most difficult to detect in the UV/optical and can even be challenging into the short wavelength near-infrared \citep[e.g.][]{williams2019, yamaguchi2019, smail2021, manning2022}.  For (proto-)cluster studies, this primarily affects our ability to establish cluster membership and measure SFRs \citep{duc2002, vulcani2010, finn2010, santos2014, ma2015}.  This challenge is strongly amplified in studies at cosmic noon ($z\sim1-3$) where star formation and black hole accretion activity peaks and the majority of star formation is obscured \citep[e.g.][Zavala \& Casey, in prep]{silverman2008, madau2014}.  Obscured AGN likewise are best identified in the MIR \citep[e.g.][]{hickox2018, alberts2020, lyu2022}.

\subsubsection{Far-Infrared to Submillimeter/Millimeter}\label{sec:intro_submm}
 
The cold molecular phase of the ISM is traced by longer-wavelength emission in the FIR-to-submm regime. However, since molecular hydrogen (H$_2$) is symmetric and therefore has no permanent dipole\footnote{H$_2$ can emit radiation through the quadrupole moment, but these transitions have low probabilities and require high excitation energies.}, optically-thin dust continuum and the rotational transitions of carbon monoxide (CO) have become the favored means of observing this gas phase; \red{see reviews by \citealt[][]{solomon2005}, \citealt{carilli2013}, \citealt{tacconi2020}, and \citealt{saintonge2022}}.
Well past the FIR dust peak ($\lambda\gtrsim250\mu$m), dust emission becomes optically-thin and is proportional to the bulk cold \citep[$\sim25$ K;][]{scoville2016} dust mass.  Given the dust temperature, dust-to-gas ratio (DGR), and dust opacity, the gas mass can be derived from this emission; a standard conversion has been calibrated for the molecular gas mass in massive (field) galaxies \citep[e.g.][]{scoville2016, scoville2017}.  This method is particularly efficient at high redshift, given the strong negative K-correction.  The submm is also home to multiple CO transitions; as the next most abundant molecule, CO serves as a robust proxy for cold H$_{2}$ emission\red{, though calibrations may change at low metallicity \citep[e.g.][]{cormier2014}}. 
In the local Universe, CO emission is often observed in conjunction with atomic \textsc{Hi} emission in the radio to track the molecular+atomic gas reservoir.  At higher redshifts ($z\gtrsim0.5$), where galaxies are more gas-rich, ISM conditions are expected to result in the molecular gas dominating over negligible atomic gas (see \citealt{schreiber2020} and references therein).  As molecular gas is the fuel for star formation, quenching processes are expected to act directly to perturb, or possibly even remove, the molecular gas reservoir in galaxies;  as such, it is\sout{the} a key observable in assessing the drivers of galaxy evolution. CO emission lines additionally provide robust spectroscopic redshifts (spec-$z$s), effective for establishing (proto-)cluster membership for dusty galaxies.

\section{Scope, Definitions, and Outline}\label{sec:scope}

This review discusses recent advances in the studies of (typical) galaxy populations in (proto-)clusters $-$ with an emphasis on high-redshift ($z\gtrsim0.5-7$) works $-$ using NIR ($\sim1-5\mu$m), MIR ($\sim5-30\mu$m), and FIR ($\sim30\mu \mathrm{m}-3$mm) observations. The range $\gtrsim500\mu$m  will often interchangeably be referred to as the submillimeter (submm) for historical reasons.  Our focus will be on progress toward understanding galaxy evolution in overdense environments, particularly quenching and the role of quenching mechanisms introduced in \S\,\ref{sec:intro}.  A notable exception: due to the explosion in progress in infrared studies of (proto-)clusters, this review cannot cover all relevant advances and we will leave a full treatment of infrared studies of cluster galaxy sizes, morphologies, and morphological transformation to another review. In addition, we will not cover Brightest Cluster Galaxies (BCGs), an evolutionarily distinct population that deserves separate consideration.  For discussions of BCGs, we refer the reader to e.g., \citealt{delucia2007a, donahue2022, overzier2016} and references therein.

An overview of environment and definitions of galaxy clusters and proto-clusters were given in \S\,\ref{sec:environment}.  Throughout this review, we will characterize clusters by their virial radius, $R_{\rm{vir}}$ ($\equiv R_{200}$, the radius enclosing 200 times the critical density of the Universe at a given redshift) and the corresponding virial mass, $M_{200}$. $R_{500}$ and $M_{500}$ are also commonly used in the literature, where $M_{200}\sim1.4\times M_{500}$\footnote{Assuming an NFW profile with a concentration $c=5$ \citep{white2001}. See \S\,\ref{sec:totallight_c} for an expanded definition of the NFW profile and concentration parameter.}. We will use $M_{500}$ where appropriate given the convention in the literature.  We adopt a concordance cosmology, ($\Omega_{\Lambda}$, $\Omega_{\rm M}$, h) = (0.7, 0.3, 0.7), and a \citealt{kroupa2001} IMF unless otherwise noted.

This review is structured as follows: \S\,\ref{sec:cluster_surveys} begins with a brief review of (proto-)cluster selection using near- and far-infrared surveys to highlight the available and upcoming datasets.  \S\,\ref{sec:nir} reviews current analyses of cluster populations using the near-infrared, covering stellar mass functions 
and quenched fractions and quenching efficiencies 
to $z\sim2$.  In \S\,\ref{sec:fir}, we present the current state of the literature regarding (obscured) star formation in (proto-)cluster galaxies from low to high redshift using M/FIR observations.  
\S\,\ref{sec:agn} diverges from this to give a brief summary of AGN activity in clusters.  Progress on FIR and submillimeter measurements of dust and molecular gas in (proto-)cluster galaxies is presented in \S\,\ref{sec:submm}.  \S\,\ref{sec:totallight} discusses the revival of the ``total light'' stacking technique -- measuring the averaged properties of large (proto-)cluster samples -- through examples in the areas of intracluster dust (ICD), dust in cluster populations, and cluster galaxy concentrations.  A discussion tying the reviewed studies to quenching in cluster galaxies is presented in \S\,\ref{sec:wrap} and a summary of open questions and important upcoming surveys and facilities is presented in \S\,\ref{sec:future}.

\section{Identifying (Proto-)clusters in the Infrared: Current and Future Large Surveys}\label{sec:cluster_surveys}

Statistical samples of galaxy (proto-)clusters $-$ covering a large range in halo mass, dynamical state, and redshift $-$ are necessary for both the use of clusters as probes of cosmology (see \citealt{allen2011} for a review) and as astrophysical laboratories for galaxy evolution \citep[][]{dressler1984, boselli2006}.  (Proto-)cluster selection is done via multiple techniques, such as tracing galaxy populations to identify galaxy overdensities, using rare sources as signposts of massive halos, or observations of the hot gas ($10^7-10^8$ K) of an established ICM.  For the latter, X-ray emission, as a direct observable of the ICM, has been a widely successful tool in building cluster samples; however, surface \red{brightness}\sout{bright} dimming results in this selection being most effective at low to moderate redshifts (see \citealt{rosati2002} for a review).  Optical imaging surveys typically identify overdensities of red early-type galaxies (ETGs) through filters that bracket the 4000\AA\ break \citep[Red Sequence (RS) selection, e.g.][]{gladders2000,gladders2005}; however, this selection is sensitive to projection effects, favors evolved clusters, lacks a direct halo mass proxy, and is limited to lower redshifts.  These drawbacks can be mitigated by incorporating imaging in the near-infrared, which can extend the selection of red galaxies to higher redshifts and/or be used to derive robust photometric redshifts (photo-$z$s) for both optically blue and red galaxies.   At longer wavelengths, the Sunyaev-Zel'dovich Effect \citep[SZ;][]{sunyaev1980} provides an indirect detection of the ICM, while at higher redshifts the nature of rare populations such as luminous IR sources can be used to select massive halos.  In this section, we give a broad overview of clusters selected using near-infrared and submillimeter imaging, including existing and future large cluster surveys.  This is followed by a discussion of infrared selection of proto-clusters. 

\subsection{Cluster Selection in the Near-Infrared}\label{sec:nir_selection}

The introduction of sensitive, wide-field near-infrared imaging surveys\footnote{For example, the NOAO Deep Wide-Field Survey \citep[NDWFS;][]{jannuzi1999}, the NEWFIRM Medium-Band Survey \citep[NMBS;][]{whitaker2011}, the IRAC Shallow Survey \citep[ISS;][]{eisenhardt2004}, the \textit{Spitzer} Deep Wide-field Survey \citep[SDWFS;][]{ashby2009}, and the \textit{Spitzer} Wide-area InfraRed Extragalactic survey  \citep[SWIRE;][]{lonsdale2003}} has greatly expanded the use of selecting clusters as NIR overdensities, with pioneering work pushing to $z>1$ \citep{stanford2006, brodwin2006, vanbreukelen2007, zatloukal2007, krick2008, eisenhardt2008, muzzin2008}.  This generally takes a few forms: optical-NIR colors can be used to span the $4000$\AA\ break at $z\gtrsim1$ \citep[][]{wilson2009,muzzin2009} expanding Red Sequence selection \citep{gladders2000} to higher redshifts, while NIR-only color cuts \citep{papovich2008} or two color optical-NIR cuts \citep[``Stellar Bump Sequence'';][]{muzzin2013} can isolate higher redshift overdensities.  With multi-band optical+NIR imaging, overdensities are now also often identified in photometric redshift space, using full photo-$z$ probability distribution functions to identify clusters and create cluster member catalogs \citep[e.g.][]{eisenhardt2008, brodwin2013}.

\begin{table}[t]
	\centering
	\caption{$-$ A(n Incomplete) List of Large Cluster Surveys Incorporating Near-Infrared Observations (\S\,\ref{sec:nir_selection})}
	\label{tbl:nir_selection}
	\resizebox{\columnwidth}{!}{ 
	\begin{tabular}{lccccccc} 
	    \hline
		\hline
Survey & Method & Cluster & Confirmed & Area & Redshift & log $M_{500}$ & References \\
 &  & Candidates & Clusters  & [deg$^2$] & (median) & [$\Msun$] (median) &  \\
 \hline
 ISCS/IDCS & Photo-$z$ Overdensities & $>300$ & $>120$ & 8.5 & $0.1-2$ & (13.8) & E08, S12 \\
 
 SHELA & RS & 1,082 & $-$ & 24 & $0.5-1.2^{a}$ & 13.9$^{a}$ & P16, F21 \\
 
 SpARCS & RS & $>200$ & $>10$ & 42 & 0.6-1.5 & $-$ & G00, G05, \red{W09, M09, M12} \\
 
 SSDF & Color Selection & 279 & $-$ & 94 & $>1.3$ & 14.1 & R14 \\
 
 HSC + unWISE & \makecell{Overdensities around \\ BGC Candidates} & 21,661$^b$ & $-$ & 800 & 0.1-2 & $\geq13.8$ & WH21 \\
 
 DES + unWISE & \makecell{Overdensities around \\ BGC Candidates} & 151,244$^c$ & $-$ & 5,000 & 0.1-1.5 (0.7) & $-$ & WH22 \\
 
 MaDCoWS & Color Selection & 2,683 & 38 & $10,000$ & $0.7-1.5$ (1.06) & (14.2) & G19 \\
 
 SDSS+\textit{WISE} & \makecell{Overdensities around \\ BGC Candidates} & 1,959 & $-$ & 10,000 & 0.7-1 & $>14.4$ & WH18 \\
 
 2MASS+\textit{WISE} & \makecell{Overdensities around \\ BGC Candidates} & 47,600$^d$ & $-$ & 28,000 &  $0.025-0.3$ & $\gtrsim14.5$ & W18\\
 
\hline
\multicolumn{8}{c}{\underline{Projected}} \\
\textit{Roman} & Photo-$z$ Overdensities & 40,000 & $-$ & 2,200 & $<3$ & $>14$ & S15 \\
MaDCoWS2 & Color Selection & $-$ & $-$ & $\gtrsim10,000$ & $\sim0.5-2$ & $-$ & T., in prep.\\
\textit{Euclid} & Photo-$z$ Overdensities & 2,000,000 & $-$ & 15,000 & $<2$ & $>13.8$ &  \makecell{S16, A17,\\R18, E19} \\

\hline
\multicolumn{8}{p{1.3\columnwidth}}{Note $-$ This list of cluster surveys incorporating NIR observations is not exhaustive and the surveys listed are not mutually exclusive. Surveys are listed in order of increasing area covered. A dash indicates information not readily available in the literature. Redshifts and masses are given as ranges and/or medians, the latter indicated by parentheses, unless otherwise noted. $^a$Redshift range and average mass for the 70 highest richness cluster candidates in SHELA, stacked in SZ \citep{fuzia2021}. $^b$15,614 previously unknown \citep{wen2021}. $^c$76,826 previously unknown \citep{wen2022}. $^d$26,125 previously unknown \citep{wen2018a}. References: A17: \citealp{ascaso2017}; E08: \citealp{eisenhardt2008}; E19: \citealp{euclidcollaboration2019}; F21: \citealp{fuzia2021}; G00: \citealp{gladders2000}; G05: \citealp{gladders2005}; G19: \citealp{gonzalez2019}; \red{M09: \citealp{muzzin2009}; M12: \citealp{muzzin2012};} P16: \citealp{papovich2016}; R14: \citealp{rettura2014}; R18: \citealp{rettura2018}; S12: \citealp{stanford2012}; S15: \citealp{spergel2015}; S16: \citealp{sartoris2016}; T, in prep: Thongkham et al, in prep.; \red{W09: \citealp{wilson2009}}; W18: \citealp{wen2018a}; WH18: \citealp{wen2018}; WH21: \citealp{wen2021}; WH22: \citealp{wen2022}} \\
	\end{tabular}
	}
\end{table}

\begin{SCfigure}[0.8][!htb]
    \parbox{0.5\columnwidth}{\centering\includegraphics[width=0.5\columnwidth]{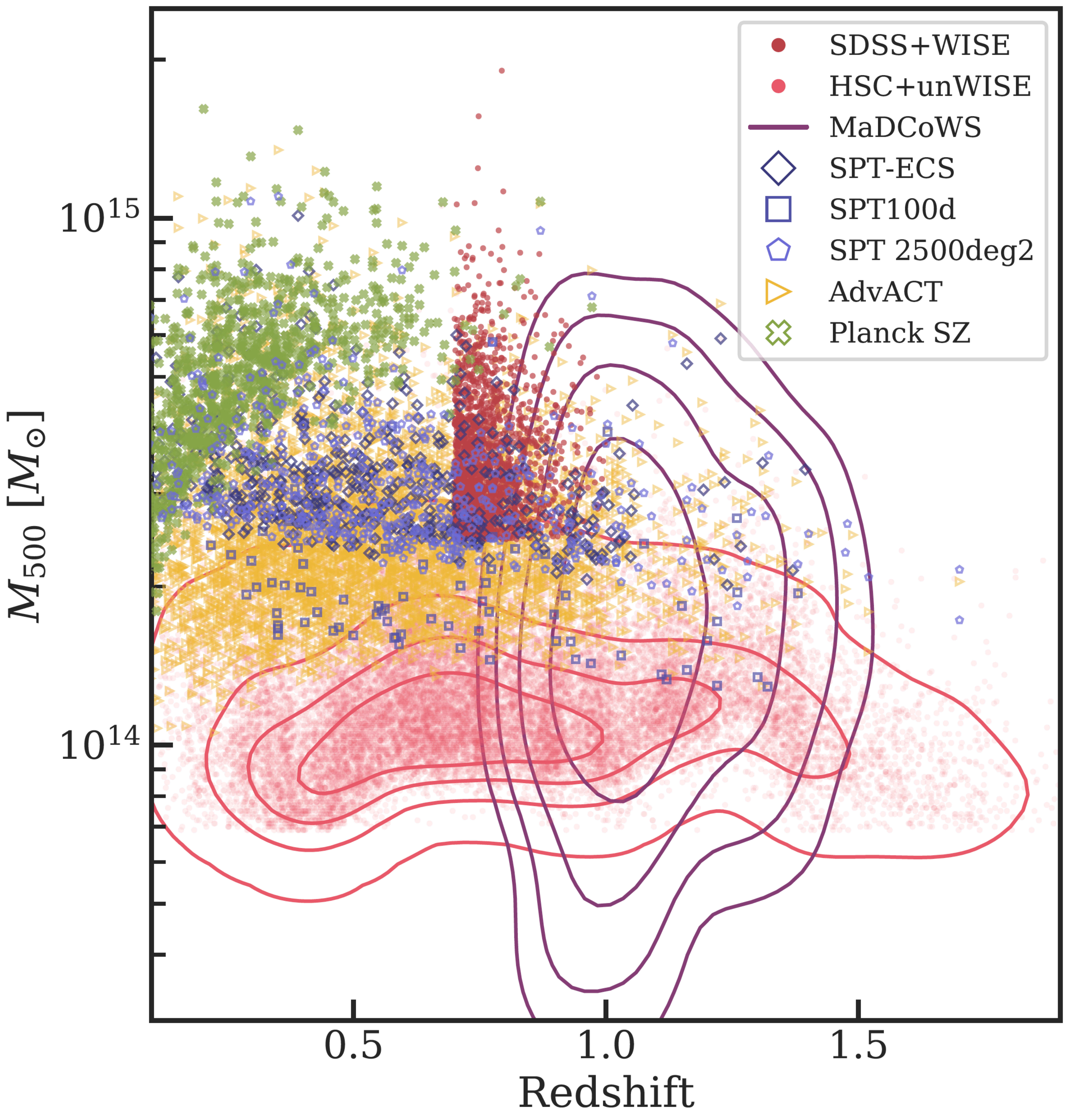}}
    \caption{ $-$ Halo mass and redshift distribution of cluster candidates identified in large ($>100$ deg$^2$) NIR-based (Table~\ref{tbl:nir_selection}) and SZ surveys (Table~\ref{tbl:submm_selection}).
    For the NIR, halo masses are largely derived from mass-richness scaling relations.  Redshifts are largely photo-$z$ based. Infrared-based cluster surveys have grown by orders of magnitude in the last several years and now span $0<z<2$ and log $M_{500}/\Msun\sim13.5-15$.  Future surveys are projected to increase this by orders of magnitude.}
    \label{fig:cluster_selection}
\end{SCfigure}

Notably, the rest NIR traces stellar mass,
allowing for mass-selected cluster catalogs. NIR selection techniques further benefit from a negative K-correction, with the observed flux density at 4.5$\mu$m nearly redshift-independent over $z\sim0.7-2.5$ \citep{eisenhardt2008, wylezalek2013}. NIR selection has successfully identified clusters out to $z\sim2$ (Figure~\ref{fig:cluster_selection}), though, like X-ray, its effectiveness starts to drop around $z\sim1-1.5$ given the current sensitivities of near-infrared surveys and the need for good coverage of stellar features.
Unlike X-ray, NIR selection may identify more disturbed and young, actively accreting clusters \citep{willis2018}.  As with optical selection, the main limitations on this method are a lack of direct halo mass proxy and projection effects \citep[e.g.][]{costanzi2019}, which can result in the false detection or mis-characterization of a cluster.
For optically or NIR-selected clusters, the halo mass-richness relation \citep{girardi2000, lin2003} $-$ where richness is the number of cluster members above some magnitude $-$ is used to infer the halo mass from the galaxy component.  Richness estimators have been developed with low scatter \citep[$\sim$0.1-0.2 dex;][]{andreon2015,rozo2015,old2015}, though contamination or incompleteness in the cluster membership can still bias the derived halo mass \citep{wojtak2018}.

Here we list some notable NIR-based cluster surveys (Table~\ref{tbl:nir_selection}). The \textit{Spitzer} Adaptation of the Red-Sequence Cluster Survey \citep[SpARCS;][]{muzzin2009, wilson2009, demarco2010} identified clusters using $z^{\prime}-3.6\mu$m color selection at $z>1$ over 42 deg$^2$ with multi-wavelength coverage and was the basis of follow-up optical spectroscopic campaigns for the well-studied Gemini Cluster Astrophysics Spectroscopic Survey \citep[GCLASS; 10 clusters at $0.85<z<1.34$;][]{muzzin2012} and Gemini Observations of Galaxies in Rich Early Environments \citep[GOGREEN; 21 groups and clusters at $1<z<1.5$;][]{balogh2017} samples, which will be discussed extensively in \S\,\ref{sec:nir}. The IRAC Shallow and Distant Cluster Surveys \citep[ISCS/IDCS;][]{eisenhardt2008, stanford2012} used extensive multi-wavelength coverage in the 8.5 deg$^2$ Bo\"{o}tes field to identify $>300$ log $M_{200}/\Msun\sim13.8$ cluster candidates from $0.1<z<2$ as overdensities using robust photometric redshifts \citep[][]{brodwin2006}. Spectroscopy from the AGN and Galaxy Evolution Survey \citep[AGES;][]{kochanek2012} and targeted follow-up \citep{stanford2005, brodwin2006, brodwin2011, brodwin2013, elston2006, eisenhardt2008, zeimann2013} were used to confirm $>120$ clusters in this sample and halo mass measurements were made using X-ray, SZ, and weak lensing as well as statistical arguments \citep{brodwin2007,jee2011, brodwin2011,brodwin2012,stanford2012,lin2013,alberts2014,brodwin2016}. Substantial M/FIR follow-up was obtained for this sample as well, adding to the available X-ray to NIR photometry \citep{chung2014}; this survey will be discussed extensively in \S\,\ref{sec:fir}.  We note two additional surveys covering up to $\sim100$ deg$^2$: $\sim1,000$ group and low-mass cluster candidates were identified using the Red Sequence cluster finder redMapper \citep{rykoff2014} in the 24 deg$^2$ \textit{Spitzer}-HETDEX Exploratory Large Area survey \citep[SHELA;][]{papovich2016} and 279 cluster candidates at $z>1.3$ were identified using color selection in the 94 deg$^2$ \textit{Spitzer} South Pole Telescope Deep Field survey \citep[SSDF;][]{rettura2014}.

More recently, substantially larger cluster surveys have been assembled taking advantage of wide-field optical surveys combined with \textit{WISE} all-sky coverage, identifying rare, massive systems over thousands of square degrees\footnote{Referenced Surveys: the Massive and Distant Clusters of \textit{WISE} Survey (MaDCoWS; \citep{gonzalez2019}, the Panoramic Survey Telescope and Rapid Response System \citep[Pan-STARRS;][]{chambers2016}), SuperCOSMOS \citep{peacock2016}, the Dark Energy Camera Legacy Survey (DECaLS; PI: D. Schlegel and A. Dey), CatWISE2020 \citep{marocco2021}, the Two Micron All Sky Survey \citep[2MASS;][]{skrutskie2006}), the Sloan Digital Sky Survey \citep[SDSS;][]{york2000}, unWISE \citep{schlafly2019}, the Hyper-Surprime Cam-Subaru Strategic Program \citep[HSC-SSP;][]{aihara2018}, and the Dark Energy Survey \citep[DES;][]{abbott2018}}.  The Massive and Distant Clusters of \textit{WISE} Survey \citep[MaDCoWS;][]{gonzalez2019} combined Pan-STARRS and SuperCOSMOS with \textit{WISE} over $\sim10,000$ deg$^2$ to identify 2,433 cluster candidates at $0.7<z<1.5$ (Table~\ref{tbl:nir_selection}).  Targeted \textit{Spitzer} follow-up was obtained for 1,723 \red{of these candidates}, from which photometric redshifts and richnesses were measured. Halo masses derived from SZ were used to calibrate the mass-richness relation. A follow-up catalog, MaDCoWS2, will expand the redshift range to $z\sim2$ using deeper DeCaLS+CatWISE2020 imaging over again $\sim10,000$ deg$^2$ (Thongkham, in prep.). Similarly, using 2MASS+SuperCOSMOS+\textit{WISE} \citep[][]{wen2018a}, SDSS+\textit{WISE} \citep[][]{wen2018}, HSC-SSP+unWISE \citep[][]{wen2021} and DES+unWISE \citep[][]{wen2022}, over 150,000 cluster candidates were identified over $0.025<z<2$ by searching for photo-$z$ overdensities around massive galaxies, presumed to be current or future BCGs.  Mass-richness relations for these surveys were calibrated using overlap with X-ray and SZ surveys.  Figure~\ref{fig:cluster_selection} shows the $M_{500}$-redshift distribution of available large catalogs, which span nearly two orders of magnitude in halo mass.  

Future faculties are projected to increase NIR-selected cluster candidates by another order of magnitude (Table~\ref{tbl:nir_selection}), in conjunction with wide-field optical surveys from, i.e. DES and the Vera C. Rubin Observatory \citep{ivezic2019}. \textit{Euclid} \citep{laureijs2011}, a wide-field UV to NIR surveyor, will image 15,000 deg$^2$ in YJH bands to moderate depths and is anticipated to identify two million cluster candidates up to $z\sim2$ \citep{sartoris2016, ascaso2017, rettura2018, euclidcollaboration2019}. The \textit{Nancy Grace Roman Space Telescope} will survey 2,200 deg$^2$ with deep (H=26.5AB) NIR imaging and slitless spectroscopy, pushing the limits of massive (log $M_{200}/\Msun\gtrsim14$) (proto)-cluster detection to $z\sim3$ \citep{spergel2015}.  

\subsection{Cluster Selection via the SZ Effect}\label{sec:sz_selection}

Similar to X-ray selection, the hot ICM of galaxy clusters can be detected, albeit indirectly, though its interaction with the Cosmic Microwave Background (CMB) via inverse Compton-scattering, termed the thermal\footnote{The SZ effect is comprised of two components: the thermal component due to the random thermal motions of electrons and a kinetic component from the bulk gas motion relative to the CMB.  For galaxy clusters, the thermal component dominates and as such as we will not discuss the kinetic SZ.} SZ Effect (see \citealt{carlstrom2002} for a review).  This interaction causes a distortion in the CMB, suppressing the spectrum below (rest) 218 GHz ($\sim1.4$ mm) and enhancing it above. The magnitude of this effect relative to the CMB is constant with redshift, meaning SZ selection of clusters has the advantage of being largely redshift independent, yielding roughly \citep[within a factor of 2-3;][]{carlstrom2002}  mass-limited cluster catalogs \citep{motl2005}.  As such, it has been proposed as a promising avenue for finding large samples of high-redshift ($z>1.5$) clusters; however, the exact cluster counts will depend strongly on instrument resolution and the nuances in the evolving relationship between the SZ observable, $Y_{500}$, and $M_{500}$ \citep[e.g.][]{andersson2011, henden2019}.  The latter arises from the SZ signal's proportionality to the ICM column density (weighted by temperature); this provides a halo mass proxy with low scatter, relatively (but not entirely) insensitive to the detailed physics of heating and cooling processes as well as structural asymmetries \citep{motl2005}.
This capability plays an important role in the calibration of mass-richness relations for the current and future optical/NIR surveys discussed in the previous section \citep{gonzalez2019, bleem2020}.  

\begin{table}[t]
	\centering
	\caption{ $-$ A(n Incomplete) List of (Proto-)Cluster Surveys Incorporating FIR and Submm Observations (\S\,\ref{sec:protocluster_selection})}
	\label{tbl:submm_selection}
	\resizebox{\columnwidth}{!}{
	\begin{tabular}{lccccccc}
	    \hline
		\hline
Survey & Method &  Cluster & Confirmed & Area & Redshift & log $M_{500}$ & References \\
 &  & Candidates  & Clusters & [deg$^2$] & (median) & [$\Msun$] (median) &  \\
		\hline
\multicolumn{8}{c}{\underline{Clusters}} \\

SPTpol & SZ & 89 & 81 & 500 & (0.6) & (14.6) & B14, H20 \\

SPT-SZ 2500 deg$^2$ & SZ & 677 & 516 & 2,500 & (0.55) & (14.4) &  B15 \\

SPT-ECS & SZ & 448 & 408 & 2,770 & (0.49) & (14.8) & B20 \\

AdvACT DR5 & SZ & 4,195 & 4,195 & 13,211 & $0.04-1.91$ (0.52) & $>14.6$ & H18, H21 \\

\textit{Planck} PSZ1/PSZ2 & SZ & 1,653  & 1,203 & 34,487 & $<1$ & 14.5 &  P14, P16a \\
\hline
\multicolumn{8}{c}{\underline{Proto-clusters}} \\

\textit{Planck} PHz & \makecell{Color \\Selection} & 2,151 & $-$ & 10,725 & ($\sim2.5^{a}$) & $-$ & P16b  \\

\hline
\multicolumn{8}{c}{\underline{Projected}} \\
SPT-3G & SZ & $\sim10,000$ & $-$ & 1,500 & (0.7) & ($>14.1$) & B14, S22\\

Simons Obs & SZ & 26,445 & $-$ & 16,500 & (0.7) & (14.3) & A19a, R22 \\

CCAT-prime & SZ & 16,000 & $-$ & $\sim20,000$ & $<2.5$ & $>14$ & C21 \\

CMB-HD & SZ & 514,530 & $-$ & 20,600 & (0.9) & (13.8) &  S19, R22 \\

CMB-S4 Wide & SZ & 107,747 & $-$ & 27,600 & (0.8) & (14.2) & A19b, R22 \\

PICO & SZ & \makecell{$150,000-$\\200,000} & $-$ & All-Sky & \makecell{$<3-4.5$} & $>14.3$ & H19 \\
		\hline
\multicolumn{8}{p{\columnwidth}}{{\bf Note} $-$ This list of cluster surveys incorporating submm observations is not exhaustive and the surveys listed are not necessarily mutually exclusive. Surveys are listed in order of increasing area covered. A dash indicates information not readily available in the literature. Confirmed clusters refers to confirmation via another wavelength regime and/or via spectroscopy. Redshifts and masses are given as ranges and/or medians, the latter indicated by parentheses. $^a$Assuming T$_{\rm dust}=35$ K \citep{planckcollaboration2016d}. {\bf References:} A19a: \citealp{ade2019}; A19b: \citealp{abazajian2019}; B14: \citealp{benson2014}; B15: \citealp{bleem2015}; B20: \citealp{bleem2020}; C21: \citealp{ccatcollaboration2021}; H18: \citep{hilton2018}; H19: \citealp{hanany2019}; H20: \citealp{huang2020}; H21: \citealp{hilton2021}; P14: \citealp{planckcollaboration2014}; P16a: \citealp{planckcollaboration2016c}; P16b: \citealp{planckcollaboration2016d}; R22: \citealp{raghunathan2022}; S19: \citealp{sehgal2019}; S22: \citealp{sobrin2022}.}
	\end{tabular}
	}
\end{table}

The last two decades have seen great strides in SZ surveys, which have now identified over 6,000 cluster candidates. Large SZ surveys have been conducted by the South Pole Telescope \citep{carlstrom2011}, ACT \citep{swetz2011} and \textit{Planck} \citep{planckcollaboration2014};
the individual surveys and their references are listed in Table~\ref{tbl:submm_selection} and the redshift and mass distributions of these surveys can be seen in Figure~\ref{fig:cluster_selection}.  Current SZ surveys identify fairly massive clusters at $z<1.5$ due to sensitivity and resolution limitations. The next major step forward will come from SPT-3G, which began a multi-year 1,500 deg$^2$ survey in 2018 and is expected to identify up to 10,000 cluster candidates above log $M_{500}/\Msun\gtrsim14$ at a median redshift $z\sim0.7$ \citep{sobrin2022, raghunathan2022}. Future improvements in sensitivity/mapping speed via the Simons Observatory \citep[][]{ade2019}, CMB-S4  \citep{abazajian2019}, and CMB-HD \citep{sehgal2019} experiments are expected to increase the number of known SZ clusters by orders of magnitude.  Projections based on expected improvements in noise and cluster masking techniques (\citealt{raghunathan2022}) predict \red{identification of} greater than $600,000$ SZ clusters with signal-to-noise (S/N) $>5$, with $>80,000$ at $z=1.5-2$ and $>20,000$ at $z>2$ over $>65\%$ of the sky, assuming a \citealt{tinker2008} halo mass function (Table~\ref{tbl:submm_selection}).  Upcoming facilities such as the ground-based telescope CCAT-prime \citep{ccatcollaboration2021} and the Probe of Inflation and Cosmic Origins (PICO) satellite \citep{hanany2019} are additionally projected to use SZ to great effect in (proto-)cluster selection to high redshift.  In addition to 150,000 SZ clusters to $z\sim3$, PICO's all-sky survey is anticipated to identify 50,000 proto-clusters to $z\sim4.5$.  

\subsection{Proto-cluster Selection in the Infrared}\label{sec:protocluster_selection}

Though proto-clusters can in principle be identified even as modest density contrasts above the general dark matter distribution out to high redshift ($z\sim6$; see \citealt{overzier2016} for a review), their nature presents several challenges to detection.  Proto-clusters are rare overdensities likely located at the conjunctions of filaments \citep[e.g.][]{boylan-kolchin2009, overzier2009, chiang2013} and can span 10-30$^{\prime}$ on the sky.  By definition, a proto-cluster will collapse into a cluster by $z=0$; however, in its pre-collapsed state, proto-clusters often have not yet heated their ICM sufficiently to detect via X-ray or SZ and have not yet established a telltale RS population. Establishing overdensities of ``normal'' galaxies in this epoch ($z>2$) requires \red{very deep surveys over 10 cMpc scales \citep{lovell2018}}\sout{very wide, very deep surveys}, and preferably spectroscopic or narrow band surveys which mitigate the risk of mistaking structures overlapping in the line-of-sight for bona fide proto-clusters. However, ``normal'' galaxies over such large scales manifest a small density contrast and the expense of such surveys can make them better suited to follow-up of likely proto-cluster candidates, which can be identified using alternative rare populations as biased\footnote{Galaxy bias is the statistical relation between the spatial distribution of a galaxy population and the underlying dark matter density field.  Bias is strongly dependent on the galaxy population being observed.} tracers. 

\subsubsection{Luminous DSFGs, obscured AGN, and ultra-massive galaxies as signposts of proto-clusters}\label{sec:dsfg_signposts}

Rare galaxy populations which preferentially inhabit the most massive halos at a given epoch, such as luminous radio galaxies \citep[e.g.][]{venemans2002, miley2008, wylezalek2013}, have long been used as relatively inexpensive tracers to identify proto-cluster candidates \citep{overzier2016}.  In the IR, DSFGs, luminous obscured AGN \citep{eisenhardt2012, wu2012, lonsdale2015, jones2017}, and, recently, Ultra-Massive Galaxies (UMGs) have been explored as such populations.  First, we will discuss DSFGs (often identified in the submm and called SMGs), which are highly star-forming (SFR$\,>500\,\Msun$ yr$^{-1}$) galaxies thought to be the progenitors of the most massive ellipticals \citep[e.g.][]{hopkins2008, toft2014, stach2021, garcia-vergara2020}, which likely formed in bursts \red{of star formation} at $z>2-3$ \citep[e.g.][]{nelan2005, delucia2006}.  Whether SMGs live in the most massive halos has been a controversial topic, however, with early studies suggesting that they live in a range of environments \citep[e.g.][]{chapman2009}.  Additional arguments have been made that the short-lived nature of the burst phase \citep[$\sim100$ Myr;][]{casey2014} and early quenching due to downsizing make SMGs a poor tracer of proto-clusters due to large scatter in the number of SMGs for a fixed dark matter overdensity \citep[][\red{but see \citep{hall2018}}]{hayward2013a, miller2015, danielson2017}.

There are two lines of observational evidence, however, that suggest DSFGs/SMGs are useful signposts of proto-clusters (see \citealt[][]{chiang2013} for arguments from the theoretical perspective). First, we can consider clustering, which links galaxy populations to their dark matter halos \citep{cole1989, mo1996, cooray2002}. The clustering of SMGs has historically been difficult to measure due to projection effects \citep[e.g.][]{williams2011, hickox2012} and the large beamsizes ($>15^{\prime\prime}$) of single-dish submm telescopes, which can blend multiple submm sources (see e.g. \citealt{casey2014, hodge2020} for reviews). Recently, however, studies with high-resolution ALMA imaging \citep[e.g.][]{garcia-vergara2020, stach2021} have mitigated the blending issue and found that the clustering of the most luminous SMGs ($S_{870}\gtrsim6$ mJy) is consistent with SMGs inhabiting very massive halos at $z>2$ \red{\citep[see also][for clustering of DSFGs around high-redshift, massive quasars, a likely tracer of massive halos]{hall2018}}. This is supported by emerging evidence of extreme massive, IR-luminous SFGs at very high redshift, which can likely only form in the largest halos in such an early epoch \citep[$z=6.9$;][]{marrone2018}.  This makes very luminous SMGs promising candidates as signposts for the most massive halos at high-$z$ \citep{blain2004}, expected to collapse into log $M_{\rm halo}/\Msun\sim15$ clusters at $z\sim0$.  The fate of halos hosting moderately luminous SMGs is less clear, however.

Second, early submm surveys quickly discovered significant DSFG overdensities in known structures such as a $z=1.99$ proto-cluster in GOODS-N \citep{blain2004, chapman2009} and SSA22 at $z=3.09$ \citep{chapman2005, geach2005, tamura2009}.  These and three other such overdensities (a proto-cluster in COSMOS\footnote{Cosmic Evolution Survey \citep[COSMOS;][]{scoville2007}} at $z=2.1$, MRC 1138-256 at $z=2.16$, and PCL2001 at $z=2.47$) discovered in $\sim3$ deg$^2$ of submm surveys, were found to have $>5-10$ DSFGs over 10-30$^{\prime}$ scales (\citealt[][]{casey2015} and references therein).  Assuming a short lifetime of 100 Myr, \citealt{casey2016} showed that the probability of randomly observing $>5$ ($>10$) rare sources over a proto-cluster volume of $10^{4}$ cMpc$^3$ is $<6\%$ ($<0.01\%$).  
They further showed that given the 3 deg$^2$ area of the submm surveys, these proto-clusters match the number density of massive clusters at $z=0$, albeit with large uncertainties.  These observations and arguments support specifically that overdensities of $>5$ DSFGs are robust signposts of proto-clusters.
Subsequently, DSFG overdensities in $>20$ proto-clusters have been identified and characterized in the literature; an incomplete list including their redshift, observing window, total SFR, projected volume, and projected halo mass is shown in Table~\ref{tbl:protoclusters}. FIR/submm surveys additionally have generated large catalogs of proto-cluster candidates identified as DSFG overdensities \citep[see the next section, e.g.][]{clements2014, planckcollaboration2015, planckcollaboration2016b, planckcollaboration2016d}. The nature of these proto-clusters will be discussed further in Section~\ref{sec:protoclusters}.

Beyond DSFGs, there are two populations identified using infrared observations that are gaining momentum as signposts for proto-clusters.  The first is extreme, hyper-luminous obscured AGN, the so-called hot dust-obscured galaxies  \citep[Hot DOGs;][]{eisenhardt2012, wu2012} and other \textit{WISE}-selected bright populations\footnote{Hot DOGs are selected as dropouts in the first two \textit{WISE} filters at 3.4 and 4.6$\mu$m \citep{eisenhardt2008, wu2012}.  Similar sources can be chosen in \textit{WISE} color-color space combined with radio detections \citep{lonsdale2015}.} \citep{lonsdale2015}.  These rare sources (there are $\sim1,000$ Hot DOGs over the full sky) appear to be signposts of massive halos, with Hot DOGs observed to reside in overdense regions as traced by Ly $\alpha$ \citep{bridge2013} and infrared/submm \citep{jones2014, jones2015, jones2017, assef2015, penney2019}.  Hot DOGs may be powered by prodigious merger activity \citep{fan2016} and have low molecular gas reserves \citep{penney2020}. This population, which can be radio-quiet \citep{jones2017, penney2020}, presents an interesting counterpoint to overdense environments which host radio-loud AGN \citep[e.g.][]{wylezalek2013}, which are associated with massive jets and thus strong feedback.  However, the populations of these proto-cluster candidates have yet to be studied in detail.

The second population is log $M_{\star}/\Msun>11$ galaxies at $z>3$, termed Ultra-Massive Galaxies.  UMGs at high redshift may be the intermediate step between DSFGs and $z=0$ cluster ellipticals and/or BCGs and are expected to inhabit massive halos given their extreme mass build-up by $z\sim3$.  Identifying these galaxies requires wide-field NIR capabilities to obtain the rest-frame optical and measure a stellar mass, with NIR spectroscopic follow-up to confirm \citep[e.g.][]{capak2011, marsan2015, marsan2017, glazebrook2017, schreiber2018, tanaka2019, valentino2020, saracco2020, ando2020}.  \citealt{mcconachie2022} recently presented spectroscopically-confirmed overdensities around two UMGs at $z\sim3.4$ from the Massive Ancient Galaxies At $z > 3$ NEar-infrared (MAGAZ3NE) survey \citep{forrest2020, forrest2020a}, which will be discussed further in \S\,\ref{sec:protoclusters}.  Additional investigation of the environments around UMGs is needed to confirm their utility as a proto-cluster signpost.  

\begin{table}[H]
\centering
\caption{\raggedright $-$ An Incomplete List of Proto-clusters in the Literature with DSFG Overdensities \label{tbl:protoclusters}}
\scriptsize
\resizebox{0.92\columnwidth}{!}{
	\begin{tabular}{lcccccccc}
		\toprule
		\toprule
		\textnormal{Name} & 
		\textnormal{Redshift} & 
		\textnormal{$N^{\rm DSFG}_{spec-z}$} & 
		\textnormal{Observing} &
		\textnormal{$\Sigma$SFR$^b$} & 
		\textnormal{ Volume$^c$} & 
		\textnormal{log $M^z_{200}$} & 
		\textnormal{log $M^{z=0}_{200}$} & 
		\textnormal{References}\\
		\textnormal{} & 
		\textnormal{} & 
		\textnormal{} & 
		\textnormal{Window$^a$} &
		\textnormal{[$\Msun$ yr$^{-1}$]} & 
		\textnormal{[cMpc$^3$]} & 
		\textnormal{[$\Msun$]} & 
		\textnormal{[$\Msun$]} & 
		\textnormal{}\\
		\midrule
GOODS-N proto-cluster & 1.99 & 6 & 10x10 & $2,600\pm300$ & 9,000 & $13.8\pm0.2$ & $\gtrsim14.5-15$ & \makecell[lt]{B04,C09,{\bf C16}}\nocite{blain2004, chapman2009, casey2016} 
\\
COSMOS proto-cluster & 2.10 & 8 & 8x20 & $5,300\pm600$ & 15,000 & $14.2\pm0.3$ & $\gtrsim15$ & \makecell[lt]{S12,Y14,H16,\\{\bf C16},Z19} \nocite{spitler2012, yuan2014, hunghongzhaoling2016, casey2016, zavala2019} 
\\ 
MRC 1138-256 (PKS1138) & 2.16 & 5 & 6x9 & $2,200\pm500$ & 8,000 & $\sim14$ & $\gtrsim15$ & \makecell[lt]{K00,P00,K11,\\V13,R14,D14,\\{\bf S14},{\bf C16},E16,\\E18, Z18,T19,\\J21}\nocite{kurk2000, pentericci2000, kuiper2011, valtchanov2013, rigby2014, dannerbauer2014, shimakawa2014, casey2016, zeballos2018, tadaki2019, jin2021} 
\\
PHz G237.0+42.5 & 2.16 & 4 & 10x11 & $1,485\pm71$ & 18,500 & $\sim14$ & $\sim15$ & \makecell[lt]{K21a,{\bf P21}} \nocite{koyama2021, polletta2021}
\\
HELAISS02 (core) & 2.171 & 4 & $\pi 0.25^2$ & $1,510\pm170$ & $-$ & $-$ & $-$ & {\bf G19} \nocite{gomez-guijarro2019}
\\
2QZCluster (core) & 2.2 & 7 & $\pi 1^2$ & 1,000 & 138 & $-$ & $-$ & {\bf K16} \nocite{kato2016}
\\
BOSS1244 (core) & 2.24 & 0 & $\pi 2^2$ & $6,720$ & \makecell{2,000} & $-$ & $-$ & {\bf Z22} \nocite{zhang2022}
\\
BOSS1542 (core) & 2.24 & 0 & $\pi 2^2$ & $6,300$ & 2,000 & $-$ & $-$ & {\bf Z22} \nocite{zhang2022}
\\
HS1700+64$^d$ & 2.3 & 4 & \makecell[t]{8x8\\$\pi1.5^2$ (core)} & \makecell[t]{$2,100\pm500$\\$4,900\pm1,200$ (core)} & \makecell[t]{$6,900$\\130 (core)} & $\sim14$ & $\gtrsim15$ & \makecell[lt]{Ch15, K16,{\bf L19},\\{\bf H19}} \nocite{kato2016, lacaille2019}
\\
PCL1002$^e$ & 2.47 & 7 & 14x14 & $4,500\pm500$ & 15,000 & $\gtrsim13.5$ & $\gtrsim14.5-15$ & \makecell[lt]{D15,C15a,{\bf C15b},\\{\bf C16},Z19,{\bf C21}} \nocite{casey2015, diener2015, chiang2015, casey2016, zavala2019, champagne2021} 
\\
HXMM20 (core) & 2.602 & 5 & $\pi 0.13^2$ & $1,700\pm200$ & $-$ & $-$ & $-$ & {\bf G19} \nocite{gomez-guijarro2019}
\\
HS1549+19$^d$ & 2.85 & 4 & \makecell[t]{50\\$\pi1.5^2$ (core)} & \makecell[t]{$2,300\pm500$\\$12,500\pm2,800$ (core)} & \makecell[t]{10,600\\240 (core)} &  $\sim14$ & $\gtrsim15$ & {\bf L19} \nocite{lacaille2019}
\\
SSA22 & 3.09 & 12 & 20x30 & $5,700\pm800$ & 21,000 & $13.9\pm0.2$ & $\gtrsim15$ & \makecell[lt]{S98,S00,H04,\\C05,G05,T09,\\L09,U12,K13,\\U14,U15,K15,\\A16,K16,{\bf C16,}, \\U17, K21b} \nocite{steidel1998, steidel2000, hayashino2004, chapman2005, geach2005, tamura2009, lehmer2009, uchimoto2012, kubo2013, kubo2015, alexander2016, umehata2014, umehata2015, umehata2017, kato2016, casey2016}
\\
SPT2018-45 (core) & 3.2 & 0 & $\pi4^2$ & 9,200 & 2,000 & $-$ & $-$ & {\bf W21} 
\\
SPT0303-59 (core) & 3.3 & 0 & $\pi4^2$ & 15,700 & 2,050 & $-$ & $-$ & {\bf W21}
\\
SPT0457-49 (core) & 3.988 & 0 & $\pi4^2$ & 7,800 & 2,600 & $-$ & $-$ & {\bf W21}
\\
Distant Red Core (core) & 4.002 & 10 & 0.6x0.7 & 6,500 & 877 & $13.7\pm0.2$ & $\gtrsim15$ & \makecell[lt]{L18,{\bf O18},L20} \nocite{lewis2018, oteo2018, long2020}
\\
SPT2052-56 (core) & 4.257 & 0 & $\pi4^2$ & 7,400 & 2,800 & & & {\bf W21}
\\
SPT2349-56 (core) & 4.302 & 23 & $\pi0.36^2$ & 4,480 & 128 & $\sim13-13.4$ & $\gtrsim15$ & \makecell[lt]{M18,{\bf H20},R21,\\W21} \nocite{miller2018, hill2020, rotermund2021}
\\
SPT2335-53 (core) & 4.756 & 0 & $\pi4^2$ & 7,000 & 3,200 & $-$ & $-$ & {\bf W21}
\\
SPT0553-50 (core) & 5.323 & 0 & $\pi4^2$ & 10,500 & 3,500 & $-$ & $-$ & {\bf W21}
\\
z57OD & 5.692 & 0 & $\pi4.2^2$ & $-$ & $-$ & $-$ & $\gtrsim14.7$ & \makecell[lt]{O05, J18,{\bf H19}} \nocite{harikane2019}
\\
SPT0348-62 (core) & 5.654 & 0 & $\pi4^2$ & 7,800 & 3,800 & $-$ & $-$ & {\bf W21}
\\
z66OD & 6.585 & 0 & $\pi4.2^2$ & $-$ & $-$ & $-$ & $\sim14.5$ & {\bf H19} \nocite{harikane2019}
\\
SPT0311-58 (core) & 6.9011 & 0 & $\pi4^2$ & 10,900 & 4,500 & $-$ & $-$ & {\bf W21}
\\
\bottomrule
\multicolumn{9}{p{0.92\columnwidth}}{{\bf Note} $-$ This list of proto-clusters with DSFG overdensities is not exhaustive. Proto-clusters are listed in order of increasing redshift. The primary references that contain the information listed in this table are bolded. Quantities measured over a limited central area designed the ``core" are labeled as such. A dash indicates information not readily available in the literature. $^a$ Observing window is listed as the full survey area or effective area (in arcmin$^2$) assumed for the proto-cluster core in the relevant reference. $^b$ Summed SFR of DSFG and/or other identified proto-clusters members, see listed references for details. $^c$ The comoving volume is derived from the listed observing window and $\delta z$ and/or taken from the relevant reference. $^d$ Proto-cluster numbers are the SFR and area of spec-z confirmed SMGs.  Core numbers are from the summed SFR of all (unconfirmed) 24$\mu$m and 850$\mu$m source in the core region as described in \citealt{lacaille2019}. $^e$ A nearby structure, CL1002-0220 at $z=2.51$ has been argued to be both a high-$z$, potentially virialized cluster \citep{wang2016,wang2018} and a filament that will merge with PCL1002 \citep{champagne2021}. Its gas properties will be discussed in \S~\ref{sec:COproto}. {\bf References:} A06: \citealp{alexander2016}; B04: \citealp{blain2004}; C05: \citealp{chapman2005}; C09: \citealp{chapman2009}; C15a: \citealp{chiang2015}; C15b: \citealp{casey2015}; C16: \citealp{casey2016}; Ch15: \citealp{chapman2015}; C21: \citealp{champagne2021}; E16: \citealp{emonts2016}; E18: \citealp{emonts2018}; D14: \citealp{dannerbauer2014}; D15: \citealp{diener2015}; G05: \citealp{geach2005}; G19: \citealp{gomez-guijarro2019}; H04: \citealp{hayashino2004}; H16: \citealp{hunghongzhaoling2016}; H19: \citealp{harikane2019}; H20: \citealp{hill2020}; K00: \citealp{kurk2000}; K11: \citealp{kuiper2011}; K13: \citealp{kubo2013}; K15: \citealp{kubo2015}; K16: \citealp{kato2016}; K21a: \citealp{koyama2021}; K21b: \citealp{kubo2021}; J18: \citealp{jiang2018}; J21: \citealp{jin2021}; L09: \citealp{lehmer2009}; L18: \citealp{lewis2018}; L19: \citealp{lacaille2019}; L20: \citealp{long2020}; M18: \citealp{miller2018}; O05: \citealp{ouchi2005}; O18: \citealp{oteo2018}; P00: \citealp{pentericci2000}; P21: \citealp{polletta2021}; R14: \citealp{rigby2014}; R21: \citealp{rotermund2021}; S98: \citealp{steidel1998}; S00: \citealp{steidel2000}; S12: \citealp{spitler2012}; S14: \citealp{shimakawa2014}; T09: \citealp{tamura2009}; T19: \citealp{tadaki2019}; U12: \citealp{uchimoto2012}; U14: \citealp{umehata2014}; U15: \citealp{umehata2015}; U17: \citealp{umehata2017}; V13:\citealp{valtchanov2013}; W16: \citealp{wang2016}; W21: \citealp{wang2021}; Y14: \citealp{yuan2014}; Z18: \citealp{zeballos2018}; Z19: \citealp{zavala2019}; Z22: \citealp{zhang2022}.  }
	\end{tabular}
	}
\end{table}

\subsubsection{Selecting DSFG Overdensities in Shallow, Wide (or All-sky) Submm Surveys}\label{sec:planck_selection}

Even given the comparatively large area over which FIR/submm data is available from ground-based facilities and \textit{Herschel}, the rarity of proto-clusters \citep{casey2016, chiang2017} presents a challenge in building statistical samples. Low-resolution, wide-field or all-sky submm surveys can integrate the emission from \red{multiple} DSFGs \citep[e.g.][]{negrello2005, miller2018, oteo2018}, while multi-band wavelength coverage can identify ``cold'' submm sources, i.e. those whose dust peak has redshifted into the $\sim$350-500$\mu$m range, placing them at $z\sim1.5-3$ \citep[modulo the dust temperature; e.g.][]{magnelli2014}. This can provide a catalog of proto-cluster candidates in a relatively unbiased way, while taking advantage of the negative K-correction in the submm \citep{franceschini1991}.  Here we present the recent results from \textit{Planck} as an example.

\begin{SCfigure}[0.8][!htb]
    \centering
    \parbox{0.5\columnwidth}{\includegraphics[width=0.5\columnwidth]{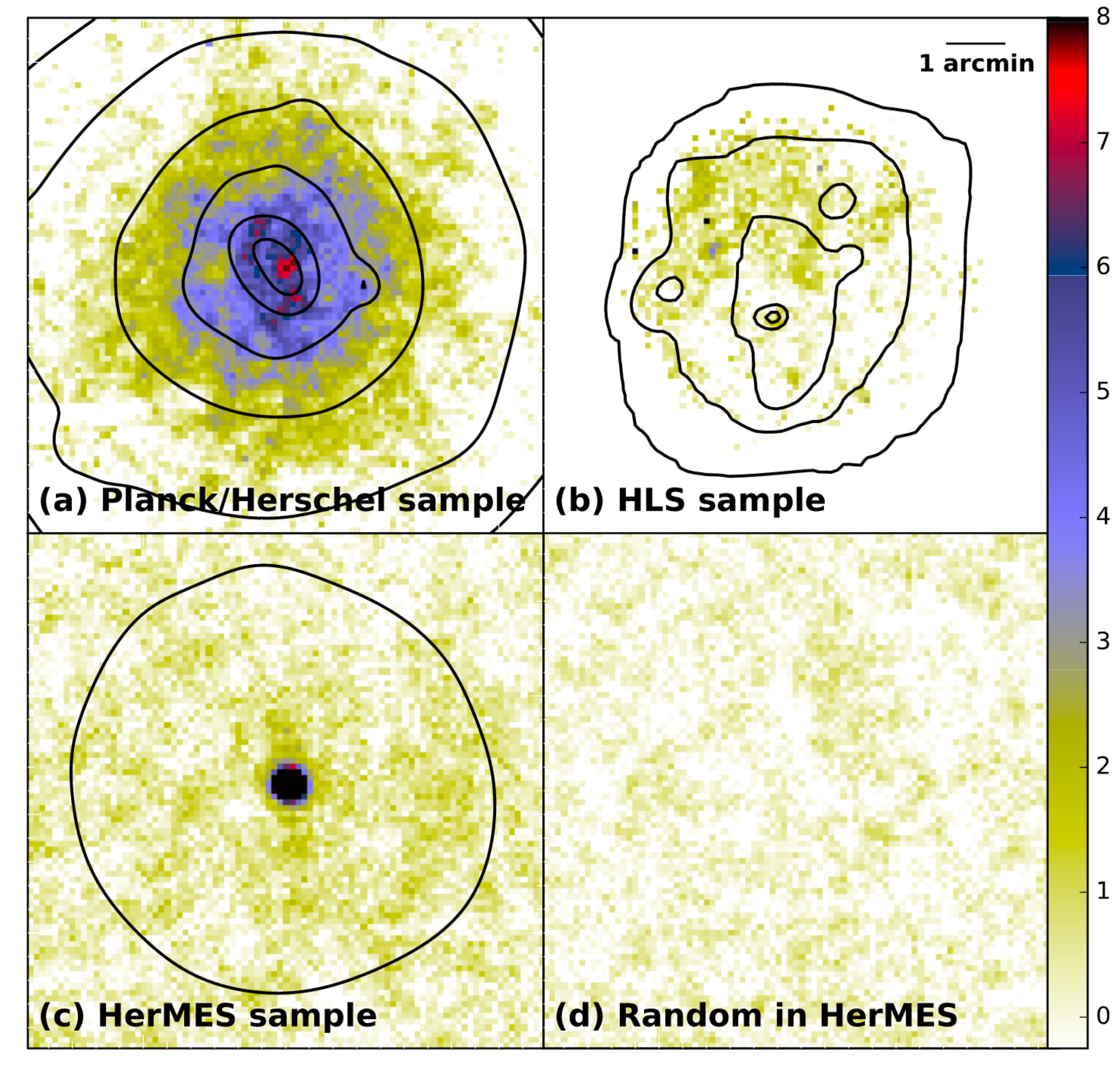}}
    \caption{ $-$ ({\bf a}) \textit{Herschel} SPIRE 350$\mu$m stacking of 228 \textit{Planck} PHz cold sources \protect\citep{planckcollaboration2016d}.  Extended emission is seen over the 8.7$^{\prime}$x8.7$^{\prime}$ cutout, consistent with these sources being proto-cluster candidates.  For comparison, panel ({\bf b}) shows the stacking of 278 $z<1$ \textit{Herschel} Lensing Survey clusters \protect\citep{egami2010}, ({\bf c}) single 500$\mu$m sources from HerMES\protect\footnote{The \textit{Herschel} Multi-tiered Extragalactic Survey \protect\citep[HerMES;][]{oliver2010}}, and ({\bf d}) shows 500 random positions in the HerMES Lockman field.  Figure reproduced from Figure 12 in \citet{planckcollaboration2015}, with permission from \copyright ESO.   }
    \label{fig:planck_spire}
\end{SCfigure}


The \textit{Planck} All-Sky survey catalogued compact sources (with a beamsize of $4-5^{\prime}$ at 100-857 GHz, $\sim2.5$ Mpc at $z\sim2$) in the \textit{Planck} Early Release Compact Source Catalog \citep[ERCSC;][]{planckcollaboration2011a}, Catalogue of Compact Sources \citep[PCCS;][]{planckcollaboration2014a}, and Second Catalogue of Compact Sources \citep[PCCS2;][]{planckcollaboration2016b}.  Early analyses of these catalogs identified numerous proto-cluster candidates which were then compared with \textit{Herschel} surveys \citep[e.g.][]{herranz2013, clements2014, baes2014}, where half of the \textit{Planck} compact sources resolved into multiple discrete \textit{Herschel} sources \citep{greenslade2018}. 
A subsequent list of 2,151 cold, compact sources was constructed via color selection using 217-857 GHz + 3 THz IRIS data \citep{miville-deschenes2005} to isolate high-redshift ($z\gtrsim1.5-4$, median $z\sim2.5$ assuming T$_{\rm dust}=35$ K) sources in the cleanest 26\% of the sky \citep[PHz catalog;][]{planckcollaboration2016d}. \textit{Herschel} follow-up of 228 of these cold sources \citep{planckcollaboration2015} revealed 93\% were associated with on average 10 red \textit{Herschel} sources (the rest were lensed DSFGs), with stacked 350$\mu$m extended emission consistent with expected proto-cluster sizes \citep[Figure~\ref{fig:planck_spire}, see][for \textit{Spitzer}/IRAC follow-up]{martinache2018}.
Unlike using a straight catalog of compact sources, this color selection rejects Galactic sources, low-$z$ contaminants, and low-$z$ clusters with strong SZ \citep[see also][]{martinache2018, greenslade2018}.

Are \textit{Planck} ``cold'' sources robust proto-cluster candidates? Several works have now confirmed submm overdensities among the PCCS sources \citep{flores-cacho2016, mackenzie2017, martinache2018, cheng2019, kaasinen2020}.  On the other hand, the source density of PHz ``cold'' sources is 0.21 deg$^{-2}$, more than three orders of magnitude higher than predicted for proto-clusters in a standard cosmological framework \citep{negrello2005,granato2015, negrello2017}.  This discrepancy can be resolved if multiple unrelated high-$z$ proto-clusters fall into the \textit{Planck} beam, as predicted by semi-analytic simulations \citep{negrello2017} and observed in limited spectroscopic follow-up \citep{flores-cacho2016, kneissl2019, polletta2021, polletta2022}. Recently, \citealt{lammers2022} cross-matched 187 PHz sources with \textit{Herschel} SPIRE\footnote{Spectral and Photometric Imaging Receiver \citep[SPIRE][]{griffin2007}} imaging compiled in the \textit{Herschel} Extragalactic Legacy Project \citep[HELP;][]{shirley2021}, finding that $21\%$ are associated with $>3\sigma$ SPIRE overdensities, a higher fraction than the PCCS sample ($8\%$). Revisiting the \citealt{negrello2017} simulations, they determined the average number of line-of-slight overdensities in a \textit{Planck} source was four, but that the ratio of the flux of the brightest overdensity was 3x higher than the second brightest overdensity in $60\%$ of cases, signaling that one candidate proto-cluster dominates.  \citealt{gouin2022} examined star formation in massive halos in IllustrisTNG \citep{pillepich2018}, finding that theoretical \textit{Planck} cold sources largely consist of one large SF halo plus smaller halos of background/foreground interlopers.  They predict $\sim70\%$ of their simulated \textit{Planck} sources will evolve into massive (log $M_{200}/\Msun>14$) clusters by $z\sim0$, though they note that the simulations continue to underestimate galaxy SFRs relative to observations \citep{granato2015, lim2021}.  In rough support of our discussion in the previous section, they find that  the number of SFGs ($>7$ with SFR$\,>10\,\Msun$ yr$^{-1}$) can discriminate \textit{Planck} sources more likely to evolve into clusters. 

The discussion is ongoing but there is compelling evidence that\sout{the} a significant fraction of submm cold sources are selecting proto-clusters at $z\gtrsim2-4$.  Techniques will have to be developed to mitigate contamination and wide-field narrow-band imaging \citep[e.g.][]{darvish2020}, spectroscopy, and/or statistical techniques (\S\,\ref{sec:totallight}) are needed to take advantage of future large proto-cluster candidate samples. In the next several sections, we turn our focus from (proto-)cluster surveys to the (proto-)cluster galaxy populations, exploring what we have learned in the near-, mid/far-IR, and submm regimes.

\section{The Near-Infrared: Stellar Mass Functions and Quenched Populations}\label{sec:nir}

Absent external influences, galaxies stop forming stars (i.e. quench) via secular processes \citep[supernovae, stellar winds, AGN feedback; e.g. ][]{oppenheimer2010, croton2006, tremonti2007}, often termed mass- (or self-)quenching.  A long standing question is whether mass-quenching is separable from environmental-quenching \citep{peng2010}: in other words, does mass-quenching operate independently of environment and does environmental-quenching operate independently of stellar mass? Mass-quenching is strongly stellar mass-dependent, with higher mass galaxies quenching first \citep[i.e. downsizing;][]{popesso2011, sobral2011, darvish2016, fossati2017}.  Low-mass galaxies mass-quench on longer timescales; at log $M_{\star}/\Msun\lesssim9$, this timescale exceeds the Hubble time and such galaxies have not yet quenched in the field \citep[with rare exceptions, e.g.][]{geha2012} as seen in the high star-forming fraction in the void regions in Figure~\ref{fig:coma}.  This makes low-mass galaxies an ideal population to address the role of environmental quenching.  Higher mass galaxies cannot be neglected, however, as they may be effected by different environmental processes.  

In this section, we review the current literature using NIR to characterize the stellar populations in clusters, with the goal of addressing how galaxies in extreme environments quench over cosmic time.  As described in the introduction, two key tools are provided by the NIR: the stellar masses of cluster members and the division of cluster members into SFG and QG populations using UVJ (or equivalent) color selection.  In \S\,\ref{sec:nir_smf}, we discuss what we know about stellar mass functions (SMFs) in clusters and as a function of local galaxy density; the latter has been pushed to low-masses.  In \S\,\ref{sec:epsilon_env}, we define and examine the environmental quenching efficiency $-$ excess quenching due to environmental processes $-$ to high redshift and beyond the virial radius.

\begin{figure}[!htb]
    \centering
    \includegraphics[width=\columnwidth]{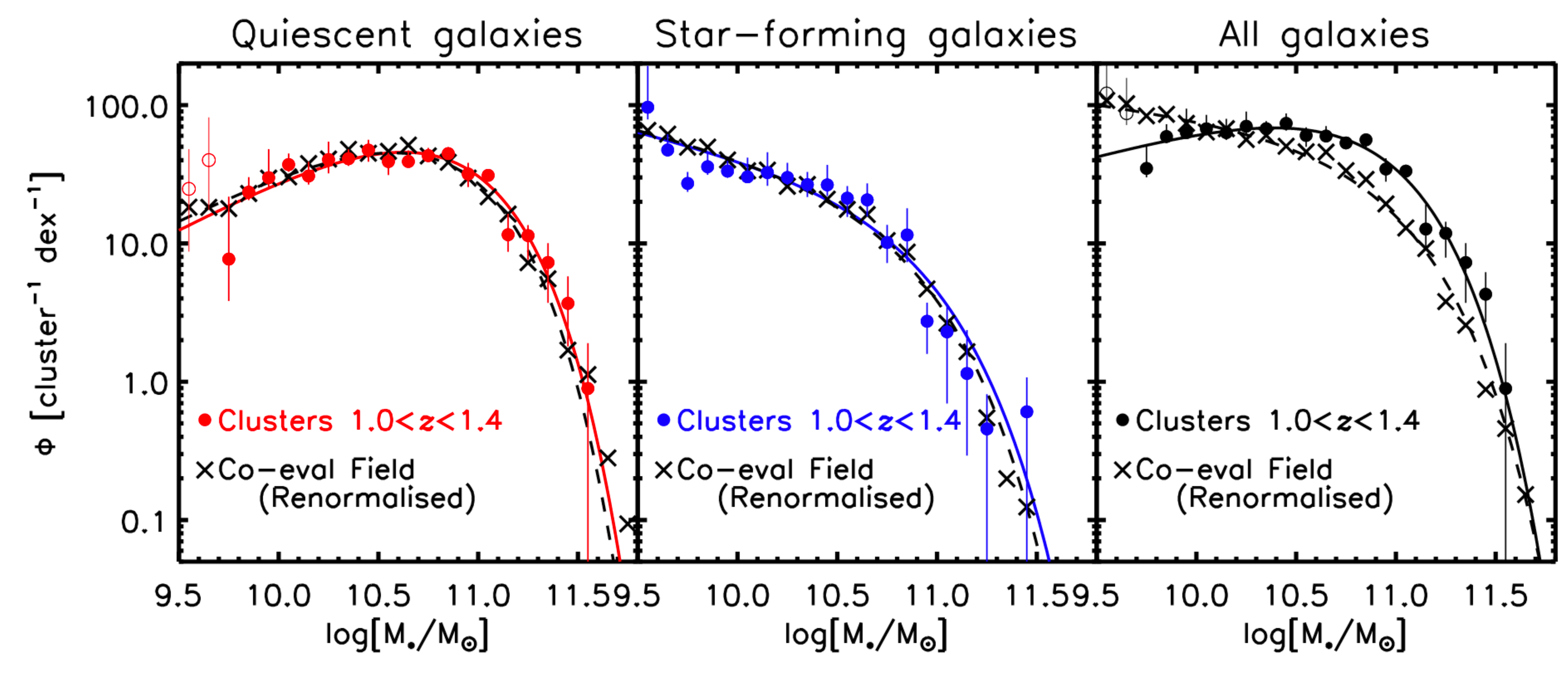}
    \caption{ $-$ The SMFs of QGs (left), SFGs (middle), and the total population (right) in 11 GOGREEN clusters at $1<z<1.4$ \citep[circles;][]{vanderburg2020}.  Renormalized field SMFs (crosses) drawn from the COSMOS/UltraVISTA survey \citep{muzzin2013a} are shown in comparison.  For QGs and SFGs separately, the shapes of the SMFs are independent of environment.  For the total population, there is an excess of high-mass  and a dearth of low-mass galaxies in the clusters. Figure reproduced from Figure 6 in \citealt{vanderburg2020}, with permission from \copyright ESO.}
    \label{fig:vdB20}
\end{figure}


\subsection{The Stellar Mass Function in Overdense Environments}\label{sec:nir_smf}

A fundamental characterization of a galaxy population is the stellar mass function, the number distribution of galaxies in bins of stellar mass, which encodes information on the processes that have contributed to and impede stellar mass growth.  The SMFs of most galaxy populations\footnote{BCGs are a notable exception \citep[e.g.][]{lidman2012, lin2017} and are typically excluded from cluster SMF measurements.} are well described by a single or double Schechter function \citep{schechter1976}, parameterized by shape [characteristic mass ($M^{\star}$), low-mass slope ($\alpha$)] and overall normalization ($\Phi^{\star}$).  In the field, the shape of the SMF changes little to $z\sim4$ while its normalization evolves, driven by mass-quenching increasing the QG population \citep[e.g.][]{perez-gonzalez2008, muzzin2013a}.

If we assume that the field SMF is dominated by mass-quenching, differences in the SMF in overdense environments can be attributed to environmental quenching. These differences have been searched for using both ``global''\footnote{For a discussion of the construction of SMFs in clusters, including necessary cluster member completeness corrections, we refer the reader to \S\,3 in \citealt{vanderburg2020}.} and ``local'' definitions of environment, as traced by different proxies (e.g. cluster-centric radius vs local galaxy density).  The distinction may not be trivial: early work in \citealt{vulcani2012} and \citealt{vulcani2013} examined the stellar mass function down to log $M_{\star}/\Msun\geq10.2$ using both local and global definitions on optical cluster surveys\footnote{the WIde-Field Nearby Galaxy-cluster Survey \citep[WINGS at $0.04<z<0.07$;][]{fasano2006}, The IMACS Cluster Building Survey \citep[ICBS at $0.25<z<0.45$;][]{oemler2013} and ESO Distant Cluster Survey \citep[EDisCs at $0.4<z<0.6$;][]{white2005}}.  They found differences from the field SMF only when considering local environment, illustrating that caution should be taken in comparing works using inhomogenous definitions of environment.

Subsequent (global) cluster studies incorporating the NIR (and often going to lower mass limits) largely find that the {\it total} SMF in clusters has a clear environmental dependence up to $z\sim1.5$  \citep[][but see \citealp{vulcani2013, calvi2013}]{annunziatella2014, annunziatella2016, vanderburg2013, vanderburg2018, vanderburg2020}.  Figure~\ref{fig:vdB20} (right panel) compares the shape of the SMF function compiled from 11 galaxy clusters at $1<z<1.4$ from the GOGREEN survey to a coeval field sample \citep{vanderburg2020}.  The cluster total SMF reflects an overabundance of massive galaxies, with a deficit at the lower-mass end relative to the field. Remarkably, however, when the galaxy populations are split into star-forming and quiescent (using UVJ colors), the shapes of the SFG and QG SMFs become independent of environment (Figure~\ref{fig:vdB20}, left and middle panels). This characteristic behavior of the SMF in overdense environments has been observed in low \citep[$z\sim0.2-0.4$;][]{annunziatella2014, annunziatella2016} and intermediate-redshift clusters\footnote{At higher redshifts, the environmental dependence of the total SMF in overdense regions is less clear \citep[][]{nantais2016, papovich2018}.} \citep[$z\sim0.5-1.2$;][]{nantais2016, vanderburg2013, vanderburg2018, vanderburg2020}, \red{in low-mass cluster/group scale halos \citep{reeves2021}}, and in local environment studies \citep{tomczak2017, papovich2018}. 
They suggest that 1) environmental quenching is mass-independent at log $M_{\star}/\Msun\gtrsim10$, in order to maintain the shape of the SFG SMF, and 2) differences in cluster total SMFs are driven by an excess in the quenched galaxy fraction. In support of the latter, QGs are found to dominate the cluster galaxy counts to survey mass limits \citep[log $M_{\star}/\Msun=9.5$;][]{balogh2016, vanderburg2018}, in sharp contrast to the field and consistent with the local SFR-density relation extending to at least $z\sim1$. 

\begin{figure}[!htb]
    \centering
    \includegraphics[width=\columnwidth]{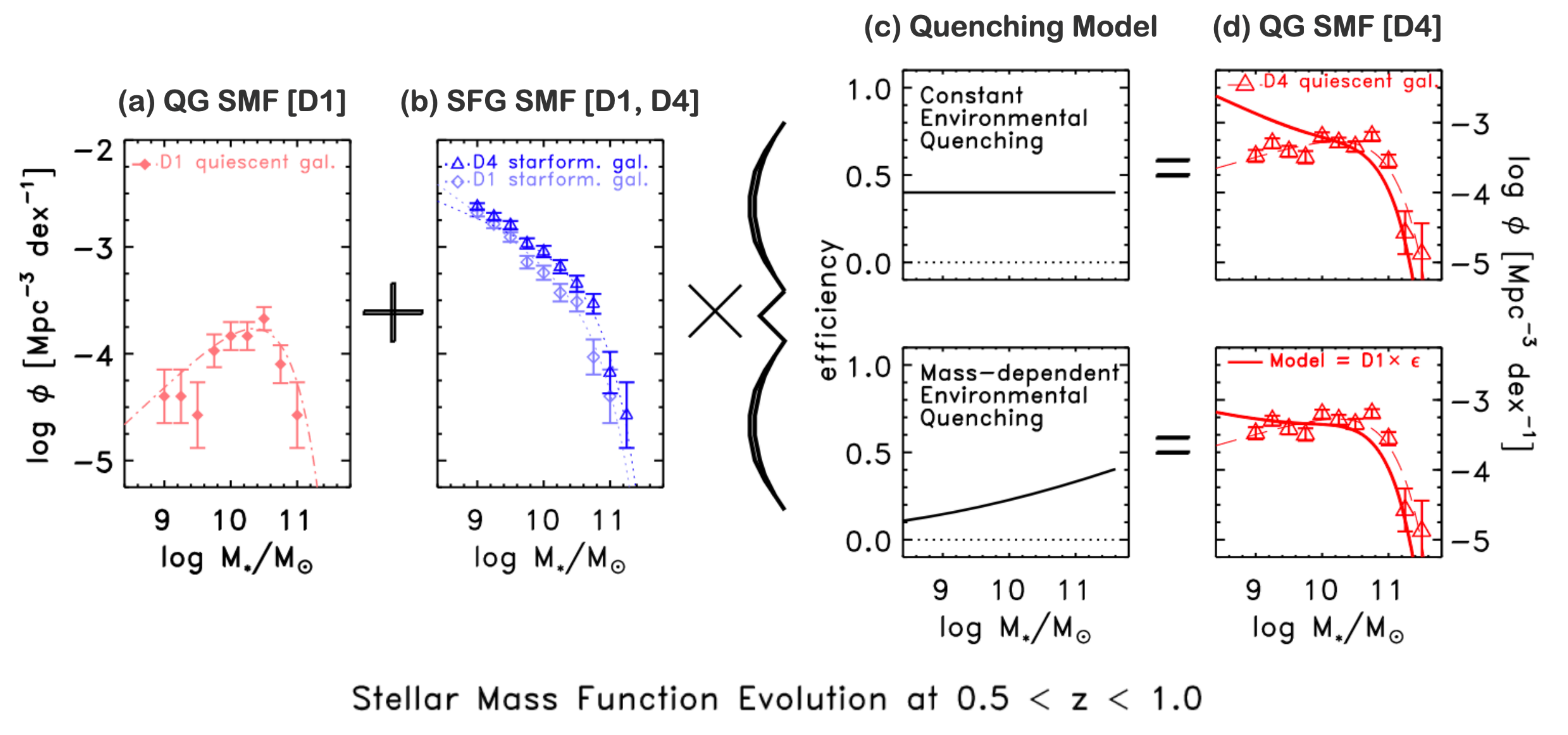}
    \caption{ $-$ A toy model showing that combining the observed QG SMF in the lowest density quartile (D1, panel a) with the steep low-mass end of the SFG SMF (panel b) observed in both the highest (D4) and lowest (D1) density quartiles requires a mass-dependent environmental quenching model (panel c, bottom) to avoid a strong excess in the low-mass quenched galaxies in overdense environments (panel d).  Both quenching models are able to reproduce the high-mass  end. Figure reproduced by permission of the AAS from Figure 5 in \citealt{papovich2018}.}
    \label{fig:papovich18}
\end{figure}


Environmental dependence in the shape of the SFG and QG SMFs, however, becomes apparent at lower stellar masses.  For example, \citealt{vanderburg2018}, observing 21 \textit{Planck}-selected clusters at $0.5<z<0.7$ to log $M_{\star}/\Msun\sim9.5$, found that the QG SMF has a significantly flatter low-mass slope in the clusters than the field, indicating an excess of low-mass quenched galaxies.  This flatter $\alpha$ was also observed for log $M_{\star}/\Msun=9-10$ QGs in group-scale overdensities over $0.2<z<1.5$ \citep{papovich2018} and up to cluster-scales in the local environment ORELSE survey\footnote{Observations of Redshift Evolution in Large-Scale Environments \citep[ORELSE;][]{lubin2009}} over $0.6<z<1.3$ \citep{tomczak2017}.  While \citealt{vanderburg2018} found no environmental-dependence in shape of the SFG SMF, \citealt{tomczak2017} reported a strong dependence in their highest density bin \citep[see also the local environment studies by][]{giodini2012, mok2013, davidzon2016}, indicating mass-dependent quenching.  The importance of quantifying the low-mass slope was demonstrated by \citealt{papovich2018}, who convolved their SFG and QG SMFs derived in regions of low galaxy density with constant and mass-dependent quenching (Figure~\ref{fig:papovich18}).  Due to the extremely steep slope of the field SFG SMF at low mass (panel b), mass-independent environmental quenching (panel c) would result in the equivalent strong upturn in the low-mass slope of the QF SMF (panel d).  Even a modest mass-dependence in environmental quenching, however, can more easily replicate the flat $\alpha$ observed in overdense environments.  This effect is even more pronounced at $z\sim1-1.5$ where the they find an even shallower low-mass slope.  Notably both constant and mass-dependent quenching toy models can reproduce the high-mass  end; as such, pushing to log $M_{\star}/\Msun=9$ and lower provides the most discriminating power in quantifying the evolution of the SMF with environment.

In summary, there is a general consensus that SMF studies reveal a clear environmental influence driving the quenched galaxy fraction to dominate at all stellar masses probed (or in other words an increase in the QG SMF normalization) which modifies the shape of the {\it total} SMF in favor of excess massive (quenched) galaxies.  At the same time, the separate QG and SFG SMFs at the high-mass  end show no evolution in shape, which can be achieved via quenching that is independent or moderately-dependent on stellar mass.  At the low-mass end, on the other hand, where \red{secular} mass-quenching is largely absent, a flattened slope in the QG SMF strongly suggests a mass-dependent \red{environmental} quenching efficiency.

\subsection{The Quenched Fraction and Environmental Quenching Efficiency}\label{sec:epsilon_env}

A more direct look at environmental quenching can be achieved using the quenched fraction ($f_{\rm q}\equiv N_{\rm QG}/N_{\rm QG+SFG}$) and the environmental quenching efficiency \citep[$\epsilon_{\rm env}$ or EQE; e.g.][]{peng2010}, 

\begin{equation}\label{eqn:env}
    \epsilon_{env} = \frac{f_{\mathrm{q,cl}} - f_{\mathrm{q,field}}}{1-f_{\mathrm{q,field}}},
\end{equation}
which quantifies how many galaxies are quenched in an overdense environment, $f_{\rm q,cl}$, that would not have been quenched in a low-density environment, $f_{\rm q,field}$.  By accounting for the field, $\epsilon_{\rm env}$ in principle removes  mass quenching \citep[though there is some evidence of excess mass quenching in overdense environments;][]{pintos-castro2019}.  This quantity has also been referred to in the literature as the ``transition fraction'' \citep[e.g.][]{vandenbosch2008}, the ``conversion fraction'' \citep[e.g.][]{balogh2016, fossati2017}, and the ``quenched fraction excess'' \citep[e.g.][]{wetzel2012, bahe2017, vanderburg2020}.  We use ``environmental quenching efficiency'' instead of ``transition fraction'' or ``conversion fraction'' as they imply a relationship between the galaxy populations being compared (i.e. that the control or ``field'' sample will become the cluster sample) that may not be accurate, as we will discuss in this section.

\subsubsection{The Multi-Dimensional Dependencies of Quenching at $z<1$}

\begin{figure}[!htb]
    \centering
    \includegraphics[width=0.46\columnwidth]{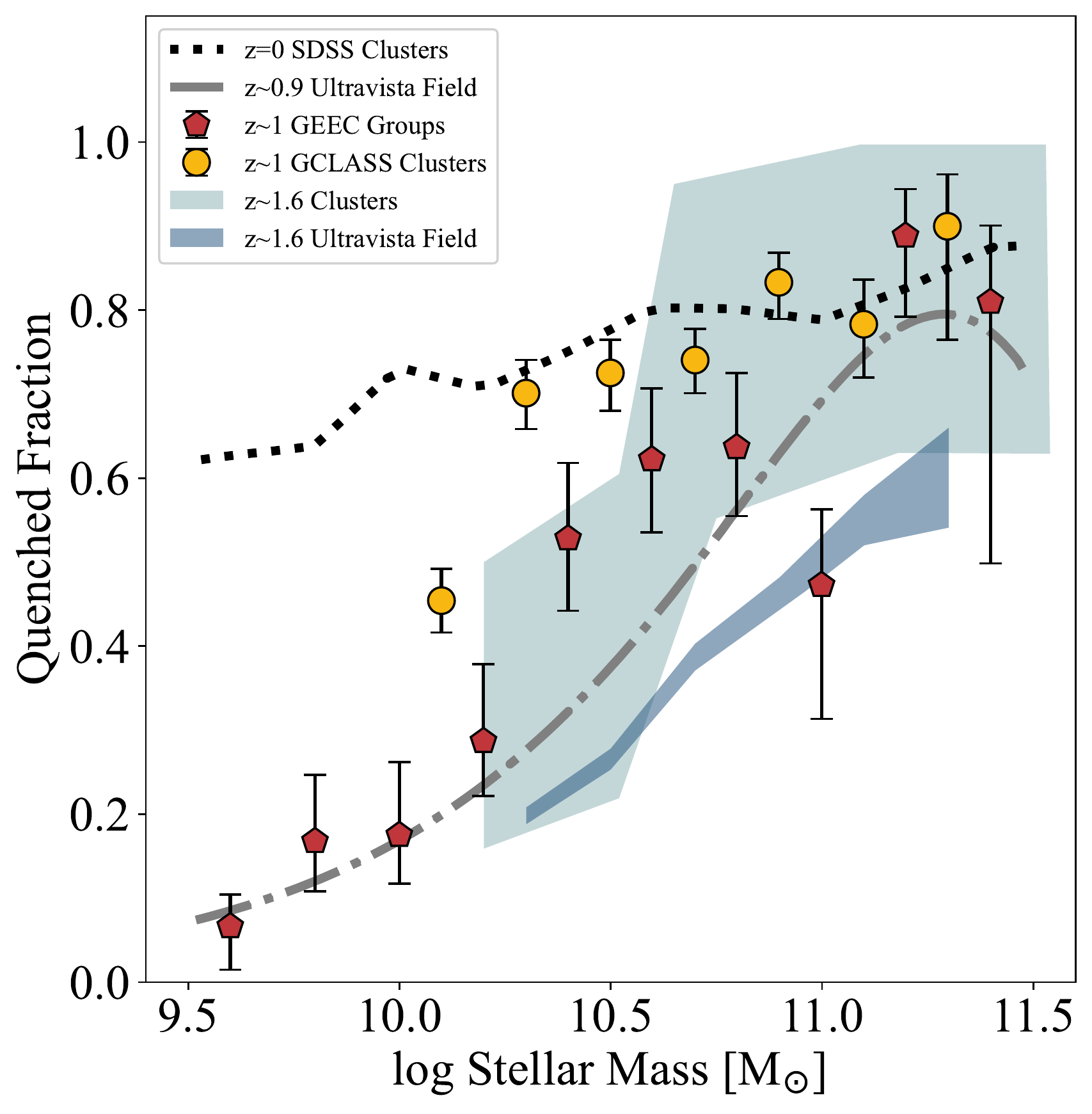}
    \includegraphics[width=0.48\columnwidth]{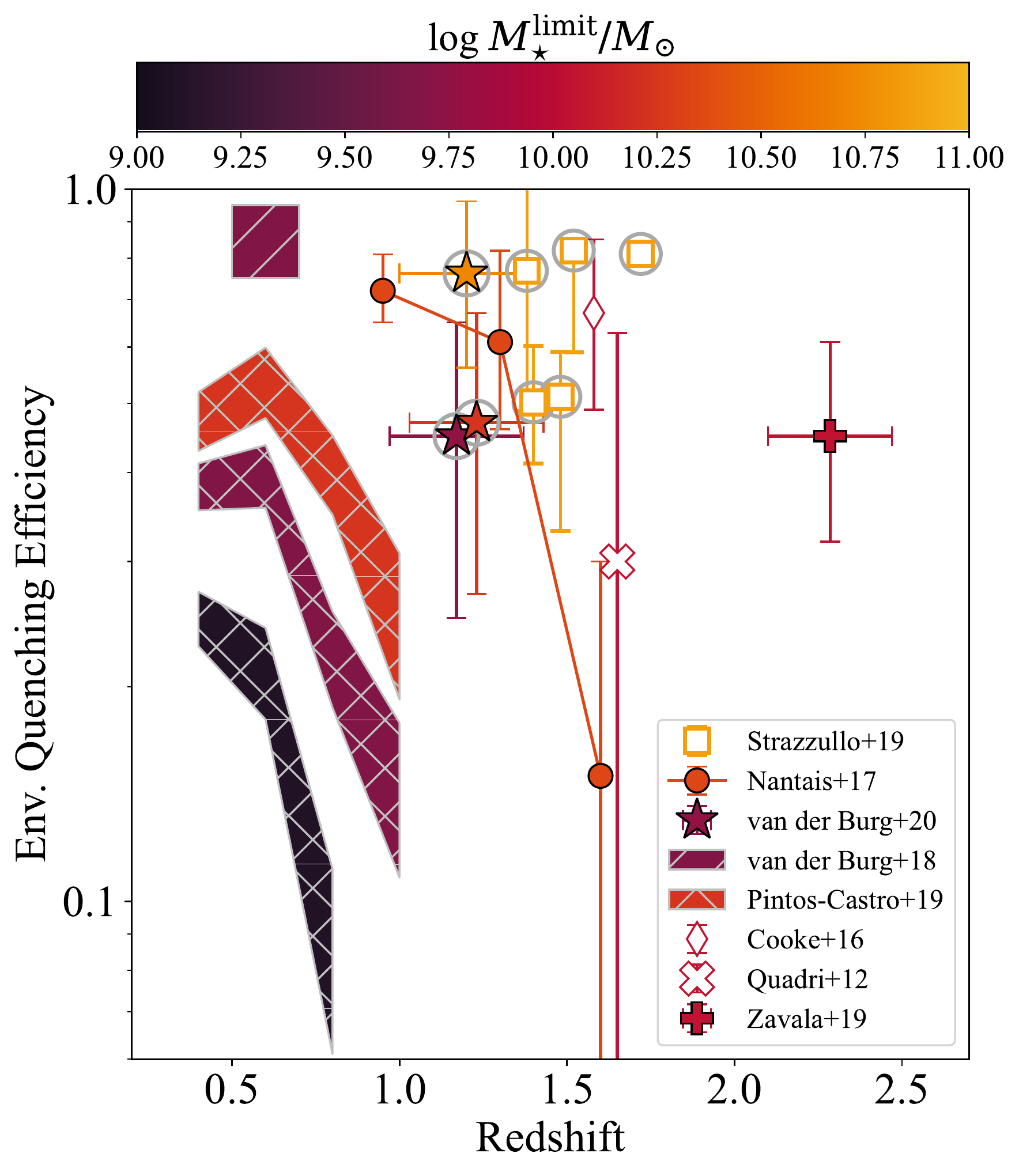}
    \caption{ $-$ {\bf (left)} The quiescent fraction versus stellar mass of GCLASS clusters and GEEC2 low-mass clusters/groups at $z\sim1$ (yellow circles and red pentagons; \citealt[][]{balogh2016}) compared to three clusters at $z\sim1.6$ (green shaded region; \citealt[][]{lee-brown2017}). Comparisons are made to coeval field samples from UltraVISTA (dash dot line at $z\sim1$ and blue shaded region for $z\sim1.6$).  Local SDSS clusters are also shown, demonstrating that the SFR-density relation is in place at $z\sim1$ for log $M_{\star}/\Msun\gtrsim10.3$ cluster galaxies. The comparison between GCLASS clusters and GEEC2 low-mass clusters/groups reveals a halo mass dependence, while the comparison of clusters at $z\sim1$ and $z\sim1.6$ shows a redshift and stellar mass dependence. {\bf (right)} The environmental quenching efficiency in \red{massive} (proto-)clusters as a function of redshift.  The stellar mass limit of each sample is denoted by the color scale. Closed symbols and regions contain multiple clusters, while open symbols are single clusters. Gray hatches and circles mark $\epsilon_{\rm env}$ measured within $r\sim0.5R_{200}$; all others are within $r\sim R_{200}$.  Strong stellar mass \citep{pintos-castro2019, vanderburg2020} and redshift \citep{nantais2017, pintos-castro2019} dependencies are observed in cluster surveys.  A halo mass dependency is also suggested by the comparison of \textit{Planck} clusters \citep[log $M_{200}/\Msun\sim15$;][]{vanderburg2018} to the SpARCS and GOGREEN surveys \citep[log $M_{200}/\Msun\sim14.5$;][]{pintos-castro2019, vanderburg2020}.  Quenched high-mass populations can be found in clusters out to $z\sim1.7$ \citep{strazzullo2019} and in proto-clusters at $z\sim2-2.5$ \citep{zavala2019}. \red{Figures adapted from Figure~3 in \citealt{balogh2016}, Figure~7 in \citealt{lee-brown2017}, and Figure~12 in \citealt{pintos-castro2019}, reproduced by permission from the AAS.}}
    \label{fig:balogh16}
\end{figure}

Quenched galaxies are well known to dominate cluster populations in the local Universe. As discussed in \S\,\ref{sec:nir_smf}, separation of the cluster SFG and QG populations using UVJ colors revealed that the total cluster SMF is dominated by quenched galaxies over a wide range in mass  up to $z\sim0.6$ \citep{vanderburg2018} and likely even up to $z\sim1$.  To demonstrate this, \citealt{balogh2016} examined the fraction of QGs and EQE in 10 GCLASS massive clusters at $z\sim1$ relative to coeval low-mass clusters/groups and the field as well as SDSS clusters  at $z\sim0$ (Figure~\ref{fig:balogh16},  left).  The SFR-density relation is clearly in place to $z\sim1$, with the log $M_{200}/\Msun\sim14.5$ GCLASS clusters showing similar quenched fractions as local clusters for log $M_{\star}/\Msun\gtrsim10.3$.  
The comparison to log $M_{200}/\Msun\sim13-14.5$ GEEC2 low-mass clusters/groups \citep{balogh2014} and the Ultravista \citep{muzzin2013a} field further illustrates the halo mass dependence of environmental quenching \red{\citep[see also][]{reeves2021, sarron2021}}, with low-mass clusters/groups showing less deviation from the field.

In addition to halo mass, EQE decreases with increasing cluster-centric radius \citep{vanderburg2018, pintos-castro2019}, a result that builds on earlier studies of the radial dependence of optically blue and red galaxy fractions in clusters \citep[e.g.][]{loh2008, raichoor2012}.  These dependencies join the possible stellar mass dependence discussed previously in the context of SMFs (\S\,\ref{sec:nir_smf}), though we note again that some works report a mass-dependence at $z\sim0.5-1$ \citep{balogh2016, papovich2018, vanderburg2020} while others report none \citep{vanderburg2018, kawinwanichakij2017, werner2022} , which may be a function of stellar mass survey limits.  Further complicating our view of quenching is the still-open question of the inter-dependency of mass- and environmental-quenching.  

There is therefore a pressing need for studies which probe these multi-dimensional dependencies simultaneously over a range in redshift.  \citealt{pintos-castro2019} recently capitalized on new Hyper Suprime Cam imaging of 209 NIR-selected SpARCS clusters at $0.3<z<1.1$ to compile a large sample of galaxies that could be binned by redshift, stellar mass, and cluster-centric radius simultaneously.  Using UVJ, they calculated the star-forming fraction ($f_{\rm SF}\equiv N_{\rm SFG}/N_{\rm QG+SFG}$) as well as $\epsilon_{\rm env}(r, M_{\star})$ as in Eqn~\ref{eqn:env}.  As expected given environmental quenching, $f_{\rm SF}$ is found to decrease from $R_{200}$ into the cluster cores. In addition, they define the mass quenching efficiency as $\epsilon_{\rm mass}(r,M_{\star})=f_q (r,M_{\star}) - f_q (r, M^{\rm SF}_{\star})/(1-f_q (r,M^{\rm SF}_{\star}))$, where $M^{\rm SF}_{\star}$ is the stellar mass at which most galaxies are still forming stars at a given radius (in practice this is driven by their stellar mass completeness limit).  Using $f_{\rm SF}$ to define the stellar mass at which quenching ``starts'' ($M^{\rm start}_{\star}[{f_{\rm SF}\sim0.8}]$) and ``ends'' ($M^{\rm end}_{\star}[{f_{\rm SF}\sim0.2}]$) in the cluster cores ($r<0.4R_{200}$), outskirts ($0.4<r/R_{200}<1$), and field ($4<r/R_{200}<6$), they observe accelerated quenching in that both $M^{\rm start}_{\star}$ and $M^{\rm end}_{\star}$ occur at lower stellar masses (by $\Delta \mathrm{log} M_{\star}/\Msun\sim0.3-0.4$) in the cores relative to the outskirts/field, a difference which increases from $z\sim1\rightarrow0.3$ (accelerated downsizing). In other words, $\epsilon_{\rm mass}$ is more efficient in the cluster cores and thus not independent of environment.  

Likewise, they find that $\epsilon_{\rm env}$ (Eqn~\ref{eqn:env}) depends on stellar mass.  Figure~\ref{fig:balogh16} (right) shows a compilation of $\epsilon_{\rm env}$ measurements in \red{massive clusters}\sout{the literature} across $0.5<z<2.5$.  Broken into multiple stellar mass bins, environmental quenching in the \citealt{pintos-castro2019} sample displays a strong dependence on {\it both} stellar mass and redshift:  for low-mass galaxies, EQE starts as relatively negligible at $z\sim0.7$ ($\epsilon_{\rm env}\lesssim 0.1$) but rises by a factor of $2-3$ by $z\sim0.5$.  EQE in higher mass galaxies starts at a higher base, but shows a similar rise over $\sim2.5$ Gyr.
At $z\sim1.2$, the GOGREEN clusters \citep[stars in Figure~\ref{fig:balogh16}, right;][]{vanderburg2020} have a comparable $\epsilon_{\rm env}$ to the SpARCS sample in matched stellar mass bins, with a higher $\epsilon_{\rm env}\sim0.8$ reported for their higher mass galaxies, demonstrating a stellar mass dependence in EQE. At redshift $0.5<z<0.7$, on the other hand, \citealt{vanderburg2018} measured $\epsilon_{\rm env}\sim0.85$ for \textit{Planck}-selected clusters using similar radial and stellar mass bins as \citealt{pintos-castro2019}.  This high quenching efficiency was observed to have no stellar mass dependence.  A possible explanation for this difference is the halo masses of the samples. The SpARCS sample covers a range in halo mass with an average log $M_{200}/\Msun\sim14.5$ based on richness measurements (I. Pintos-Castro, private communication), comparable to the GOGREEN clusters. The \textit{Planck}-selected sample has a higher typical halo mass of log $M_{200}/\Msun\sim15$. \red{Furthermore, low-mass cluster/group studies find lower, stellar-mass dependent EQEs than either cluster survey \citep{sarron2021, reeves2021}.} Could different quenching mechanisms be dominating in the most massive halos?  This comparison underscores the need for\sout{follow-up} analysis that\sout{also} controls for halo mass as well as stellar mass, cluster-centric radius, and redshift, as quenching processes likely depend on all four parameters. 

\subsubsection{The role of pre-processing to high redshift}\label{sec:nir_preprocessing}

So far the results we've discussed have focused on the quenched fraction and EQE within $R_{200}$ to $z\sim1$. In the local Universe, however, quenching and morphological transformations have been observed well beyond the cluster virial radius \citep[$3-5R_{\rm vir}$;][]{lewis2002, vonderlinden2010, chung2011, patel2009, wetzel2012, cybulski2014, haines2015,just2019,ayromlou2022}, which requires environmental effects to start the quenching processes during infall.  This is generally termed pre-processing \citep[e.g.,][]{zabludoff1996,zabludoff1998,kodama2001,fujita2004,berrier2009,mcgee2009,dressler2013,cybulski2014,weinzirl2017}, where quenching begins in the lower-density group environment\footnote{For higher redshift clusters, pre-processing may occur in the proto-cluster environment through top-heavy halo and stellar mass functions \citep[e.g.][]{cooke2014, chiang2017, muldrew2018}.} \citep{bianconi2018, papovich2018}. For context, $20-40\%$ of local clusters' stellar mass is expected to accrete in the form of groups \citep{mcgee2009}.  Outside the local Universe, a non-zero EQE has been observed beyond the virial radius in clusters at $z\sim0.5-1$, converging on a value as high as $\epsilon_{\rm env}\sim0.35$ in very high-mass  clusters \citep{vanderburg2018,vanderburg2020}. \citealt{pintos-castro2019} found that $f_{\rm SF}$ was flat from $r\sim R_{200}\rightarrow 6R_{200}$ for all stellar masses up to $z\sim1$ \citep[see also][]{haines2015}, suggesting that if pre-processing is occurring, it starts at very large radii.

Recently, a study by \citealt{werner2022} demonstrated that the choice of the ``field'' in global environment studies can obfuscate environmental signatures.  Using GOGREEN and GCLASS clusters at $0.8<z<1.4$, they compared the quenched fraction and $\epsilon_{\rm env}$ in the cluster ($r<R_{200}$), infall regions ($1<r/R_{200}<3$), and field (outside the cluster and infall regions).  They note that the backsplash\footnote{Backsplash galaxies are gravitationally-bound cluster members that have completed their first pass of the cluster center and are on orbits taking them back into the cluster outskirts and infall regions \citep{diemer2014}.} population is expected to be small at these redshifts \citep[e.g.][]{haggar2020}.  They found that the infall region contains a higher fraction of massive QGs and twice as many satellites per central galaxy at fixed stellar mass compared to the field region, indicating it is populated by more massive halos. Comparing the cluster and infall regions suggests that nearly all log $M_{\star}/\Msun>11$ and half of all log $M_{\star}/\Msun=10-11$ galaxies are quenched prior to crossing $R_{200}$ by $z\sim1$. Similarly, local environment studies find high quenched fractions in group-scale overdensities, supporting groups as the dominant site for quenching \citep{mcgee2009, fossati2017, papovich2018}. \citealt{werner2022} calculated the EQE separately using the field plus infall regions (probing pre-processing) and the infall plus cluster regions (probing cluster quenching), finding that the former is strongly mass-dependent, while the latter shows only a weak stellar mass dependence. This suggests different processes dominating in the group vs cluster environments and that the ``field" control samples chosen impact the measured EQE.  This has yet to be reconciled with the studies discussed in the previous section.

Can all pre-processing be attributed to group processes? A major hindrance to current discussions of pre-processing is understanding the range of influence of the primary halo relative to the effects of the local (group) environment. An open question is where to place the boundary of a cluster; this is commonly assumed to be the virial radius for convenience but recent works suggest a more physically motivated ``edge" in the splashback radius \citep{adhikari2014, diemer2014, mansfield2017, more2015}, based on the fall-off of the matter density profile.  And the influence of the ICM may reach even further as suggested by recent simulations \citep[e.g.][]{zinger2018, mostoghiu2021,ayromlou2021} which show gas stripping at large radii, likely associated with virial accretion shocks \citep{birnboim2003, dekel2006}.
This adds to the multi-dimensional quenching dependencies (stellar mass, halo mass, cluster-centric radius, redshift) outlined in the previous section, indicating simultaneous measures of both the global and local environment may be necessary to disentangle the quenching processing arising from cluster and group environments. Quenching in the group environment and this evidence of extended cluster influence will be discussed in more detail in \S\,\ref{sec:beyond}.

\subsubsection{Strongly evolving quenching efficiency at high redshift?}\label{sec:nir_highz}

Moving further into the epoch of $z=1-2$, the picture of quenching in individual clusters becomes one of large variation from system to system over a relatively small set of observations.  There are several examples of clusters with \red{substantial}\sout{significant} quenched populations at $z\sim1.5-2$ \citep{grutzbauch2012, quadri2012, stalder2013, santos2013, strazzullo2013, newman2014, cooke2016, lee-brown2017}, which contrasts with increasing evidence for significant star formation activity in cluster cores during this epoch (see \S\,\ref{sec:fir}). \citealt{lee-brown2017} analysed IRC 0218, a low-mass cluster at $z=1.62$, in comparison to a similarly low-mass cluster 
at $z=1.58$ \citep{cooke2016}, and a high-mass  X-ray cluster JKCS 041 at $z=1.8$ \citep{newman2014}.  With large uncertainties, all three clusters show a stellar mass-dependent quenched fraction (Figure~\ref{fig:balogh16}; left), with nearly all high-mass galaxies (log $M_{\star}/\Msun>10.85$) quenched.  The quenched fraction drops to the field level by log $M_{\star}/\Msun\sim10$.  Comparing with clusters at $z\sim1$ from \citealt{balogh2016}, \citealt{lee-brown2017} concluded there is at most a modest evolution in environmental quenching efficiency (at the low-mass end) from $z\sim1.6\rightarrow1$ indicated by these three systems.

Conversely, statistical evidence has pointed to a strong evolution. Construction of the IRAC $3.6\mu$m and $4.5\mu$m luminosity functions (LFs) of ISCS cluster galaxies over $0.3<z<2$ found a strong deviation from the characteristic luminosity predicted from passive evolution models at $z\gtrsim1.4$ \citep{mancone2010, mancone2012}, indicating an era of substantial stellar mass growth above this redshift.  \citealt{nantais2017} later found a strong rise in the EQE from $z\sim1.6\rightarrow0.9$ using 14 confirmed SpARCS clusters.  For cluster galaxies with log $M_{\star}/\Msun\geq10.3$, the average $\epsilon_{\rm env}$ in three $z\sim1.6$ clusters was consistent with zero (though with large variation between the clusters).  The environmental quenching efficiency then rises abruptly (Figure~\ref{fig:balogh16}, right) and the quenched fraction increases from 42$\%$ at $z\sim1.6$ (consistent with the field) to 80$\%$ by $z\sim1.3$ \citep[see also][for elevated environmental quenching at $z\sim1.3$]{vanderburg2013, vanderburg2020, reeves2021}, an evolution requiring rapid quenching over $<1$ Gyr.   These results suggest that some massive clusters may undergo a significant transition between $z=2\rightarrow1$, where environment-specific quenching mechanisms ``turn on'' and then efficiently produce substantial quenched populations by $z\sim1$. What drives this transition is not yet known, though the variation from cluster-to-cluster appears to be intrinsic rather than a selection effect. This is demonstrated in Figure~\ref{fig:balogh16} (right, open squares) through a sample of 5 SPT clusters at relatively fixed halo mass (log $M_{\star}/\Msun\sim14.5$) at $1.4<z<1.7$, where significant variation  is observed in the environmental quenching of a specific population: high-mass  (log $M_{\star}/\Msun>10.85$) galaxies in the cluster cores \citep[$r<0.3R_{200}$;][]{strazzullo2019}.

The strong redshift evolution in $\epsilon_{\rm env}$ in the \citealt{nantais2017} sample is reminiscent of the redshift evolution seen earlier in the \citealt{pintos-castro2019} clusters (and seen in cluster SFGs, see \S\,\ref{sec:sf_transition}), with a significant shift in the epoch in which quenching ramps up.  Massive halos are more likely to complete their collapse and virialization during the epoch of $z\sim1.5-2$ \citep{muldrew2015, chiang2017} and it is not unreasonable to assume the ramp up of efficient quenching starts earlier in more massive halos.  Although both drawn from the SpARCS survey, the higher redshift \citealt{nantais2017} clusters are likely already of similar mass as the lower redshift \citealt{pintos-castro2019} sample on average, consistent with this hypothesis. \red{A more direct demonstration was presented in \citealt{reeves2021}, which found a strong dependence of EQE on halo (and stellar) mass by comparing X-ray and spectroscopically-selected low mass clusters/groups to GOGREEN clusters at fixed redshift ($1<z<1.5$). }
This again stresses the need to control for halo mass in these analyses.

\subsection{Summary}

In summary, NIR studies of SMFs, quenched fractions, and environmental quenching efficiencies in clusters and high-density environments confirm an excess in quenched populations attributed to environmentally-driven processes.  The picture of quenching is complex, however, and different processes may mix or dominate in different epochs, as well as depend on halo and stellar mass.  Stellar mass dependence in particular can constrain specific quenching mechanisms (see \S\,\ref{sec:wrap}) and stellar mass-dependent quenching is strongly supported by the low-mass slope of cluster SMFs and EQE analysis in large studies where environment and stellar mass can be treated simultaneously.  This is not universally observed, however, and can be complicated by our still developing understanding of the role of pre-processing.  In terms of redshift, quenched fractions among massive galaxies remain comparable to the local Universe up to $z\sim1$, though an evolving EQE is observed over large redshift baselines, increasing by a factor of $\sim3$ over $z\sim1\rightarrow0.5$.  Above $z\sim1$, a large variation is seen in the quenched populations in (small numbers of) massive clusters, with again evidence for an evolving EQE where a large redshift range is probed.  A field-like quenched fraction and EQE consistent with zero in clusters at $z\sim1.5$ suggests a transition epoch, wherein a rapid ramp up in environmental quenching processes occurs in massive halos completing their initial collapse during this epoch.  

As a final note, rest-frame UVJ colors have been invaluable (and relatively inexpensive method) for separating the star-forming and quenched populations in overdense environments.  However, this separation is not without contamination and results can be sensitive to the exact UVJ boundaries chosen \citep{pintos-castro2019}. UVJ also does not provide a robust proxy for (specific)-SFR without additional (in)direct tracers at UV and/or IR wavelengths \citep{leja2019}.   In the next section, we examine (obscured) star formation directly from the M/FIR regime.

\section{The Mid- to Far-Infrared: Dust-obscured Star Formation and AGN}\label{sec:fir}

As stated in the introduction, it has long been established that the morphology-density relation observed in local clusters \citep{dressler1980} is accompanied by a SFR-density relation \citep{dressler1984}, whereby cluster populations have a significantly lower SFG fraction than populations in lower density environments.  Subsequently, this relation was found to evolve over time, with the optically blue cluster SFG fractions increasing with redshift \citep[the Butcher-Oemler Effect;][]{butcher1984}. This evolution was later observed in IR-selected populations as well, including a rise in the (U)LIRG fraction \citep{coia2005, geach2006, geach2009, marcillac2007, saintonge2008, koyama2008, muzzin2008, oemler2009, smith2010a, koyama2010, tran2010, finn2010, chung2011, popesso2011, haines2009, haines2013, haines2015}, mirroring the steep rise in SF in the field to $z\sim1-3$ \citep[e.g.][Zavala \& Casey, in prep.]{madau2014}, though with a lower normalization.  The nature of this evolution provides a vital complement to the studies focused on quenched populations discussed in \S\,\ref{sec:nir}.

In this section, we break the discussion of M/FIR observations of (proto-)cluster galaxies into a few topics. In \S\,\ref{sec:sf_lowz} and \S\,\ref{sec:sf_highz}, we examine galaxy clusters at $z<1$ and $1<z<2$, respectively, in terms of the observed (obscured) star-forming fraction ($f_{\rm SF}$) and cluster galaxy SFRs and specific-SFRs (SSFR$\equiv \mathrm{SFR}/M_{\star}$) as a function of cluster-centric radius (out to the infall regions well beyond $R_{200}$) and normalized by halo mass.  In general, these sections will be cast in a framework of looking for evidence of slow  (few to several Gyr) versus rapid ($<100-500$ Myr) environmental quenching.  The former will produce a signature of suppressed SSFRs, distinguishing cluster SFGs from the field MS, while the latter will manifest itself primarily in $f_{\rm SF}$.  Environmental quenching mechanisms and their timescales were introduced in \S\,\ref{sec:environment}. Additionally, our discussion of $1<z<2$ clusters will consider whether the SFR-density relation is weakened or even reversed during this epoch.

In \S\,\ref{sec:protoclusters}, we explore the current M/FIR observations of proto-cluster galaxies at $z>2$, outlining the challenges in working with limited data and resolution over the large areas and volumes typical of proto-clusters \citep[$>10-30^{\prime}$, $\sim10^{3}-10^{4}$ cMpc$^{3}$;][]{chiang2013, chiang2017, lovell2018}. From these observations, vigorous total SFRs have been measured and we place them in context of halo mass (observed and at $z=0$) and the cosmic SFRD, with the caveats of large uncertainties.  We then discuss DSFG-rich proto-clusters in the context of the full proto-cluster population,
relative to our understanding of gas availability at high redshift and emerging evidence for early quenching populations.  Finally, \S\,\ref{sec:agn} presents a brief overview of AGN in (proto-)clusters and the importance of identifying obscured AGN using current and future MIR capabilities. M/FIR emission due to intracluster dust is saved for \S\,\ref{sec:icd}.

\subsection{Low redshift ($z<1$): evidence for multiple quenching mechanisms operating in clusters}\label{sec:sf_lowz}

\subsubsection{The IR Butcher-Oemler Effect}

\begin{figure}[!htb]
    \centering
    \includegraphics[width=0.9\columnwidth]{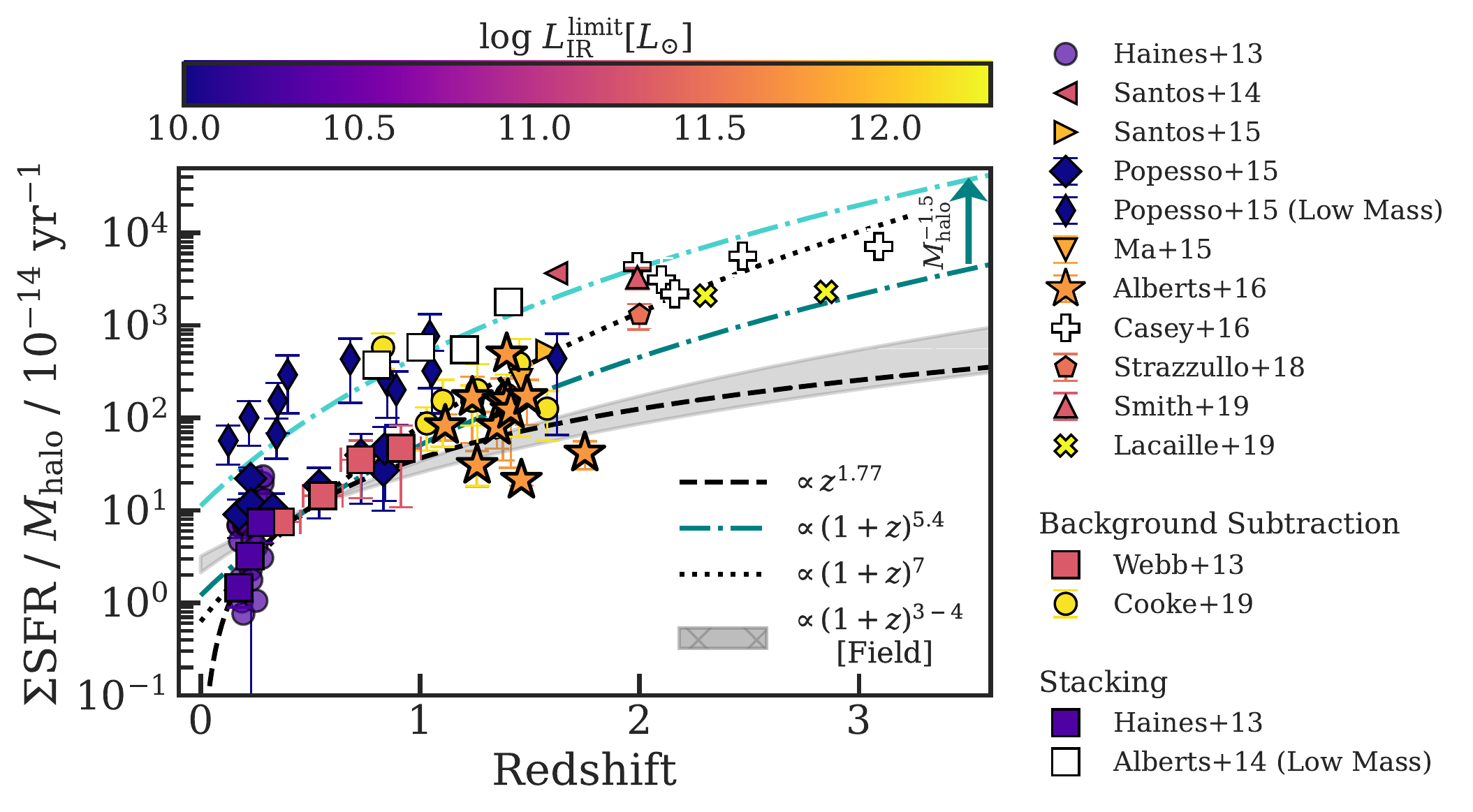}
    \caption{ $-$ The halo mass-normalized total SFR from infrared observations of (proto-)clusters in the literature as a function of redshift. Colors indicate the reported limit in $L_{\rm IR}$ for each study (white indicates that the limit was not available).  Studies that use background subtraction rather than spec-$z$ or photo-$z$ cluster membership are indicated in the legend. Catalog or imaged stacked samples are shown as squares.  All SFRs were converted to a \citealt{kroupa2001} IMF and the \citealt{kennicutt1998} conversion from $L_{\rm IR}$ to SFR.  The lines show different redshift evolution relations for $\Sigma \mathrm{SFR}/M_{\rm halo}$: $\propto z^{1.77}$ \citep[black dashed line;][]{popesso2011}; $\propto (1+z)^{5.4}$ \citep[dark teal dash-dot line;][]{bai2009,webb2013,alberts2014}; $\propto(1+z)^{7}$ \citep[black dotted line;][]{cowie2004,geach2006,biviano2011}, normalized at $z=0.5$. The general evolution for field galaxies $\propto (1+z)^{3-4}$ \citep{lefloch2005,rujopakarn2010,sargent2012,ilbert2015} is shown in the gray region.  The light teal dash-dot line shows the $\propto(1+z)^{5.4}$ relation scaled from log $M_{200}/\Msun=14.5\rightarrow14$ following $\Sigma \mathrm{SFR}/M_{\rm halo}\propto M^{-1.5}_{\rm halo}$ \citep{webb2013}, see text for details.}
    \label{fig:sfrmhalo}
\end{figure}

We start by looking at SFG populations in clusters at low to intermediate redshift for evidence of environmental processes.  A natural first question is whether the IR Butcher-Oemler Effect is driven by the evolution in the (obscured) star formation of the infalling field population, which strongly decreases over $z\sim1\rightarrow0$. \citealt{webb2013} explored this issue using \textit{Spitzer}/MIPS\footnote{Multi-Band Imaging Photometer for \textit{Spitzer} \citep[MIPS;][]{rieke2004}} 24$\mu$m imaging to study the IR population in 42 massive (log $M_{200}/\Msun=14-15$) red-sequence selected clusters \citep{gladders2000, gladders2005} at $0.3<z<1$. Using statistical background subtraction $-$ which avoids selection bias due to requiring an optical counterpart to identify cluster membership $-$ they quantified the redshift evolution of the halo mass-normalized total SFR ($\Sigma \mathrm{SFR}/M_{\rm halo}$) in $L_{\rm IR}>2\e{11}$ $\Lsun$ cluster galaxies, fitting a redshift evolution $\propto(1+z)^{5.4\pm1.9}$ (Figure~\ref{fig:sfrmhalo}). Given the uncertainties, this is consistent with the evolution of field galaxies \citep[$\propto(1+z)^{3-4}$; e.g.][]{lefloch2005,rujopakarn2010,sargent2012,ilbert2015}.  A similar analysis and conclusion was reached in \citealt{haines2009}, studying the star-forming fraction in the lower redshift ($0.1<z<0.3$) LoCuSS\footnote{The Local Cluster
Substructure Survey \citep[LoCuSS;][]{smith2010}} cluster sample using background subtraction. The same was again concluded by \citealt{popesso2015a}, 
who used group \citep{popesso2015} and cluster \citep[e.g.][]{bai2007a, haines2010, haines2013, finn2010} IR LFs to integrate to low 
luminosities (log $L_{\rm IR}/\Lsun\sim7$). 

An evolution in $\Sigma \mathrm{SFR}/M_{\rm halo}$ that is the same between clusters and the field, however, is inconsistent with the non-zero environmental quenching efficiency discussed in \S\,\ref{sec:nir}.  Follow-up studies of the LoCuSS clusters using spectroscopy (identifying $L_{\rm IR}>5\e{10}\,\Lsun$ cluster members via optical emission lines) found a steeper evolution in the (obscured) star-forming fraction \citep[$\propto(1+z)^{6-7}$; see also][]{geach2006, koyama2010, biviano2011, shimakawa2014, smail2014}, which they attributed to more accurate sampling of the cluster population over their previous study using background subtraction \citep{haines2015}.  After controlling for the field, a residual evolution on order (1+z)$^{2-4}$ remained up to $z<0.4$, indicating sub-dominant but significant environmental quenching. Similarly, a residual evolution in the average cluster galaxy SFR on order (1+$z)^2$ was found in \citealt{alberts2014} over $z=0.3\rightarrow1.5$. Figure~\ref{fig:sfrmhalo} presents a compilation of these studies compared with example redshift evolutions $\propto(1+z)^{5-7}$ and a shallower relation $\propto z^{-1.77}$ proposed by \citealt{popesso2012}.  The latter better captures the uniquely steep evolution at low redshift observed by \citealt{haines2015}, which provides the strongest evidence for a deviation from the evolution of the field (infalling) population up to $z\sim1$.

In addition to studying the evolution with redshift, \citealt{webb2013} examined the dependence of the mass-normalized total SFR on halo mass, finding $\Sigma \mathrm{SFR}/M_{\rm halo}\propto M_{\rm halo}^{-1.5\pm0.4}$ \citep[see also][]{popesso2015a}.  In Figure~\ref{fig:sfrmhalo}, we scale the $(1+z)^{5.4}$ relation (dark teal dot-dash line) by this mass dependency to low-mass clusters (log $M_{200}/\Msun\sim14$; light teal dot-dash line), assuming the original \citealt{webb2013} sample has an average mass of log $M_{200}/\Msun\sim14.5$ \citep[as do][on average]{haines2013, popesso2015a}.  The result is roughly consistent with the low-mass cluster samples presented in \citealt{popesso2015a} (thin diamonds) and stacked in \citealt{alberts2014} (white squares), confirming qualitatively a halo mass dependence.  As in the measurement of quenched populations (\S\,\ref{sec:nir}), halo mass should therefore not be ignored in examining the total SFR budget in clusters.  

\subsubsection{The global (radial) dependence of the obscured SFR}

\begin{figure}[!htb]
    \centering
    \includegraphics[width=0.475\columnwidth]{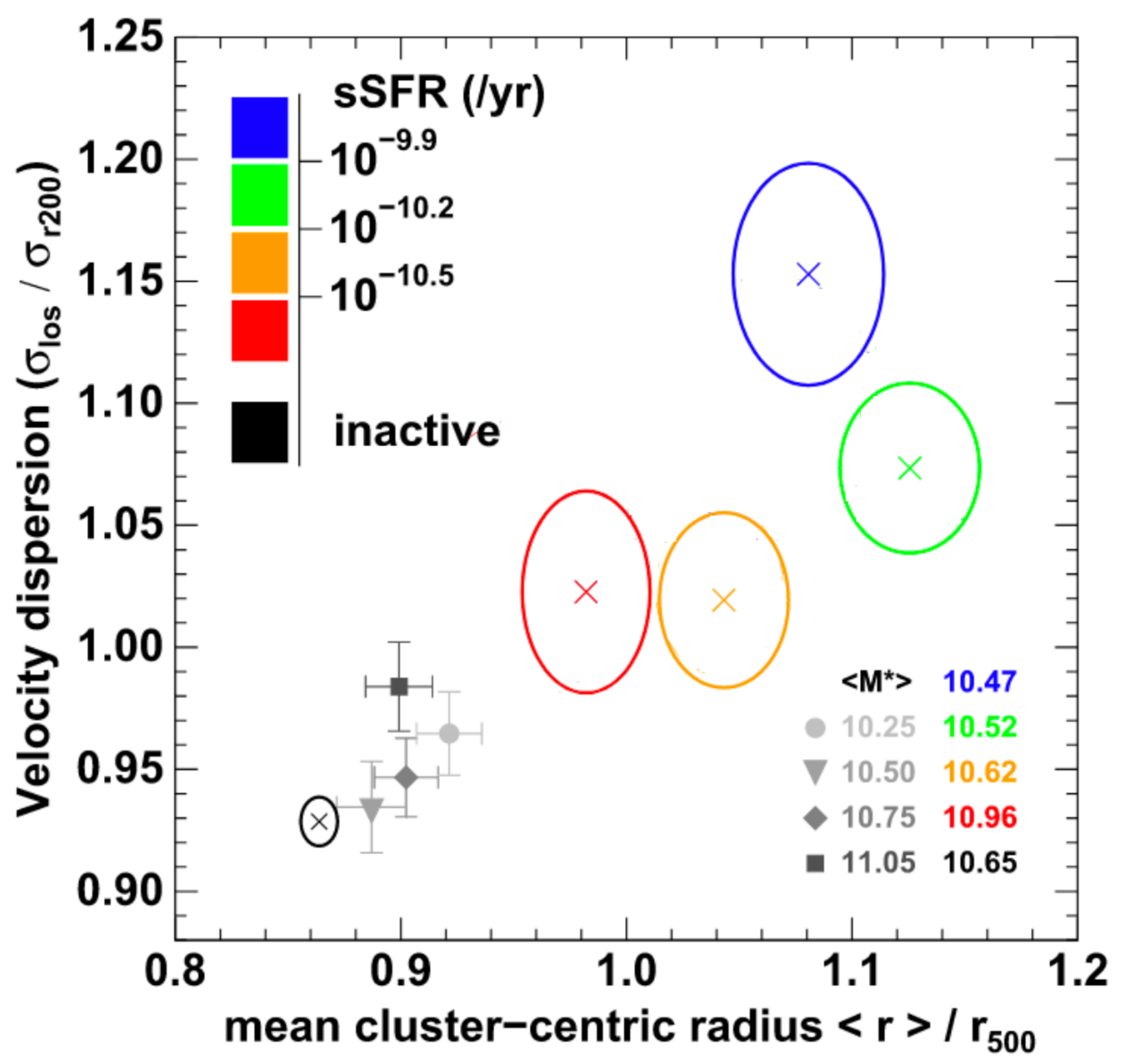}
    \includegraphics[width=0.43\columnwidth]{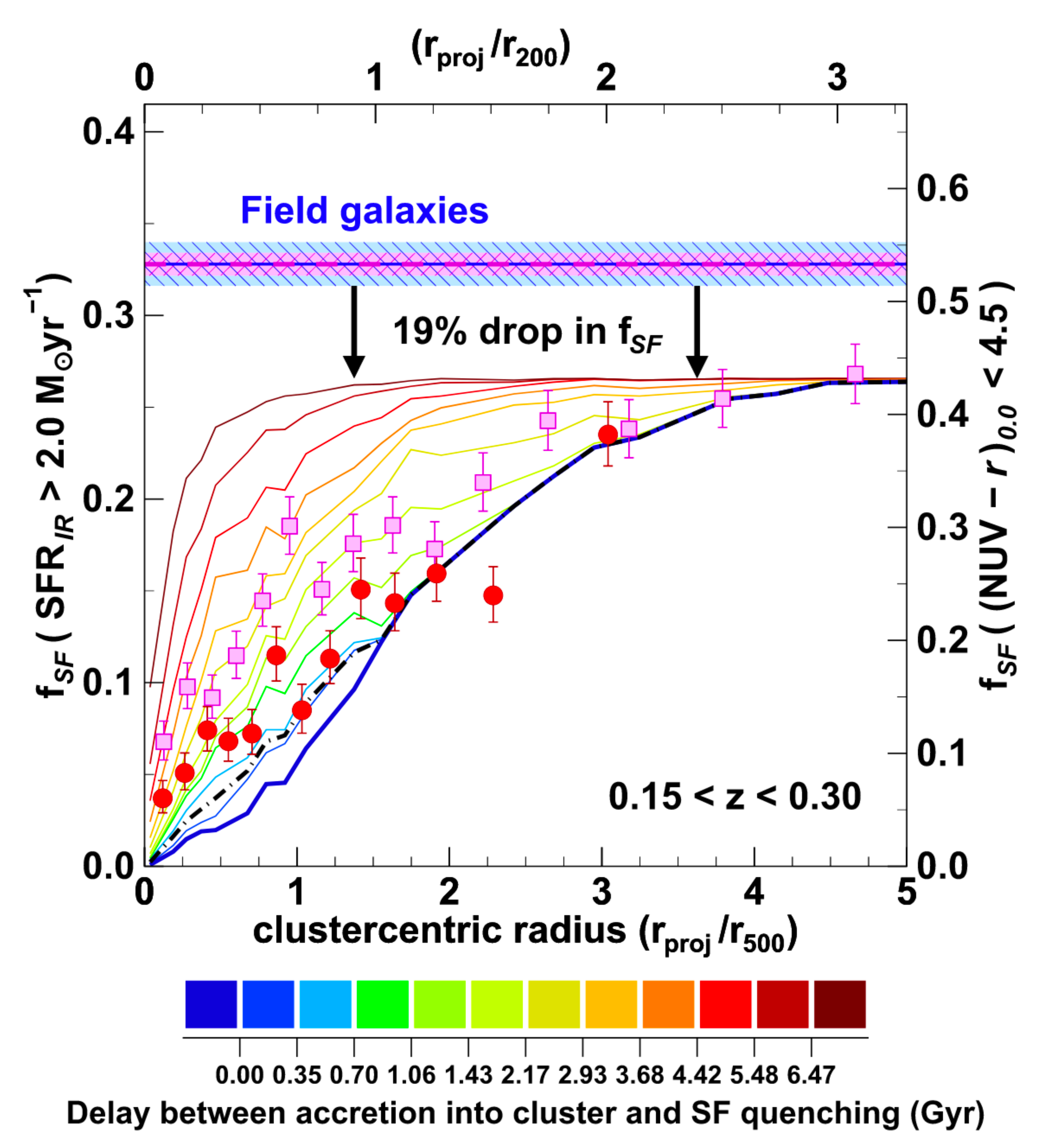}
    \caption{ $-$ {\bf (left)} Phase space (caustic) diagram of SFGs in the LoCuSS sample split into four bins of SSFR.  X's indicate the mean radii and line-of-sight velocity dispersions in each bin, with the ellipses showing the $1\sigma$ uncertainties.  Gray points are QGs.  SFGs with suppressed SSFRs are kinematically separated from those without.  {\bf (right)} The star-forming fraction from obscured (red) and unobscured (pink) SF tracers. A sharp decrease is seen into the cluster cores. Delayed then rapid models of quenching are shown in the solid lines; a delay time of $\sim1$ Gyr best \red{models}\sout{reproduces} the IR $f_{\rm SF}$.  A $\sim20\%$ suppression below the field to large radii requires pre-processing to reproduce.  The dot dash line indicates the model where all star formation is instantaneously quenched when galaxies reach pericenter. Figures adapted from Figure 8 and Figure 16 in \citealt{haines2015}, reproduced by permission of the AAS.}
    \label{fig:haines}
\end{figure}

In further support of environmental-specific quenching mechanisms are two observational characteristics of cluster SFG populations at $z<1$.  First, the overall fraction of IR luminous galaxies is a strong function of cluster-centric radius and suppressed below the field fraction out to large radii \citep[$3-5R_{\rm vir}$;][]{finn2010,patel2009,chung2011,noble2013,haines2013,haines2015}. Second, systematic suppression of SSFRs at fixed mass and redshift in massive (log $M_{\star}/\Msun>10$) {\it star-forming} cluster galaxies is observed at the level of $\sim0.2-0.3$ dex up to $z\sim1.4$ \citep{vulcani2010, haines2013,haines2015, noble2013, alberts2014, rodriguez-munoz2019}.  In the LoCuSS sample, these suppressed SFGs are kinematically separate, intermediate in phase space\footnote{An observational (projected) caustic or phase space diagram uses the line-of-sight velocities relative to the cluster velocity dispersion and cluster-centric radii relative to the virial radius of a cluster population to kinematically separate galaxies by their time since infall \citep[e.g.][]{mahajan2011,hernandez-fernandez2014,muzzin2014}.   Simulations show projected phase space is in good agreement with theoretical full 3D phase space diagrams  \citep[e.g.][]{oman2013, rhee2017}.}  between quenched cluster members in the cores and those forming stars at unsuppressed rates in the outskirts \citep[Figure~\ref{fig:haines}, left;][]{haines2015}.  Such separation could point to a slow quenching mechanism such as starvation, the gradual stripping of the diffuse hot halo and cessation of new gas accretion on entering the cluster ICM \citep{larson1980, bekki2002}. Conversely, a strong radial gradient in $f_{\rm SF}$ in the same sample is best reproduced with a delayed quenching model\footnote{A ``delayed, then rapid'' model, in which SFRs are unaltered for some delay time upon infall followed by  quenching on short timescales, was first proposed in \citealt{wetzel2013}.} with a high quenching efficiency, in which SF was able to continue for 0.3-2 Gyr on crossing $R_{200}$ before quenching \citep[Figure~\ref{fig:haines}, right;][]{haines2015}.   Given that both suppressed SFRs and a strongly spatially evolving $f_{\rm SF}$ are seen, both slow and rapid quenching  
might be operating simultaneously. Out to large radii ($r\gtrsim2R_{200}$), a persistent 20$\%$ deficit in $f_{\rm SF}$ \citep[see also][discussed in \S\,\ref{sec:epsilon_env}]{pintos-castro2019} requires pre-processing, likely in the group environment \citep[Figure~\ref{fig:haines}, right; see][for a LoCuSS group study]{bianconi2018}.  As such, the \citealt{haines2015} analyses indicate environmental quenching is occurring via multiple processes in the LoCuSS clusters. 

At higher redshift, a MIPS-selected population in SpARCS J161314+564930 \citep[$z=0.872$;][]{noble2013} was observed to have a bimodal nature, including both normal SFGs on the star-forming Main Sequence as well as sub-Main Sequence SFGs with optical spectra more consistent with passive galaxies \citep[see also][]{patel2009}.   When placed on a phase space diagram, the sub-MS SFGs occupy the \red{same region as}\sout{space of} a virialized or backsplash population\red{.  Compared to recently-accreted SFGs, their SSFGs are suppressed by $\sim0.9$ dex,}\sout{, with a suppression in their SSFRs of $\sim0.9$ dex below recently\red{-accreted}\sout{accretion} SFGs,} echoing the kinematic separation \red{by SSFR} found by \citealt{haines2015}.  While these suppressed SFRs again point to slow quenching, a strong radial dependence in the MIPS-detected fraction suggests rapid quenching is also occurring.  The combination of slow and rapid quenching is further supported over a long redshift baseline ($0.3<z<1.5$) in local environment studies with ORELSE \citep{tomczak2019} and in the ISCS cluster sample \citep[][see also \citealp{rodriguez-munoz2019}]{alberts2014}.  Stacked mass-limited ISCS cluster catalogs (containing both SFGs and quiescent galaxies) reveal a strong evolution relative to the field (Figure~\ref{fig:sfrmhalo}).  However, by splitting off the optically blue population (an incomplete stand-in for the SFG population only), \citealt{alberts2014} found suppressed SFRs with at most a weak trend relative to the field evolution with redshift.  This indicates that the change in the mass-limited sample is driven by a swiftly changing $f_{\rm SF}$, found by \citealt{alberts2014} to be roughly consistent with RPS models \citep{tecce2010} up to $z\sim1$, occurring simultaneously with slower quenching that suppresses SFRs.  

\subsection{Intermediate Redshift ($z\sim1-2$): a transition epoch for massive clusters}\label{sec:sf_highz}

Despite the strong evolution in SF activity seen relative to the field discussed in the previous section, the quenching efficiency remains high for massive galaxies in massive clusters to $z\sim1$ (\S\,\ref{sec:nir}) and the local SFR-density relation appears to be in place \citep{muzzin2012, webb2013, noble2013, balogh2016}.  Early optical and NIR analysis of $z<1$ cluster populations predicted that cluster galaxies formed at high redshift in short bursts, quenched, and evolved passively since $z\sim2$ \citep{bower1992, aragon-salamanca1993,stanford1998, blakeslee2006, mei2006, mei2006a, mei2009, eisenhardt2008, muzzin2008}.
Infrared studies moving to $z>1$, however, quickly provided evidence of a deviation from passive evolution in the NIR LFs (see \S\,\ref{sec:nir}) and significant SF activity, weakening (or even reversing) the SFR-density relation, in both local environment studies \citep{elbaz2007, cooper2008} and down into cluster cores \citep[e.g.,][]{hayashi2010, hilton2010, tran2010, lemaux2010, fassbender2011, hayashi2011, tadaki2012, bayliss2014, zeimann2013,fassbender2014,santos2014,ma2015,santos2015,stach2017}.  However, evolved massive clusters with quenched cores have also been identified at these high redshifts \citep[e.g.][]{grutzbauch2012, stalder2013, strazzullo2013, santos2013, newman2014,cooke2016, lee-brown2017, zavala2019}.

\subsubsection{Is there a(n infrared) reversal in the SFR-density relation?}\label{sec:sfr-density_relation}

A true reversal in the SFR-density relation at high redshift would signal that environmental processes are capable of driving an {\it excess} of star formation in addition to quenching it---a distinct scenario from the disappearance of the SFR-density relation due to the ramping down of environmental quenching as seen in \S\,\ref{sec:nir}. An enhancement of SF in massive structures is consistent with hierarchical growth \citep{moster2013} and galaxy interactions and/or RPS could drive instabilities that trigger nuclear starbursts \citep[][and references therein]{boselli2022}. Early evidence for a reversal was presented in local environment studies \citep{elbaz2007, cooper2008} and for massive (log $M_{200}/\Msun>13.8$) clusters XMMXCS J2215.9-1738 \citep[$z=1.46$;][]{hilton2010} and ClG J0218.3-0510 ($z=1.62$; \citealt{tran2010}, but see \citealt{tran2015}) based on an increase in the fraction of DSFGs with increasing local density, into the cluster cores.  Conversely, XMMU J2235.2-2557 at $z=1.393$ \citep{santos2013}, a more massive cluster, was found to show no reversal.  

The difficulty in establishing a change in the SFR-density relation is illustrated nicely by ClG J0218.3-0510 \citep{papovich2010, tanaka2010, tran2010}.  \citealt{quadri2012} examined this cluster using a mass-limited (log $M_{\star}/\Msun>10$) catalog and defining the environment using the nearest neighbors technique.  Quiescent galaxies were separated from SFGs using color-selection plus MIPS 24$\mu$m imaging and, from this, the quenched fraction and EQE (see \S\,\ref{sec:nir}) were both found to be in excess over the field, albeit with large uncertainty, signaling the SFR-density relation is still in place \citep[see also][using UVJ selection]{lee-brown2017}.  By way of contrast, \citealt{santos2014} examined the same cluster as a function of cluster-centric radius out to large radii ($r\sim10$ Mpc) in optical to FIR imaging, with membership based on spec-$z$s and photo-$z$s.  Quiescent cluster galaxies were identified using a cut on SSFR, from spectral energy distribution (SED) fitting, with a field sample drawn from the same data.  From this, they found that the $f_{\rm SF}$ was statistically identical between the cluster ($r<1\,$Mpc) and field, with an excess of SFGs in the infall region (1-3 Mpc).  This excess may be due to an overabundance of high-mass  SFGs in the infall region \citep[][]{werner2022}, while the lack of excess within 1 Mpc \citep[or a potential deficit in the core, see][]{smail2014} may be consistent with the positive quenching efficiency noted by \citealt{quadri2012}.  However, the lack of environmental dependence of the SFG SMF (\S\,\ref{sec:nir_smf}) and the field-like $f_{\rm SF}$ in the cluster, drawn from the same data, suggests the SFR-density relation is no longer in place \citep{santos2014}. These studies, mass-limited versus luminosity-limited, with different definitions of environment and different field comparison samples, demonstrate the challenges in interpreting the data.

Like \citealt{santos2014}, \citealt{ma2015} and \citealt{santos2015} interpreted a high halo mass-normalized SFR in XMMXCS J2215.9-1738 ($z=1.46$) and XDCP J0044.0-2033 ($z=1.58$) as evidence for a reversal in the SF-density relation (Figure~\ref{fig:sfrmhalo}).  \citealt{smith2019} examined a low-mass cluster with an emerging Red Sequence at $z=1.99$ \citep{gobat2013} using SCUBA-2, ALMA, and JVLA imaging.  They found the SFR per unit area in the cluster core to be orders of magnitude greater than the field; together with a high mass-normalized SFR, they reported a reversal.  However, the increase in SFR density paralleled the increase in galaxy density, so this effect can likely be explained without environmental enhancement of star formation. This was also demonstrated for ClG J0218.3-0510 in \citealt{tran2015}; using H$\alpha$ SFRs, they found that the SFR {\it per galaxy} still decreased slightly into the cluster cores, ruling out a reversal.  Taken together, though individual cluster studies prove that vigorous star formation is starting to be evident in some $z=1-2$ massive clusters, perhaps weakening the SFR-density relation, differences in cluster selection and cluster membership identification, constraints on SF, and intrinsic variations make it difficult to establish or refute a true reversal.

\subsubsection{The transition to efficient (rapid) quenching at $z\sim1.4$ in clusters}\label{sec:sf_transition}

Both targeted and survey imaging with \textit{Spitzer}/MIPS, \textit{Herschel}/PACS\footnote{Photodetector Array Camera \& Spectrometer \citep[PACS;][]{poglitsch2010}} and SPIRE, and SCUBA-2 have been used to move beyond single cluster M/FIR studies in a limited number of cluster samples at $z=1-2$.  \citealt{noble2016} obtained deep PACS imaging of 3 SpARCS/GCLASS clusters at $z\sim1.2$, using optical spectroscopy to place IR luminous cluster galaxies in phase space and look at trends in SFR.  A significant drop in the SFR, SSFR, and $f_{\rm SF}$ of the (optically-confirmed) IR population was found in the intermediate (between infalling and core) and cluster core regions of phase space.   This is consistent with the environmental signatures observed in optical studies of the full GCLASS samples (10 clusters at $z=0.85-1.2$) in \citealt{muzzin2012}.

\begin{figure}[!htb]
    \centering
    \includegraphics[width=0.8\columnwidth]{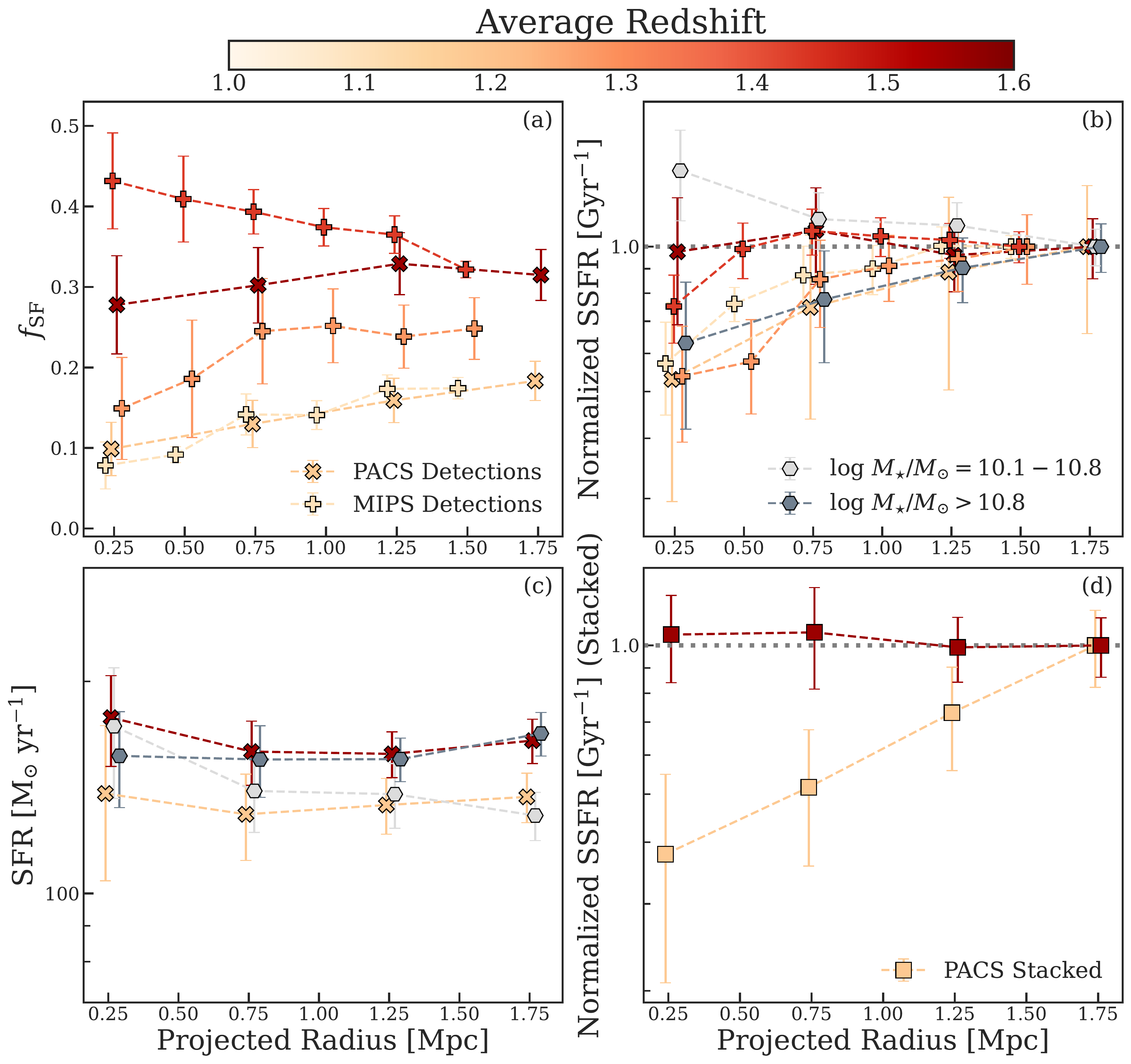}
    \caption{ $-$ Star formation properties of ISCS/IDCS cluster galaxies at $z>1$ as a function of \red{projected,} cluster-centric radius.  ({\bf a}) The star-forming fraction for MIPS- \citep{brodwin2013} and PACS-detected \citep{alberts2016} populations. By $z\sim1.4$, $f_{\rm SF}$ is flat or even rising into the cluster cores. ({\bf b}) The field-normalized SSFRs, where the field is taken from the largest radial bin. Light and dark gray hexagons show the PACS-detected sample divided into log $M_{\star}/\Msun=10.1-10.8$ and log $M_{\star}/\Msun>10.8$ bins, respectively, instead of redshift. The lower mass bin shows evidence of a rise in SSFR into the cores. ({\bf c}) The average SFR of the PACS-detected sample, which show no radial trends. ({\bf d}) Mass-limited cluster members stacked on PACS 100$\,\mu$m imaging.  No radial trend is evident at $z>1.4$, while a strong radial trend appears by $z\sim1.2$. }
    \label{fig:sf_radial}
\end{figure}


Very few cluster samples with FIR covering a longer redshift baseline are available. Extensive M/FIR imaging in the Bo\"{o}tes field was used to examine the $\sim300$ ISCS/IDCS cluster candidates over $z\sim0.3-2$ \citep{eisenhardt2008, stanford2012}, including \textit{Herschel}/SPIRE from HerMES over nearly the full survey and follow-up deep \textit{Spitzer}/MIPS and \textit{Herschel}/PACS for a confirmed subset at $z\sim1-2$ (see \S\,\ref{sec:nir_selection} for further details on the ISCS/IDCS).  
Analyzing 16 spectroscopically-confirmed massive (log $M_{200}/\Msun\sim14$) ISCS clusters at $z=1-1.5$ with MIPS 24$\mu$m imaging, \citealt{brodwin2013} measured the field-relative $f_{\rm SF}$ and SSFR of log $L_{\rm IR}/L_{\odot}>11.5$ cluster galaxies as a function of cluster-centric radius, using their largest radial bin ($r\sim1.5$ Mpc) as a proxy for the field \citep[see also][]{wagner2016}.  Figure~\ref{fig:sf_radial} (panel a) shows a rapid evolution over the redshift range probed ($\sim1.5$ Gyr), with the lower redshift end showing the expected decrease in $f_{\rm SF}$ with decreasing radius.  At $z\gtrsim1.4$, however, the ISCS clusters reveal an {\it increase} in the MIPS-derived $f_{\rm SF}$ and a flattening of the field-relative SSFR (panel b). Supporting this, \citealt{alberts2014} stacked on mass-limited (log $M_{\star}/\Msun\geq10.1$) cluster member catalogs from 274 ISCS clusters from $z=0.3-1.5$ using shallow \textit{Herschel}/SPIRE imaging, finding that the stacked (average) 250$\mu$m luminosity in their highest redshift bin draws even with and may even be in excess over the stacked (average) 250$\mu$m luminosity of coeval field galaxies.

Subsequent studies of the ISCS/IDCS clusters paint a similar picture. Looking at visually-classified early-type galaxies, \citealt{wagner2015} observed significant residual star formation at $z\sim1.5$, which declined by $z\sim1.25$. Using deep \textit{Herschel}/PACS imaging of 11 ISCS/IDCS clusters over $1<z<1.75$, \citealt{alberts2016} established that this transition holds at the bright end. Figure~\ref{fig:sf_radial} (panels a-c) shows again that the $f_{\rm SF}$ and average SSFR of PACS-detected cluster (U)LIRGs are independent of environment at $z>1.4$\red{, though the uncertainties and use of projected radii could hide a weak trend}.  Stacking on mass-limited cluster member catalogs on PACS images confirms no radial dependence at $z>1.4$ (panel d).  The average SFR is likewise independent of environment at all redshifts probed, giving no indication that slow quenching has had time to act (panel c). Accounting for the IR luminous population (log $L_{\rm IR}/\Lsun\geq11.7$) only, the halo mass-normalized total SF draws even with the field at these redshifts (Figure~\ref{fig:sfrmhalo}), though with significant cluster-to-cluster variation even at fixed halo mass \citep{brodwin2013, alberts2016} that spans the range of suggested redshift trends $\sim(1+z)^{5-7}$.  An analysis of SCUBA-2 850$\mu$m sources in 8 X-ray selected clusters at higher mass (log $M_{200}/\Msun\gtrsim14.5$) likewise shows large scatter at the same epoch \citep[][]{cooke2019}.   Weak excesses in the SFR at high redshift ($z>1.4$) or in $10.1< \mathrm{log}\, M_{\star}/\Msun<10.8$ galaxies (Figure~\ref{fig:sf_radial},panel b) in the ISCS/IDCS cluster cores may support a reversal; however, this sample is not large enough to split by redshift and stellar mass simultaneously. 

An important consideration in redshift evolution studies is whether one is working with progenitor samples, at fixed halo mass, or with no halo mass accounting.  The ISCS/IDCS studies just discussed are at roughly fixed halo mass; the full ISCS catalog has roughly log $M_{200}/\Msun\sim13.8$ over $0.3<z<1.5$ while the targeted ISCS/IDCS clusters at $1<z<1.75$ are more massive at log $M_{200}/\Msun\sim14.3-14.7$.  The wide redshift baseline studies of both the full catalog and targeted clusters reveal that the ISCS clusters are undergoing a transition at roughly $z\sim1.4$ {\it at approximately fixed halo mass}, above which quenching is inefficient and a true SFR-density reversal is possible, suggesting the quenching ramp up is not solely a function of halo collapse or growth.  On the other hand, a comparison of the halo mass-normalized SFR budget (Figure~\ref{fig:sfrmhalo}) demonstrates a dependence of the bulk star formation properties on halo mass.  As we have seen that $\Sigma \mathrm{SFR}/M_{\rm halo}$ is inversely correlated with halo mass \citep[Figure~\ref{fig:sfrmhalo}; see][]{webb2013}, we can speculate that this transition will be earlier (later) for higher (lower) mass halos.  The redshift bins in the ISCS targeted and full catalog studies, however, prevent a direct comparison and this type of analysis has not been done in any progenitor samples. A transition scenario, however, mirrors NIR studies of environmental quenching efficiency, which see a drop at $1.4<z<1.65$ \citep[][see \S\,\ref{sec:nir}]{nantais2017}.

So has a true reversal in the SFR-density relation been observed?  Both NIR and M/FIR studies have found that the local SFR-density relation is significantly weakened, or even negligible, at $z\sim1.4-1.6$ in samples of massive clusters, though with significant scatter likely including an order of magnitude intrinsic variation in the total SF at relatively fixed cluster properties (selection, halo mass, epoch; Figure~\ref{fig:sfrmhalo}). Evidence for an excess of (obscured) SF in the cluster environment is more tenuous, owing to small samples and large uncertainties that preclude i.e. the ability to separately account for stellar mass and environment over the relevant redshift range. Compelling evidence for a reversal has been recently provided, however, by both local environment studies and simulations.  \citealt{lemaux2022} examined the SFR-density relation over $z\sim2-5$ in spectroscopic samples divided by local density from the VIMOS Ultra Deep Survey \citep[VUDS;][]{lefevre2015}.  Probing up to proto-cluster core densities \citep[including the area containing the Hyperion supercluster;][]{cucciati2018}, they found a significant trend of increasing SFR (derived from UV-NIR SED) with increasing local density at $z\gtrsim3$, driven mostly by an excess of log $M_{\star}/\Msun\gtrsim10$ galaxies in overdense regions.  The trend was still significantly detected, however, when stellar mass differences across density bins were accounted for, signaling a true SFR-density reversal \citep[but see][for a conflicting result using an overlapping dataset]{chartab2020}.  While a flattening, not a reversal, was observed at $z\sim2-3$, the VUDS spectroscopic sample lacks DSFGs, which dominate the total star formation during this epoch, and so the onset of this reversal remains an open question.
On the simulation front, \citealt{hwang2019} used the {\tt IllustrisTNG} \citep{pillepich2018} cosmological hydrodynamic simulation to study the SFR-density relation over $0<z<2$ from the cluster (log $M_{\rm halo}/\Msun\sim14$ at $z\sim1.5$) to the field environment \citep[see also][]{delucia2004, chiang2017}. A true reversal was observed in all intermediate- to high-density bins, with excess star formation in intermediate-density (group) environments at $z\gtrsim1$ and in clusters from $z\gtrsim1.5$ \citep[see also][]{tonnesen2014}.  This simulated reversal holds at fixed stellar mass, indicating it is environment driven.  Conversely, the (simulated) molecular gas mass always decreases with increasing local density, suggesting that mechanisms that accelerate the consumption of gas may drive the steep evolution in the star formation in clusters.  

\subsubsection{Summary}

To summarize these two sections on galaxy clusters, obscured SFGs in galaxy clusters at $z<1$ provide evidence for a mix of slow and rapid environmental quenching in the suppression of their (S)SFRs and $f_{\rm SF}$. These effects depend on cluster-centric radius (or location in phase space) and halo mass, but in general widespread quenching of the cluster SFG population persists to $z\sim1$.  In the $1<z<2$ era, however, the SFR-density relation is observed to weaken in some cluster samples, though quenched clusters are also observed throughout this epoch, signaling a large intrinsic cluster-to-cluster variation (often compounded by non-intrinsic variation caused by inhomogeneous datasets and techniques).  At these redshifts, evolutionary trends of the total SFR normalized by halo mass with redshift start to diverge and there are indications of an accelerated evolution compared to the field (Figure~\ref{fig:sfrmhalo}); however, limited cluster samples with IR data that cover a range in redshift limit our ability to robustly pin down the redshift evolution for the dusty populations that make up the bulk of the total SFR.  Evidence for a true reversal of the SFR-density relation is likewise tenuous at best, though it is supported by simulations. Nevertheless, in good agreement with NIR studies that see a drop in the EQE at $z\sim1.5$ (\S\,\ref{sec:nir}), M/FIR studies of cluster samples over a similar redshift range find field-like star formation activity at this epoch, necessitating rapid quenching of their populations by $z\sim1$.  This indicates that the dominant quenching mechanism(s) likely evolve over cosmic time (see \S\,\ref{sec:wrap}).  

\subsection{High redshift ($z>2$): the realm of proto-clusters}\label{sec:protoclusters}

Now we move into the realm of infrared studies of proto-clusters. In \S\,\ref{sec:protocluster_selection}, we discussed the evidence for luminous DSFGs (often synonymous with SMGs) as signposts of early massive halos, potentially proto-clusters in the phase of vigorous star formation that builds up massive cluster ellipticals \citep[e.g.][]{stevens2003}.  Significant DSFG overdensities have been identified in over 20 individual proto-clusters (confirmed and candidates) from $z\sim2-7$ (Table~\ref{tbl:protoclusters}) and wide, shallow submm surveys are a promising source of new proto-cluster candidates. In this section, we examine the nature of (obscured) star formation in proto-clusters (see  \citealt[][]{overzier2016} for a general proto-cluster review).

\subsubsection{The Nature of (Obscured) Star Formation in Proto-clusters}\label{sec:sf_protoclusters}

Proto-clusters subtend large areas on the sky \citep[$>10^{\prime}$;][]{casey2016, chiang2013, chiang2017, lovell2018} and correspondingly fill large cosmic volumes (10$^{3-4}$ cMpc$^3$), theoretically existing at the nodes of filaments.  Recent semi-analytic modeling from \citealt{chiang2017} provides a rough framework for how star formation may proceed in these early environments: at $z\sim10\rightarrow5$, proto-clusters experience inside-out growth, establishing a core as the primary halo(s) reach $10^{12}\,\Msun$ by $z\sim5$, the halo mass at which star formation efficiency (SFE) peaks (see \citealt{wechsler2018} for a review).  This is followed by a period of more spatially extended star formation over the full proto-cluster volume, as secondary halos grow to peak SFE.  High SFRs are enabled in this environment ($\sim1,000\,\Msun$ yr$^{-1}$),  building $\sim65\%$ of the stellar mass in $z=0$ massive clusters \citep{chiang2017}.  The epoch of $z\sim1.5\rightarrow0$ then sees the final transition to collapsed clusters and widespread quenching, roughly consistent with the observational literature presented in previous sections.

Testing this framework has made slow \red{progress}\sout{process} due to the large areas and high redshifts involved. For DSFGs, wide-field surveys with single-dish submm telescopes and/or \textit{Herschel} have paved the way for targeted follow-up with ALMA and spectroscopic campaigns.  FIR or submm spectroscopy is often necessary to confirm the most dusty sources \citep[e.g.][]{daddi2009, walter2012}.  Early work associated single or a few extreme DSFGs with galaxy overdensities at $z\sim4-5$ such as GN20  \citep{pope2005, daddi2009, hodge2013}; HDF850.1 \citep{walter2012}; AzTEC-3 \citep{riechers2010}; and CRLE \citep{pavesi2018}. Subsequently, DSFGs have been both associated with known proto-clusters and used directly in proto-cluster identification when detected in large numbers ($>5$) over an appropriate area (\S\,\ref{sec:protocluster_selection}).  As an example of the former, a correlation between Lyman-$\alpha$ Emitters (LAEs) and red SMGs has tentatively been detected in an LAE overdensity at $z=5.692$ \citep{harikane2019}.  And for the latter, spectroscopic follow-up of SPT-2349-56 confirmed 14 DSFGs at $z=4.3$ using [CII] and CO emission lines, with a combined SFR of $6,000\pm600\,\Msun$ yr$^{-1}$ \citep{miller2015}. Ten of these DSFGs occupy a region with a 19$^{\prime\prime}$ (130 kpc) diameter, signaling a core in the process of rapid assembly, which will likely collapse to form the BCG. Single-dish submm and wider ALMA observations suggest a surrounding extended structure \citep{hill2020}, consistent with our nominal theoretical framework.

\begin{figure}[!htb]
    \centering
    \includegraphics[width=0.9\columnwidth]{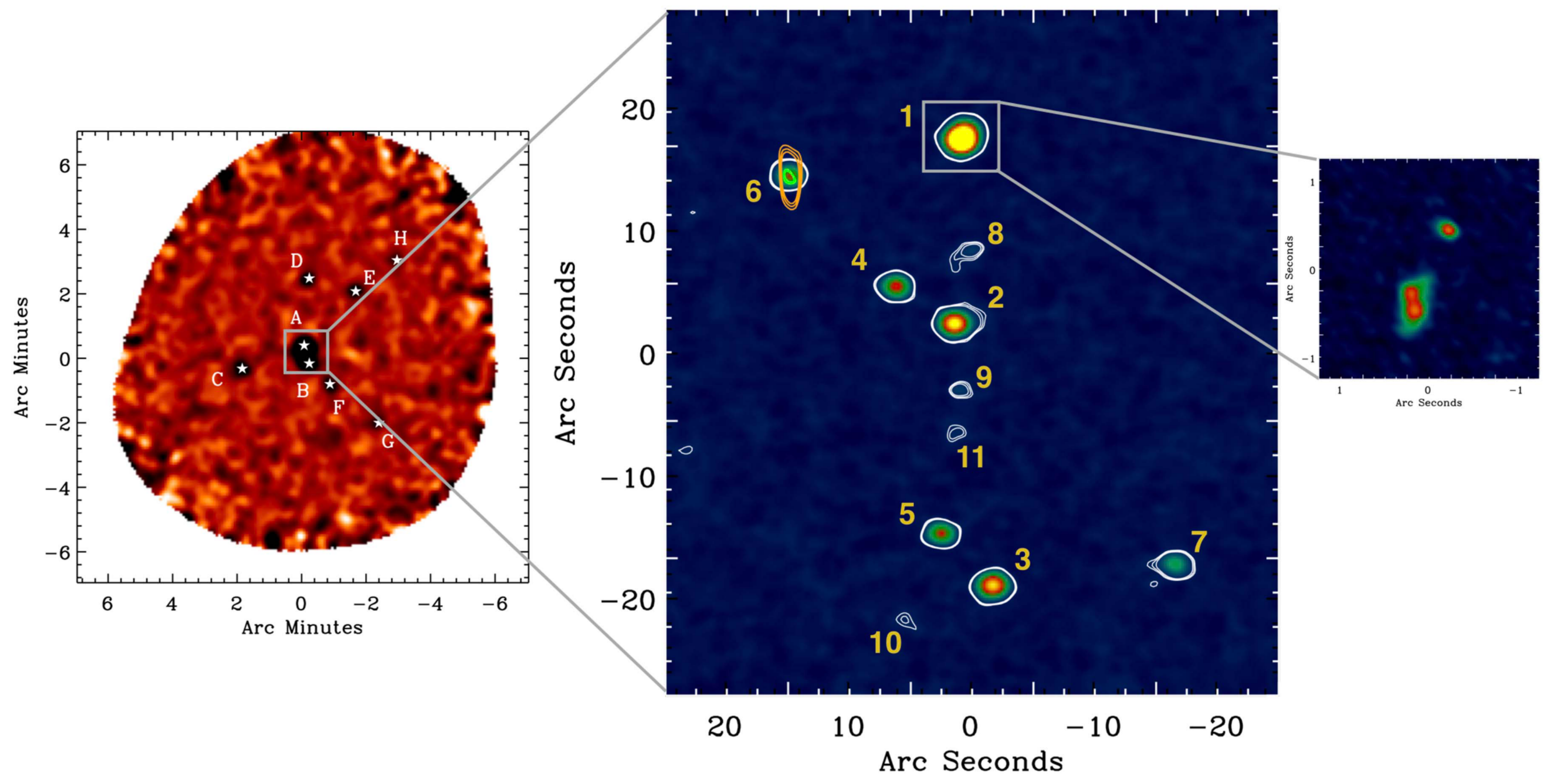}
    \caption{ $-$ The Distant Red Core, a proto-cluster at $z=4.002$.  (left) The 870$\mu$m LABOCA map (smoothed to a resolution of 27$^{\prime\prime}$) reveals 8 sources, an overdensity relative to field submm number counts \citep[e.g.][]{geach2017}.  (Middle) ALMA 2mm continuum map of LABOCA sources A and B, which resolves into 11 discrete sources, 10 of which are in the proto-cluster.  Orange and green contours show radio continuum from ATCA and JVLA, respectively. (right) A high-resolution (0.12$^{\prime\prime}$) ALMA 870$\mu$m map of source DRC-1, which resolves further into three star-forming clumps.  Figure adapted from Figure 1 in \citealt{oteo2018}, reproduced by permission of the AAS.}
    \label{fig:drc}
\end{figure}


Teasing out the detailed nature of these DSFG overdensities requires high-resolution follow-up.  Figure~\ref{fig:drc} illustrates the power of ALMA in characterizing these proto-cluster cores as well as the difficulties in characterizing the full structure.  The Distant Red Core \citep[DRC;][]{lewis2018,oteo2018,long2020} was targeted as the brightest source (A/B) in an overdensity of LABOCA 870$\mu$m sources.  Moving from 27$^{\prime\prime}$ to $\sim1.5^{\prime\prime}$ resolution reveals the extended LABOCA emission to be comprised of 11 blended DSFGs, 10 of which form a proto-cluster core at $z=4.001$ \citep[see also SSA22;][]{umehata2017}.  Occupying an area of 260 kpc x 310 kpc, these DSFG have a combined $\mathrm{SFR}\sim6,500\,\Msun$ yr$^{-1}$, with the majority of this provided by just three members.  Furthermore, imaging at $0.12^{\prime\prime}$ resolution reveals that even these galaxies may be blends; DR-1 is comprised of three clumps with extreme SFR densities of $\sim800-2,000\,\Msun$ yr$^{-1}$ kpc$^{-2}$.  Expanded follow-up of the full structure could recover up to $14,400\,\Msun$ yr$^{-1}$ if all LABOCA sources are associated with the proto-cluster. 
Interestingly, MUSE\footnote{Multi-Unit Spectroscopic Explorer \citep[MUSE;][]{bacon2010}} observations of the central 1$^{\prime}$x1$^{\prime}$ area found an associated Lyman $\alpha$ blob but no LAEs; it is unclear if this is due to i.e. suppression of Ly$\alpha$ in proto-cluster cores  \citep{shimakawa2017} or a dearth of unobscured SFGs. As an example with wider coverage, the core of SSA22 ($z=3.01$) hosts an overdensity of DSFGs, at the node of 30 Mpc-scale filaments traced by LAEs in narrow band imaging \citep[e.g.][]{hayashino2004, matsuda2004}.  Again, this appears consistent with our theoretical framework, but these studies also serve to emphasize that the available current rich datasets still give an incomplete view.

Large surveys over areas of the expected size of proto-clusters ($\sim10^{\prime}$) have been conducted in a few fields with good spectroscopic coverage. This has yielded proto-clusters with total SFRs of  $\sim1,500-6,500 \Msun$ yr$^{-1}$ at $z\sim2-4$, with $\sim4-14$ spectroscopically-confirmed DSFGs per structure \citep[Table~\ref{tbl:protoclusters}; e.g.][]{casey2015, casey2016, miller2018, oteo2018, polletta2021}.  Importantly, these total SFRs are formally lower limits due to spectroscopic incompleteness in lower luminosity SFGs.  Individually, the DSFGs are forming stars at hundreds to thousands of solar masses per year.  Despite this, DSFGs in proto-clusters largely lie on the field MS at $z\sim2$ \citep{hunghongzhaoling2016, zavala2019, polletta2021} and $z\sim4$ \citep{long2020}, with few examples of true starbursts (4x above the MS).  

Ideally, we would like to understand this activity in the context of a proto-cluster's current structure and its evolution to $z\sim0$; however, this is a challenging task.  As described in \citealt{long2020}, halo masses of proto-clusters are estimated using various methods based on summing  the halos of individual members using halo abundance matching \citep{behroozi2013}.  This includes a host of uncertainties; for example, understanding the bias of the population being observed, double counting halos due to overlap, and corrections for spatial and/or spectral incompleteness.  Some attempts that have been made in the literature for dusty proto-clusters are listed in Table~\ref{tbl:protoclusters} and in Figure~\ref{fig:sfrmhalo}, we show the halo mass-normalized total SFRs in comparison with lower redshift clusters.  The dusty proto-clusters largely favor the steepest trend in $\Sigma\mathrm{SFR}/M_{\rm halo}$ of (1+$z$)$^{7}$ (see discussion in \S\,\ref{sec:sf_lowz}-\ref{sec:sf_highz}); however, are these proto-clusters representative?  Given a halo mass and redshift, the final $z=0$ halo mass of a proto-cluster can be estimated from simulations \citep[albeit subject to the uncertainties outlined above as well as uncertainties from intrinsic scatter in proto-cluster properties and the simulation assumptions; e.g.][]{chiang2013, lemaux2014, muldrew2015, muldrew2018, lovell2018}.  Following \citealt{chiang2013}, most of the proto-clusters with infrared observations discussed here will evolve into Coma-like structures with log $M^{z=0}_{200}/\Msun\gtrsim15$.  It remains to be shown whether lower mass proto-clusters with DSFG overdensities are common and also favor a steep rise in the halo mass-normalized total SFR.

\begin{SCfigure}[0.8][!htb]
    \centering
    \parbox{0.6\columnwidth}{\centering\includegraphics[width=0.6\columnwidth]{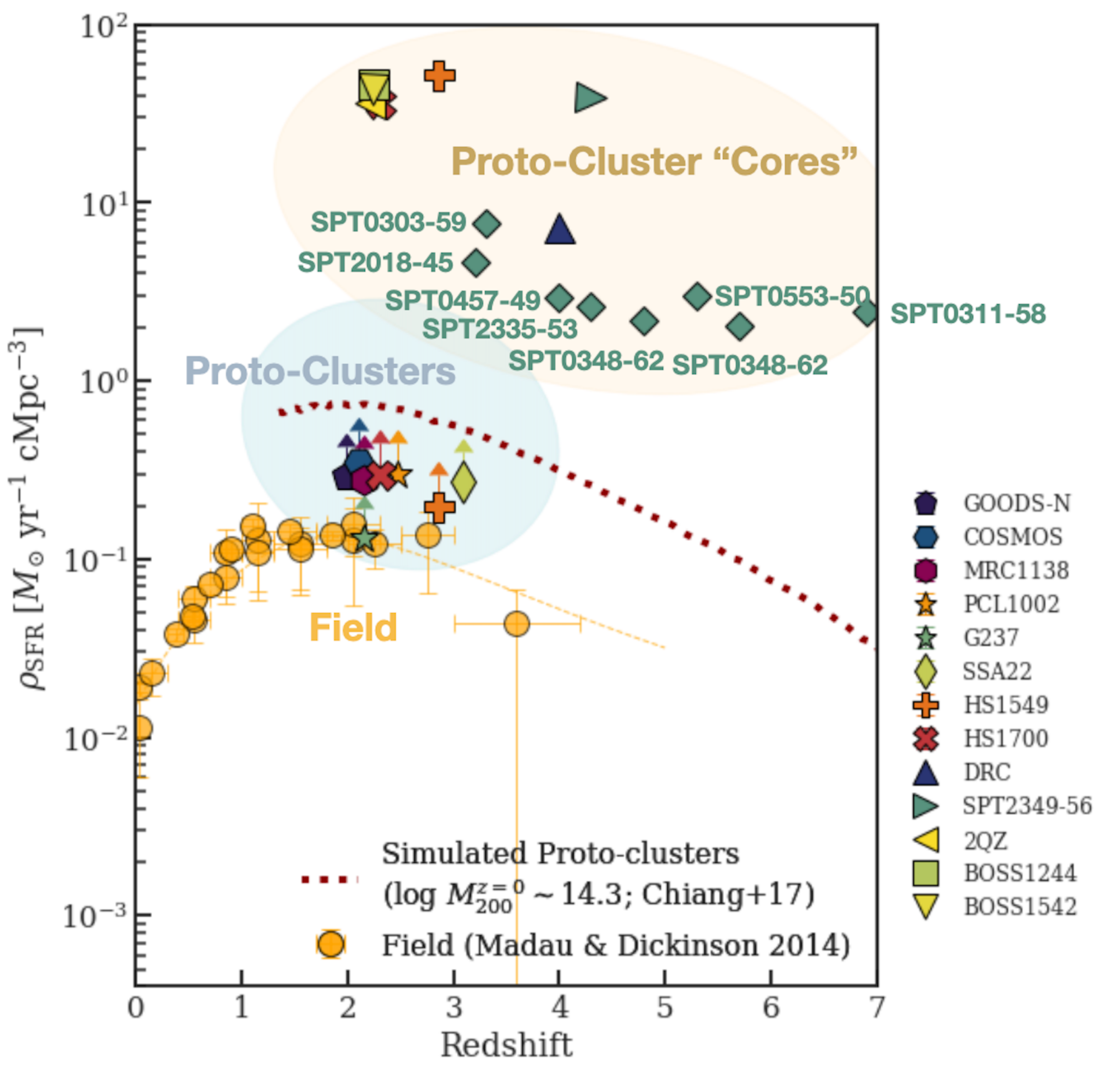}}
    \caption{ $-$ The SFRD of proto-clusters and proto-cluster cores with DSFG overdensities at $z\sim2-7$ (Table~\ref{tbl:protoclusters}) in comparison with the field \citep{madau2014}.  Most of these proto-clusters will have a $z=0$ halo mass comparable to the Coma Supercluster (log $M^{z=0}_{200}/\Msun\gtrsim15$).  {\bf Caution: the volumes assumed are highly uncertain and inhomogeneously derived; a robust analysis requires more uniform volume definitions be adopted in the literature. The total SFRs are also often underestimated due to membership/spectroscopic incompleteness.} For comparison, we show the projected SFRD for proto-clusters with a $z=0$ halo mass of log $M_{200}/\Msun\sim14.3$ from \citep[][]{chiang2017} (red dotted line).  Coma progenitors sit above the field but as lower limits, a comparison to theory is limited.  Proto-cluster cores, however, are up to a few orders of magnitude in excess of the field when considering the central DSFG overdensity.  See Table~\ref{tbl:protoclusters} for the relevant references.}
    \label{fig:sfrd}
\end{SCfigure}


An alternative measure of star formation activity that has become 
common in the literature is to estimate the SFR density compared to that of the field.  This has been done in an inhomogeneous way, however, with some studies adopting the survey area to estimate the proto-cluster volume \citep[e.g.][]{casey2016} and others adopting a smaller effective area around the main DSFG overdensity (Table~\ref{tbl:protoclusters}).  We will refer to the latter as proto-cluster ``cores'', given that the volumes estimated are much smaller than the predicted total volume \citep{chiang2013, chiang2017} and likely only represent the primary halo.  In Figure~\ref{fig:sfrd}, we compile the SFRDs of the proto-clusters (shown as lower limits due to spec-$z$ incompleteness) and proto-cluster ``cores'' listed in Table~\ref{tbl:protoclusters}.  The volumes used have been converted to comoving where necessary for a fair comparison to the field SFRD, which is dominated by halos that have not detached from the Hubble flow.  

In addition to the observations, we include the theoretical SFRD predicted for a log $M^{z=0}_{200}/\Msun\sim14.3$ halo from  \citealt[][]{chiang2017} (dashed red line), perhaps representing more typical proto-clusters to our Coma progenitors.
The dusty proto-cluster SFRDs largely sit above the coeval field relation in Figure~\ref{fig:sfrd} but being lower limits, we are prevented from a meaningful comparison with the simulated lower mass halo.  The proto-cluster ``cores'' largely sit 1-2 orders of magnitude above the proto-clusters, though their specific relative positions are likely inaccurate due to inhomogeneous definitions of effective area and again spectroscopic incompleteness (or uncertain or no background subtraction when spec-$z$s are unavailable). High-density central halos appear to persist to $z\sim7$ \citep{wang2021}. Deriving a robust, comparable measure of proto-cluster volume and improving spectroscopic (or narrow band imaging) coverage would allow us to assess the star formation in these central halos relative to the extended proto-cluster structure.  This would be a compelling tool in constraining the build-up of stellar mass in proto-clusters.  

\subsubsection{Placing Dusty Proto-clusters in Context: a More Ubiquitous or Atypical Phase?}\label{sec:protocluster_disc}

As discussed in the previous section, the DSFG-rich proto-clusters most well studied to date likely inhabit the most massive halos in their respective epochs, with projected masses at $z\sim0$ comparable to the Coma Supercluster.  Placing these dusty proto-clusters into the broader context of all proto-clusters and ultimately structure growth is complicated by several factors.  Primarily, we are limited by our ability to spectroscopically confirm the rising number of DSFG-rich proto-cluster candidates from shallow, wide field surveys (\S\,\ref{sec:planck_selection}).  Similarly, not all known proto-clusters have the infrared data necessary to identify or rule out a DSFG population.  

\begin{SCfigure}[0.8][!htb]
    \parbox{0.6\columnwidth}{\centering\includegraphics[width=0.6\columnwidth]{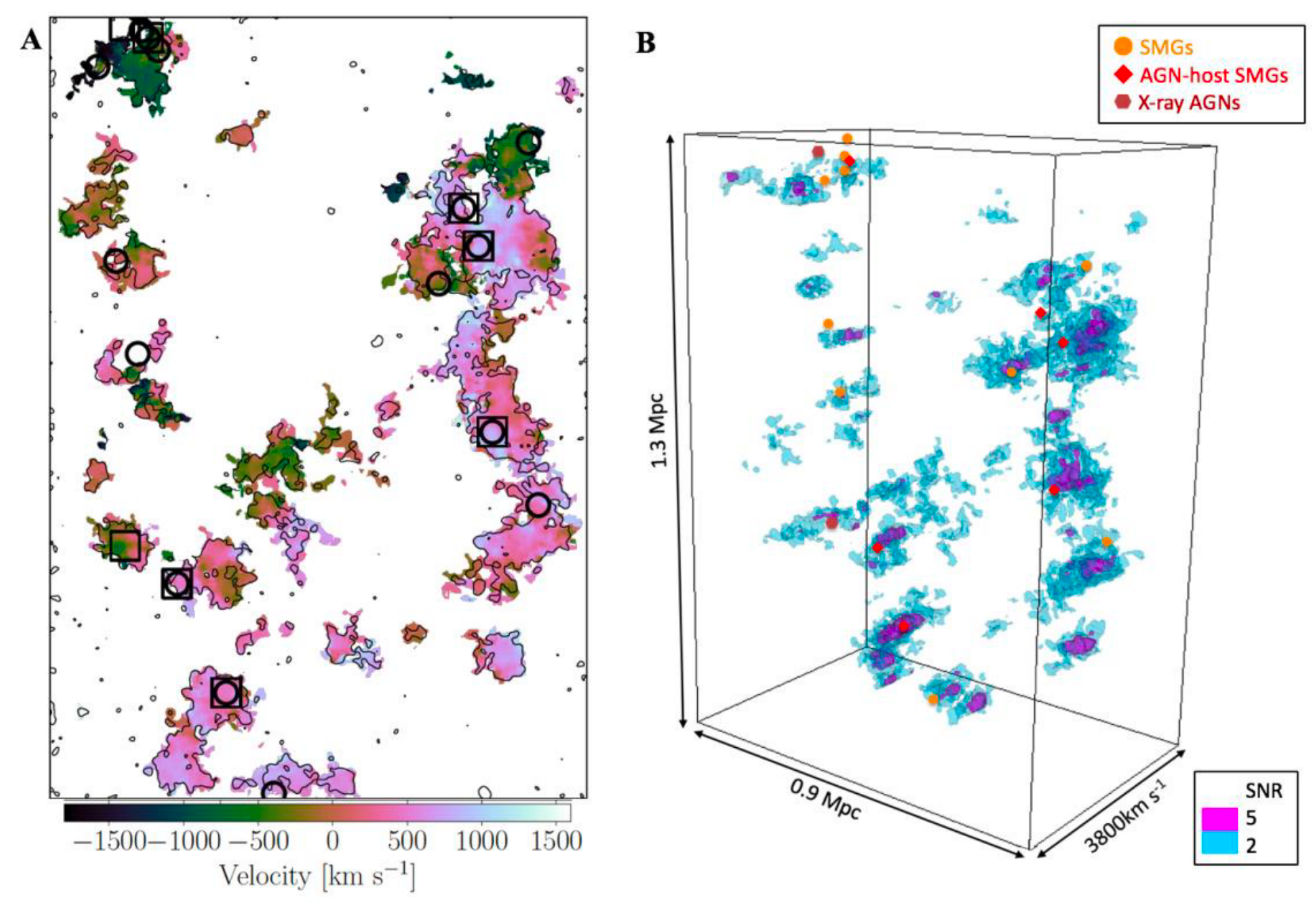}}
    \caption{ $-$ {\bf (left)} Velocity map of Ly$\alpha$ emission from MUSE.  Contours show Ly$\alpha$ surface brightness greater than $0.3\e{-18}$ erg s$^{-1}$ cm$^{-2}$. Circles (squares) indicate the positions of SMGs (X-ray AGN). {\bf (right)} The Ly$\alpha$ filaments in three-dimensions.  Blue and magenta denote $\mathrm{SN}>2$ and $>5$ voxels.  The locations of SMGs, AGN-host SMGs, and X-ray AGN are shown in orange circles, red diamonds, and brown hexagons.  The DSFGs and AGN are cospatial with the Ly$\alpha$ filaments on megaparsec scales.  Figure reproduced from \red{Figure 3 in} \citealt{umehata2019} and reprinted with permission from AAAS.}
    \label{fig:filaments}
\end{SCfigure}

Despite these limitations, we can consider what the nature of DSFGs themselves tells us about the prevalence of DSFG-rich proto-clusters. As alluded to in \S\,\ref{sec:dsfg_signposts}, the typical lifetimes of rare sources provide key constraints.
In the field, DSFGs are thought to be a phase of short ($10-150$ Myr) bursty star formation, triggered by gas-rich mergers/interactions or disk instabilities, and regulated by the resulting feedback \citep[e.g.][]{mihos1996, swinbank2014}. 
However, steady gas infall may prolong this phase, sustaining high SFRs for almost a gigayear \citep[e.g.][]{narayanan2015}.  As discussed earlier, DSFG-rich proto-clusters  may be preferentially located at the nodes of massive filaments, which are predicted to contain $\sim60\%$ of all gas at $z\sim3$ \citep{martizzi2019}.  An example of this might be the DSFG-rich proto-clusters PCL1002 \citep[$z\sim2.5$;][]{casey2015, diener2015, chiang2015, casey2016, zavala2019, champagne2021} and a secondary structure containing DSFGs at $z=2.51$  \citep[CL J1001+0220\footnote{CL J1001+0220 was originally identified as a high-$z$ cluster due to associated extended X-ray emission \citep{wang2016}.  Its status as a cluster is unclear, however; subsequent analysis in \citealt{champagne2021} argues that the X-ray emission originates from a radio relic rather than an ICM.};][]{wang2016, wang2018, gomez-guijarro2019, champagne2021}, which are part of the proto-super cluster ``Hyperion'' \citep{cucciati2018}, a structure containing seven high-density peaks connected by filaments.  Recently, \citealt{umehata2019} directly traced such filaments via Ly$\alpha$ emission in the intergalactic medium of SSA22, finding that SMGs and AGN were coincident with the gas in both projected and velocity space (Figure~\ref{fig:filaments}).  These filaments, which connect the DSFGs and AGN in SSA22 over large scales, may be capable of funneling gas into these active galaxies, prolonging their lifetimes. Direct evidence in support of such inflowing gas into massive halos was presented in \citealt{daddi2021}, which analyzed three Ly$\alpha$ filaments leading into the center of mass of a DSFG-rich group at $z=2.91$.  Definitively identifying inflows is difficult, however \citep[see the discussion in][]{daddi2021}.  Likewise, filamentary structures have only been mapped in a few systems, and as such we can not yet rule out DSFGs outside these concentrated nodes or determine the importance of filaments in extending DSFG activity. See \S\,\ref{sec:COproto} for further discussion of submm observations of molecular gas in proto-cluster galaxies relative to filaments.

In favor of shorter lifetimes, molecular gas studies of proto-cluster DSFGs (see \S\,\ref{sec:COproto}) point to short gas depletion timescales\footnote{We note the uncertainties that arise from the CO-H$_2$ conversion could result in increased $\tau_{depl}$, see \S\,\ref{sec:submm}.} \citep[$\lesssim200-500$ Myr;][but see \citealp{tadaki2019}]{oteo2018, zavala2019, polletta2021} and therefore rapid quenching in the absence of new (cold) gas accretion and/or gas recycling \citep[e.g.][]{lilly2013, christensen2016, walter2020}.  Sustained gas accretion (via filaments or otherwise) is at odds with models that project $z\sim2$ massive halos enter a shock-heated regime at $\gtrsim10^{12}\,\Msun$ \citep{dekel2006}, which throttles fresh gas accretion via starvation (but see the work by \citealt{daddi2021} on gas inflows into massive halos discussed in the previous paragraph). Even at its high redshift, the mass of the DRC puts it in a ``cold in hot'' regime \citep[see Figure 9 in][as well as Figure 19 in \citealp{polletta2021} for more examples]{long2020} where cold gas streams established prior to virial shock heating would need to penetrate an otherwise hot halo \citep[][and references therein]{dekel2006, keres2005}.  Consistent with this picture, \textsc{[Ci]} and CO mapping of the molecular gas halo around the Spiderweb Galaxy (central to the Spiderweb proto-cluster) support growth via recycled gas rather than pristine gas inflows \citep{emonts2018}.  This would seem to favor DSFGs as a short-lived phase.

As discussed extensively in \citealt{casey2016}, \red{statistical}\sout{statistically} arguments and predictions can be made for both possibilities.  As mentioned in \S\,\ref{sec:dsfg_signposts}, they found the volume density of DSFG-rich proto-clusters to be consistent with the population of low-redshift, massive ($>10^{15}\,\Msun$) clusters, albeit with large uncertainties.  If accurate, and assuming a long lifetime due to cold gas accretion, they predict that most proto-clusters should be observed in a DSFG-rich phase\footnote{Given a high individual SFRs ($>1,000\,\Msun$ yr$^{-1}$, discussed in \S\,\ref{sec:protoclusters}), a ``long'' lifetime would be limited to $<500$ Myr so as not to exceed a final mass of log $M_{\star}/\Msun=11.5$.}.   Conversely, they estimated that $20-40\%$ of proto-clusters should be observed in a DSFG-rich phase in a short lifetime scenario, assuming a 100 Myr lifetime and $4-8$ bursts of $5-10$ DSFGs to build the $40\pm10$ log $M_{\star}/\Msun\sim11$ cluster galaxies observed at $z\sim1$ \citep{vanderburg2020}.  An interesting question posed by \citealt{casey2016} is whether, in the scenario of observing $>5$ short-lived DSFG simultaneously, a common triggering mechanism is required. Hypothetically, star formation triggering on large scales could be facilitated by filaments \citep{umehata2019} or by optimal conditions for mergers and interactions \citep{umehata2017}; however, both the ability of these processes to trigger star formation (\S~\ref{sec:interactions}) and the nature of DSFGs in proto-clusters need further investigation. 

Are DSFGs (observationally) ubiquitous in proto-clusters given our current datasets?  Recently, \citealt{mcconachie2022} identified two proto-clusters at $z\sim3$ as overdensities associated with UMGs (see \S\,\ref{sec:dsfg_signposts}) in a survey covering 0.84 deg$^2$. Using UVJ color selection to separate star-forming and quiescent galaxies, they established an elevated quenched fraction over the field for log $M_{\star}/\Msun>11$ galaxies \citep[see also][]{chartab2020, shi2021, kubo2021}, suggesting (in the absence of submm data) that these proto-clusters are being observed post- or in between bursts, given that the UMG was likely assembled in a dusty star-forming phase \citep{forrest2020, forrest2020a}.  On the other hand, non-detections in ALMA established significant gas-poor, massive galaxy populations in two proto-clusters {\it with} DSFG overdensities \citep[COSMOS and PCL1002, Table~\ref{tbl:protoclusters},][]{zavala2019}.   Along with the vigorous dusty star formation, these proto-clusters have an EQE of $\sim0.45$ (Figure~\ref{fig:balogh16} [right]), comparable to clusters at lower redshift. Understanding the universality of the DSFG-rich phase in proto-clusters will require large, unbiased proto-cluster surveys combined with improved sensitivity, resolution, and wide-field capabilities in the submm (such as with LMT/TolTEC, see \S\,\ref{sec:future}).

\subsection{Obscured AGN in clusters}\label{sec:agn}

\red{Supermassive black holes are thought to be ubiquitous in massive galaxies, and in their actively accreting phase }\sout{Actively accreting supermassive black holes, termed Active Galactic Nuclei }\red{(termed Active Galactic Nuclei or AGN)}\sout{are thought to be ubiquitous in massive galaxies  and} may play\sout{an} a crucial role in the regulation of galaxy growth via feedback \citep{fabian2012, u2022}.  In overdense environments, AGN may influence both (proto-)cluster galaxy evolution and the halo ecosystem through injection of energy into the ICM \citep[][]{voit2005a}.  Hyper-luminous, heavily obscured AGN may also serve as signposts for proto-clusters, as discussed in \S\,\ref{sec:dsfg_signposts}. 

The current literature on AGN in (proto-)clusters has established a deficit of luminous AGN in local and low-redshift clusters relative to the field \citep[i.e.][]{kauffmann2004, popesso2006, lopes2017}.  The fraction of AGN in clusters rises rapidly with redshift, however, reaching field levels by $z\sim1-2$ \citep{galametz2009, fassbender2012, martini2009, martini2013, alberts2016, bufanda2017}.  In fact, there is some evidence that AGN fractions may even be enhanced in high-redshift clusters \citep{fassbender2012, alberts2016} and in proto-clusters \citep[][but see \citealp{macuga2019}]{lehmer2009, kocevski2009, digby-north2010, kubo2013, kubo2019}.  This suggests an important role for AGN in overdense environments, however, our current picture is hindered by both the current unclear nature of the relationship between AGN and galaxy growth \citep{alexander2012, brandt2015} and our ability to account for all AGN in a given population.

AGN are notoriously difficult to survey in a complete manner; they can be identified by their emission in the X-ray through the radio, though no one single wavelength selection returns a full census of AGN\sout{in a given population} \citep[e.g.][]{mendez2013, azadi2017, lyu2022a}.  By far the most elusive AGN population is that of heavily dust-obscured and Compton-thick (CT) AGN.  These AGN can be so deeply embedded in dust (columns of $N_{\rm H}\gtrsim10^{24}$ cm$^{-2}$) that they can be missed even in the hard X-ray bands. Their emission, however, cannot help but escape in the mid-infrared. With current capabilities, the luminous end of this population has been identified using infrared colors via \textit{Spitzer} \citep[e.g.][]{lacy2004, stern2005} and \textit{WISE} \citep{stern2012} or through SED fitting of the optical-infrared \citep[e.g.][]{assef2008, assef2010, chung2014, alberts2016, alberts2020, lyu2022a}.  These techniques have not captured the faint end of the population, however, and estimates based on local samples poorly constrain the percentage of heavily obscured AGN at $10-50\%$ of the total AGN population \citep[e.g.][]{treister2009, akylas2012, akylas2016, georgantopoulos2019}.  For a recent review of what we know about heavily obscured AGN, see \citealt{lyu2022} in this special issue.

Few cluster studies have examined MIR AGN. Using \textit{WISE} color selection, \citealt{mishra2020} confirmed a lower fraction of MIR AGN in clusters than in the field at low redshift, in good agreement with AGN studies in other wavelength regimes.  The clustering of MIR AGN is likewise known to be weak relative to X-ray and radio AGN at low redshift \citep[$z<0.8$; ][]{hickox2009}, but rises rapidly to $z\sim2$, suggesting that MIR AGN are increasingly associated with high-$z$ clusters \citep{brodwin2008, dey2008}.  The fraction of color-selected MIR AGN in the ISCS cluster sample was found to rise to field levels by $z\sim1-1.5$ \citep{galametz2009, martini2013}.  \citealt{alberts2016} expanded this analysis using SED fitting to classify galaxies as AGN or host-dominated as a function of cluster-centric radius, finding \red{at $z>1$} an excess of AGN-dominated galaxies in the cluster cores\sout{at $z>1$} relative to the field drawn at $R\sim3R_{\rm vir}$.  This suggests triggering of AGN in the cluster environment, possibly through increased merger activity.  This connection has not been established however, as the relationship between AGN and mergers in the field is still unclear at high redshift \citep[e.g.][]{shah2020}, despite compelling evidence in the local Universe \citep[e.g.][]{ellison2011, ellison2013, ellison2015, ellison2019, weston2017}. It is important to note that mergers may be preferentially associated with the obscured AGN phase \citep[e.g.][]{urrutia2008, glikman2015, donley2018, blecha2018} and so our understanding in both the field and in clusters will likely undergo a dramatic shift with upcoming surveys by \textit{JWST}, which can robustly identify obscured AGN to low luminosities \citep[e.g.][]{alberts2020}.

\section{The Far-Infrared to Submillimeter: Dust and Gas Measurements}\label{sec:submm}

While our previous sections have primarily focused on the consequences of environmental quenching $-$ as observed through changes to the stellar populations and star formation rates at near-IR and mid-IR wavelengths $-$ we now shift our focus to the underlying cause for these changes, namely the role of the cold molecular phase of gas in cluster galaxies. As discussed in \S\,\ref{sec:learn}, cold H$_2$ cannot be observed directly and therefore $^{12}$CO rotational transitions and dust continuum emission, observed at FIR-to-submm wavelengths, are instead used as proxies.
This important gas phase provides the direct fuel for star formation and is linked to the evolution of the cosmic SFRD \citep[i.e.][]{madau2014}, with the cosmic molecular gas density mirroring the shape of the SFRD, peaking at $z=1-3$ and declining by almost an order of magnitude to the present day \citep[e.g.,][]{decarli2019, decarli2020, scoville2016}. This strongly suggests that molecular gas is fundamental to the process of quenching and therefore a key component in investigating the link between quenching and environment.
However, whether the molecular gas in galaxies is affected by the cluster environment has long been debated.



In this section, we discuss the progress that has been made over the last 40 years on quantifying the (molecular) gas content in cluster galaxies.  As context for these studies, we start by briefly reviewing the state of gas scaling relations in the field and the major techniques and uncertainties in molecular gas analysis.  In \S\,\ref{sec:environment_gas}, we cover gas measurements from low-$z$ clusters to $z>2$ proto-clusters. We start from a historical perspective:  early integrated CO studies in cluster galaxies suggested that the dense molecular gas phase was impenetrable to environmental effects. More recent work, however, using the revolutionary capabilities of ALMA, NOEMA, and the JVLA over a wider range of redshifts, cluster environments, and gas proxies has shown a broader spread of behaviors, from depleted to enhanced molecular gas contents. Finally, we end with a discussion of spatially-resolved gas studies, which have the potential to provide the greatest insight into quenching mechanisms. In low-redshift cluster galaxies, environmental effects on the molecular gas have become evident through perturbed and asymmetric gas features, but spatially-resolved molecular gas studies are still few and far between beyond the local Universe.



\subsection{Gas Scaling Relations}\label{sec:scale}



The cold interstellar medium, as observed through dust and molecular gas, has emerged as a primary focus of galaxy evolution studies. 
Before we jump into environmental effects, however, we first briefly review efforts using field galaxies to quantify the global correlations between SFR and molecular gas content, both of which have been shown to be redshift- and mass-dependent \citep[e.g.,][]{lilly2013, whitaker2014, speagle2014, geach2011, genzel2015, tacconi2018}.  There has been considerable effort in trying to understand the origin and evolution of the star-forming MS, in particular through the ``bathtub" equilibrium model \citep[e.g.,][]{bouche2010, dave2012}, where gas content is regulated through inflows, outflows, and star formation.  For example, the tightness of the MS can be attributed to variations in the molecular gas content and the SFE (or its inverse, the gas depletion timescale  ($t_{\rm depl} = M_{\rm gas}$/SFR)).  In other words, a galaxy's SFR offset from the MS correlates with these quantities, with galaxies above the MS having shorter depletion timescales and higher molecular gas-to-stellar mass ratios \citep{saintonge2016, saintonge2017}.  

At low redshift, the global scaling relations\footnote{For spatially-resolved scaling relations, including the definition of a molecular gas Main Sequence, see \citet{bolatto2017}, \citet{lin2019}, and \citet{ellison2020}.} between star formation and the cold molecular gas content are primarily due to large-scale efforts to measure these properties over a wide dynamic range within a homogeneous, unbiased sample of galaxies.  The IRAM 30-m telescope\footnote{Institut de Radioastronomie Millim\'{e}trique \citep[IRAM; ][]{baars1987}} has been critical in this effort through mass-selected surveys such as the CO Legacy Database for GASS  \citep[COLD GASS;][]{saintonge2011} and its extension, xCOLD GASS, \citep{saintonge2017}, with CO ($1-0$) observations of $\sim500$ galaxies down to $M_{\star}>10^9\,\Msun$ in the local Universe \citep[see also][for a tabulated list of field surveys using CO and dust continuum]{tacconi2020}.

At higher redshifts (typically $z>0.5$), increased sensitivity and resolution are paramount, and therefore more distant surveys require the use of interferometers, such as PdBI, NOEMA, and ALMA.  The Plateau de Bure High-z Blue Sequence Surveys \citep[PHIBSS1 and 2;][]{tacconi2013, freundlich2019} represent the largest pointed CO surveys in the distant Universe, consisting of $\sim130$ SFGs (SFR $>30\,$\myr) with CO ($2-1$) or CO ($3-2$) detections in three distinct redshift slices over $0.5<z<2.5$. Averaging on-source integration times of 25 and 12 h, respectively \citep{tacconi2020}, PHIBSS1 and 2 targeted sources on or above the MS in order to ensure a high probability of a CO detection, and  thus have a slight bias against sub-MS galaxies with potentially lower CO luminosities.  Moreover, ALMA observations of the Rayleigh–Jeans dust continuum emission ($\lambda\sim1\,$mm) exist for $\sim600$ galaxies in the COSMOS field, selected from $24\mu$m and \textit{Herschel} priors, with a similar bias toward more massive and IR-bright SFGs \citep{scoville2017}.   Nevertheless, these surveys, along with a host of supplementary CO and dust continuum samples, have provided the best means to empirically extend the integrated molecular gas scaling relations to $z\sim3$ using hundreds of galaxies \citep[][and references therein]{genzel2015, tacconi2018, scoville2016, scoville2017}. This has yielded functional forms for both the molecular gas-to-stellar mass ratio and molecular gas depletion timescale in terms of the products of power laws with three separable variables: redshift, stellar mass, and offset from the MS.  The outcome of these scaling relations has revealed that the depletion timescale and molecular gas-to-stellar mass ratio vary slowly and steeply with redshift, respectively, as well as strong dependencies for both with location on the star-forming MS plane \citep{koyama2017, tacconi2020}. Despite this progress, there remain gaps in our understanding due to observational limitations (e.g. sample incompleteness at low SFR and stellar mass) and uncertainties associated with the H$_2$ proxies used to measure the gas mass.  We discuss the latter briefly below.





  

\subsubsection{Caveats for Molecular Gas Mass Measurements}\label{sec:caveats}
The exploitation of CO transitions and dust continuum observations has paved the way for molecular gas studies, but not without several limitations and assumptions.  The derivation of a total molecular gas mass from either the CO line luminosity or dust mass depends on various parameters, namely the CO excitation and the CO-to-H$_2$ conversion factor ($\alpha_{\rm CO}$) for the former, and the mass-weighted dust temperature ($T_{\rm dust,mw}$),\footnote{The mass-weighted dust temperature represents the temperature of the dominant (by mass) cold dust component.  It is typically lower than the luminosity-weighted temperature measured from the peak of the FIR SED. } the dust opacity, and the dust-to-gas ratio for the latter. A detailed discussion of the assumptions used to estimate these parameters can be found in \citealt{bolatto2013}, \citealt{carilli2013}, \citealt{genzel2015} and \citealt{scoville2017}; here we summarize the main caveats and current outlook.

As briefly introduced in \S\,\ref{sec:intro_submm}, carbon monoxide is the second most abundant molecule in the ISM, with a ratio of one CO molecule to every $\sim$10,000 hydrogen molecules. Given its low excitation potential, which primarily occurs through collisions with H$_2$, it is easily observed from ground-based facilities, thus making it a convenient molecular gas tracer. The global CO excitation produces a ladder of populated energy levels, with the relative strengths of the observed rotational transitions quantified in CO spectral line energy distributions (SLEDs).  The SLED is dependent on gas properties such as temperature and critical density, with higher rotational transitions tracing denser molecular gas.  In the case of thermalized excitation, the CO line luminosity is constant for all energy levels; however, in the more likely case of subthermalized excitation, a correction is required to account for the conversion of a $J_{\rm upper}>1$ line intensity to the ground state in order to infer a total gas mass.  This is typically expressed as a CO line brightness temperature (e.g., r$_{\rm J1}$ = $L^{\prime}_{\rm CO(J-(J-1))}/L^{\prime}_{\rm CO(1-0)}$).  There are large uncertainties associated with excitation corrections, particularly as the rotational number $J$ increases, with the exact SLED shape strongly dependent on the galaxy type and gas excitation properties \citep[see detailed discussions in ][]{bolatto2013, carilli2013,narayanan2014}.  Indeed, beyond the local Universe, most surveys rely on these mid-to-higher order CO rotational transitions, due to their frequency and brighter flux at a given redshift ($\propto \nu^2$), and thus suffer from potentially large systematic uncertainties when estimating the ground-state emission from higher excitation lines.  

An even greater source of uncertainty arises from the conversion of the CO luminosity into a molecular gas mass, known as $\alpha_{\rm CO}$, where $M_{\rm mol} = \alpha_{\rm CO} L_{\rm CO}$.  This factor depends on a host of ISM conditions, such as the gas density, temperature, and velocity dispersion, as well as a strong dependence on metallicity \red{\citep[e.g.][]{cormier2014}}.  In the Milky Way, $\alpha_{\rm CO}$ can be determined directly through the virial mass technique with the observation of spatially-resolved giant molecular clouds. By measuring their sizes and kinematics, and under the common assumption that the CO ($1-0$) line emission is optically thick, one can obtain a relationship between CO luminosity and gas mass.  This has yielded a Galactic value of $\alpha_{\rm CO}$ = 4.3\alphaUnits
, with $\sim0.1$ dex uncertainty (see detailed review by \citealt{bolatto2013}).  

Outside the local Universe, direct determination of $\alpha_{\rm CO}$ is challenging, and studies typically rely on spectral line modeling of multiple transitions and isotopologues of CO \citep[such as $^{13}$CO, e.g.,][]{papadopoulos2012} to determine gas conditions, or use optically-thin dust emission to estimate a gas mass (see below).  Therefore, it is common for the Galactic $\alpha_{\rm CO}$ to be adopted for star-forming disk galaxies with approximately solar metallicity on the MS. However, various factors could contribute to deviations from this value.  For example, galaxies with molecular gas at higher temperatures or with increased velocity dispersion (e.g., due to turbulent motion, outflows, ram-pressure, mergers, starbursts) would require a lower value of $\alpha_{\rm CO}$ as higher energy states are excited, yielding larger CO luminosities \citep{narayanan2012}.  These factors will somewhat influence the $\alpha_{\rm CO}$ in high-metallicity galaxies, with a much more dramatic influence on 
lower-metallicity galaxies, as less dust is available to shield CO from photodissociating, necessitating a higher value for $\alpha_{\rm CO}$ \citep{genzel2012}.  Many studies attempt to account for this with a metallicity-dependent correction, though often based on a mass-metallicity relationship. Thus, measurements for $\alpha_{\rm CO}$ can vary from $\sim$1 to 4 to 12 \alphaUnits\
for starbursts, Milky Way-like, and low-metallicity galaxies, respectively, making direct comparisons of gas masses over heterogeneous samples difficult.

A comparatively inexpensive alternative to CO, submm continuum emission in the optically-thin regime ($\lambda_{\rm rest}>250\mu$m) is proportional to the total dust mass, which can be related to the gas mass\footnote{Local studies have shown that the dust mass is a good tracer of \textit{total} (\textsc{Hi}+H$_2$) gas mass in \textsc{Hi} dominated galaxies \citep{janowiecki2018}.  As \textsc{Hi} cannot yet be observed beyond low redshift, high-redshift calibrations of dust as a gas proxy are largely based on CO and are often stated to represent the molecular component.  This is likely reasonable as high-$z$ galaxies are thought to be dominated by H$_2$ (see \citealt{schreiber2020} and references therein); however, this remains an unknown systematic.}
via a minimal number of parameters.  
The first, the mass-weighted dust temperature, has an observed range of 15-30 K in local to high-redshift (MS to starburst) field galaxies \citep[e.g.][]{draine2007}.  While it can be robustly measured in individual galaxies with adequate coverage of the FIR, studies are often working off of a single continuum band, particularly beyond the low-redshift Universe.  In these cases, it is common to adopt a constant $T_{\rm dust,mw}=25$ K\footnote{\red{This calibration may no longer be valid at high redshifts where the cosmic background boosts galaxy dust temperatures.}} \citep[e.g.][]{scoville2016}.  This assumption was recently tested in \citealt{dunne2022}, which found that, despite a significant correlation between $L_{\rm IR}$ and $T_{\rm dust,mw}$, using a constant and individually-measured $T_{\rm dust,mw}$ produced consistent results for massive, MS galaxies. 



In addition to $T_{\rm dust,mw}$, deriving a gas mass from $M_{\rm dust}$ requires the dust opacity \citep[a function of grain properties;][]{galliano2018} and the dust-to-gas ratio \citep[proportional to metallicity; e.g.][]{draine2007, leroy2011}, both of which are often unavailable.  This has limited most studies to massive, high-metallicity galaxies, for which the DGR can be reasonably assumed to be in the range 100-150 \citep[e.g.][]{draine2007, leroy2011, sandstrom2013}.  Dust opacity, on the other hand, is difficult to measure outside the Milky Way and so, given that we expect reasonably small ranges for $T_{\rm dust,mw}$ and the DGR, empirical calibrations between $L_{\rm dust}$ and $M_{\rm gas}$ have been derived through comparisons with CO both locally and at high redshift \citep{eales2012, scoville2016, scoville2017, kaasinen2019, tacconi2020}, with the uncertainties for CO-based gas masses described above.  Using these calibrations, dust continuum and CO- (and \textsc{[Ci]}-)based gas masses are found to be in good agreement across a range of redshift and luminosity \citep[e.g.][]{groves2015, tacconi2020, dunne2022}, with dust-based gas masses having modest systematics (secondary dependencies on e.g. $M_{\textsc{Hi}}$/$M_{H_2}$ ratio, SFR) at the 20\% level \citep{janowiecki2018, dunne2022}.  We note that these calibrations should only be applied to the global submm flux; $T_{\rm dust,mw}$ and the DGR ratio vary on smaller spatial scales \citep{roman-duval2017, popping2022}.  

Thus, while systematics remain in our gas measurements and we can expect our calibrations to break down in the low-mass, low-metallicity regime, both CO and dust provide robust, comparable gas tracers in the field.
Whether this holds as a function of environment, however, is largely untested.  Particularly at high redshift, it is unknown if cluster galaxies have similar dust properties as their counterparts in the field.  Likewise, we can expect that environmental quenching processes effect both gas and dust and potentially in different ways.  Dust and molecular gas are often co-spatial, as H$_2$ requires dust grain surfaces to form \citep{gould1963, lebourlot2012}.  However, ram pressure stripping, for example, could strip dust and molecular gas at different rates, depending on their distribution within the disk \citep{cortese2016}.  In the other direction, preliminary studies suggest that RPS can also compress \textsc{Hi} within a galaxy, triggering the transformation to H$_2$.  The enhanced molecular gas fractions \citep[e.g.][]{ nakanishi2006} observed may require commensurate grain growth during this process \citep{henderson2016}.  Until we do a thorough, simultaneous study of CO and dust across environments, our interpretation of gas masses will be limited.

\subsection{Environmental Effects on the Galaxy-Integrated Molecular Gas Content}
\label{sec:environment_gas}
\subsubsection{An Historical Perspective}

Studies of environmental effects on the cold ISM of galaxies were pioneered at radio wavelengths, focusing on atomic hydrogen (\textsc{Hi}). This is the dominant gas component of the ISM at low redshifts, often extending beyond the optical disk in typical galaxies and therefore not as strongly bound to the gravitational potential well \citep[e.g.,][]{haynes1984, boselli2006}.  The resounding consensus from these studies was that cluster environments host a unique population of ``anemic'' or \textsc{Hi}-deficient galaxies; this depletion of atomic hydrogen presented the first observational signature of the removal of a galaxy's ISM due to environmental effects \citep[e.g.,][]{davies1973, haynes1984a, giovanelli1985}.  As technological advances paved the way for shorter wavelength observations in the infrared and submillimeter regimes, molecular gas studies within clusters searched for similar environmental signs through CO emission lines. Although molecular gas makes up a smaller budget of the ISM than \textsc{Hi} in low-redshift disk galaxies, it is more directly linked to the formation of stars \citep[e.g.,][]{wong2002}, and therefore could be a more straightforward probe into environmental quenching.  However, it is also a denser phase of gas and more tightly bound to the inner disk.  

The Virgo cluster, nearby and containing a significant number of \textsc{Hi}-deficient galaxies, was the breeding ground for some of the first studies on environmental effects of molecular gas.  
Following in the footsteps of atomic gas cluster studies,
the molecular gas content has often been analyzed within the context of the \textsc{Hi}-deficient cluster galaxies. Initial studies found this population to exhibit high molecular to atomic gas mass ratios and contain typical (field-like) CO gas reservoirs \citep{kenney1989, stark1986, boselli1994}, even after accounting for mass and morphology dependencies \citep{kenney1989} and probing to low luminosities \citep{Boselli2002}.  Other cluster studies followed \red{suit}\sout{suite}, finding comparable CO gas contents in isolated field galaxies to those associated with the rich Coma supercluster, where environmental effects might be more extreme \citep{Casoli1996}. This was confirmed not only in FIR-selected galaxies, which could be biased against low star formation rates \citep{Casoli1991}, but also in optically-selected galaxies in both the Coma \citep{boselli1997} and Fornax  \citep{horellou1995}\footnote{There was some evidence that Fornax cluster galaxies in \citet{horellou1995} had weaker CO emission compared to the atomic gas content, but the amount was typical given their low star formation rates.} clusters. Conversely, after reanalyzing distances to Virgo galaxies and comparing to a small field CO sample from \citet{young1989}, \citet{rengarajan1992} concluded there was evidence for H$_2$ deficiencies in the \citet{kenney1989} cluster data, though the field galaxies were infrared-selected and thus favored a more gas-rich population given the strong correlation between gas mass and infrared luminosity \citep{boselli1997}.  Indeed, \citet{fumagalli2009} later quantified that the average infrared luminosity of gas-deficient galaxies (defined as having both low \textsc{Hi} and H$_2$ contents) was $\sim2\times$ lower than that of normal galaxies, signifying the importance of probing down the luminosity function to the same flux completeness when comparing isolated and cluster galaxies.

These initial studies were also plagued by a lack of large-field CO surveys, rendering comparisons between cluster and isolated populations incomplete.  In 1995, the Five College Radio Astronomy Observatory Extragalactic CO Survey \citep{young1995}\footnote{See also \citealt{young1991} for a compilation of individual detections at that time.} presented one of the first CO surveys that sampled $\sim300$ galaxies spanning a range of parameters (e.g., morphologies, optical sizes, environments), \red{though the survey}\sout{albeit} was not complete in a flux-limited sense \citep{young1995} and favored FIR-bright (and thus CO-bright) galaxies \citep{boselli1997, casoli1998}. Utilizing this survey along with other nearby field galaxies from \citet{Sage1993}, and compiling molecular gas observations of cluster galaxies from the literature, \citet{casoli1998} investigated statistical comparisons of environmental effects on molecular gas.  They defined a CO deficiency parameter normalized to the optical size of the galaxy and analogous to prior \textsc{Hi} studies.  Even after attempting to account for non-detections of CO using a survival analysis, they still found no evidence for significant CO deficiencies in nearby cluster cores. These early results thus lead to the general interpretation that processes like strangulation and ram-pressure stripping can effectively remove the atomic gas on relatively short timescales of $\sim$ a few hundred Myr \citep[e.g.,][]{vollmer2004, crowl2008}, while the denser, and often centrally peaked, molecular gas \citep{leroy2008} is left unaltered and shielded from hydrodynamical effects \citep{kenney1989, young1991, young1995} over $\sim$Gyr periods, comparable to cluster crossing times \citep[e.g.,][]{kenney1989}.

\subsubsection{Low-redshift ($z<1$) Cluster Trends}
\label{sec:gas_lowz}
As sample sizes have increased and sensitivity limits have decreased, the picture has now become more nuanced, with low-redshift cluster studies typically finding depleted molecular gas reservoirs in cluster galaxies.  Many of the aforementioned pioneering studies suffered from small sample sizes, biased selections, heterogeneous data sets, and large uncertainties on the total CO fluxes and gas masses \citep[see also discussions in][]{fumagalli2009, boselli2006}.  
With recent technological advances, there has been a resurgence of CO surveys over the last $\sim$15 years, allowing environmental comparison studies to partially mitigate some of these issues in the nearby Universe. For example, \citealt{fumagalli2009} attempted to homogenize the recovered CO flux measurements in 47 nearby spiral galaxies within Virgo, Coma, groups, and in the field, after finding a weak correlation between molecular and atomic gas depletion in Virgo-only cluster galaxies \citep{fumagalli2008}.  They utilized the BIMA\footnote{Berkeley Illinois Maryland Association \citep[BIMA; ][]{helfer2003}} and Nobayama CO \citep{kuno2007} surveys, which had similar sensitivities and analysis techniques that yielded estimated gas mass dispersions of $<0.4$ dex.  Despite missing gas-poor galaxies due to infrared selections, they detected a global depletion of molecular gas in a significant subset of \textsc{Hi}-deficient galaxies, and a notable lack of H$_2$ deficiency in \textsc{Hi}-normal galaxies; this presented some of the first evidence for indirect environmental effects on the molecular gas in cluster galaxies.  A similar trend was observed in Abell 1367, part of the Coma Supercluster, using 19 optically-selected spiral galaxies with CO (1-0) and (2-1) detections from the IRAM 30-m telescope \citep{scott2013}.  Classifying the galaxies according to their evolutionary stage, with more evolved states defined as higher levels of \textsc{Hi} deficiency and redder optical colors \citep{scott2010}, they found larger H$_2$ deficiencies in spirals with more advanced evolutionary states, in addition to a few abnormal gas morphologies (see also \S\,\ref{sec:resolved}).

The launch of \textit{Herschel}  
ushered in a number of dedicated surveys at far-infrared wavelengths, spanning galaxies over a wide range of environments, morphologies, and masses. As in the pioneering molecular gas studies from the 1980s, the Virgo cluster has been a ubiquitous target, for example, in both the \textit{Herschel} Virgo Cluster Survey \citep[HeViCS,][]{Davies2010} and the volume-limited \textit{Herschel} Reference Survey \citep[HRS,][]{boselli2010}.  Providing estimates of obscured SFRs and dust masses, these surveys offered suitable samples for studying environmental trends on dust and molecular gas in combination with new CO follow-up studies and archival programs.  With a magnitude-limited sample of 35 Virgo spiral galaxies from HeViCS, \citealt{corbelli2012} investigated correlations between stellar mass, \textsc{Hi}, cold dust, and molecular gas contents.  They found that as \textsc{Hi} deficiency increases, the dust-to-stellar mass and molecular gas-to-stellar mass ratios decrease, while the dust-to-total gas and molecular-to-total gas ratios increase, with a more pronounced correlation in the dust component.  This provides evidence that both dust and molecular gas can indeed be affected by the cluster environment, but to a lesser extent than the \textsc{Hi} gas, likely due to H$_2$ being more tightly confined in the central disk.  
Studies using the mass-selected HRS confirmed this result and further differentiated Virgo cluster galaxies from a field sample selected with similar criteria,  totaling $\sim200$ galaxies for statistical comparisons.
Homogenizing various CO measurements and defining a control sample of field galaxies with low \textsc{Hi}-deficiencies, \citealt{boselli2014a} found that \textsc{Hi}-deficient cluster galaxies also have depleted molecular gas contents for a given stellar mass, with the level of H$_2$-deficiency weakly but significantly increasing with \textsc{Hi}-deficiency.  
\citealt{cortese2016} similarly measured a notable depletion of \textsc{Hi}, H$_2$, and dust contents in the Virgo cluster population. Each of these results was robust to systematic uncertainties in the CO-to-H$_2$ conversion factor (\S\,\ref{sec:caveats}), using both a constant and a luminosity- or metallicity-dependent $\alpha_{\rm CO}$. Taken together, these Virgo studies hint at a possible differential quenching or stripping efficiency of each ISM component \citep[see also ][]{zabel2021}, from molecular gas to dust to atomic gas, likely due to their decreasing concentration within the disk (i.e., increasing scale length).  This is consistent with outside-in quenching, as expected from hydrodynamical mechanisms such as ram-pressure stripping \citep{boselli2014a, cortese2016}.

\begin{figure}[!htb]
    \centering
    \includegraphics[scale=0.23]{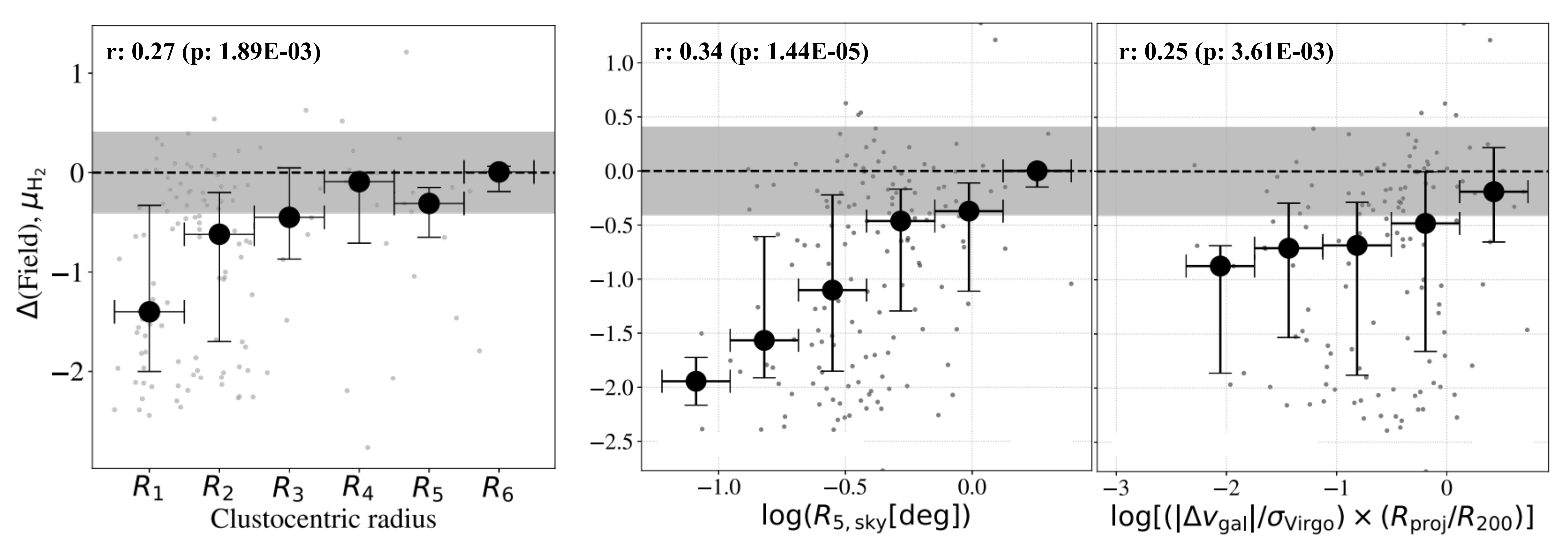}
    \caption{ $-$ The radial trends of the molecular gas-to-stellar mass ratio ($M_{\rm H_2}/M_{\star}$) for Virgo cluster galaxies, where Mstar-dependencies of field
galaxies have been subtracted to expose field offset quantities.  The median values are plotted as large black circles, with individual points shown as gray circles.  The dashed line at zero represents the field value with 1$\sigma$ uncertainties indicated by the shaded gray region.  The Spearman's rank order correlation coefficient and corresponding p-value are displayed in \red{black text at the top of each panel.}\sout{purple text.}  There is a trend toward larger offsets of molecular gas-to-stellar mass ratio with decreasing clustercentric radius  (left panel), as well as with other environmental tracers, such as increasing projected local galaxy density (middle panel), and earlier accretion epoch (right panel). Figure adapted from \red{Figures 6, 7, and 8 in}  \citealt{morokuma-matsui2021}, reproduced by permission of the AAS.}
    \label{fig:virgo}
\end{figure}

\begin{figure}[!htb]
    \centering
    \begin{minipage}{1.\textwidth}
    \includegraphics[scale=0.43]{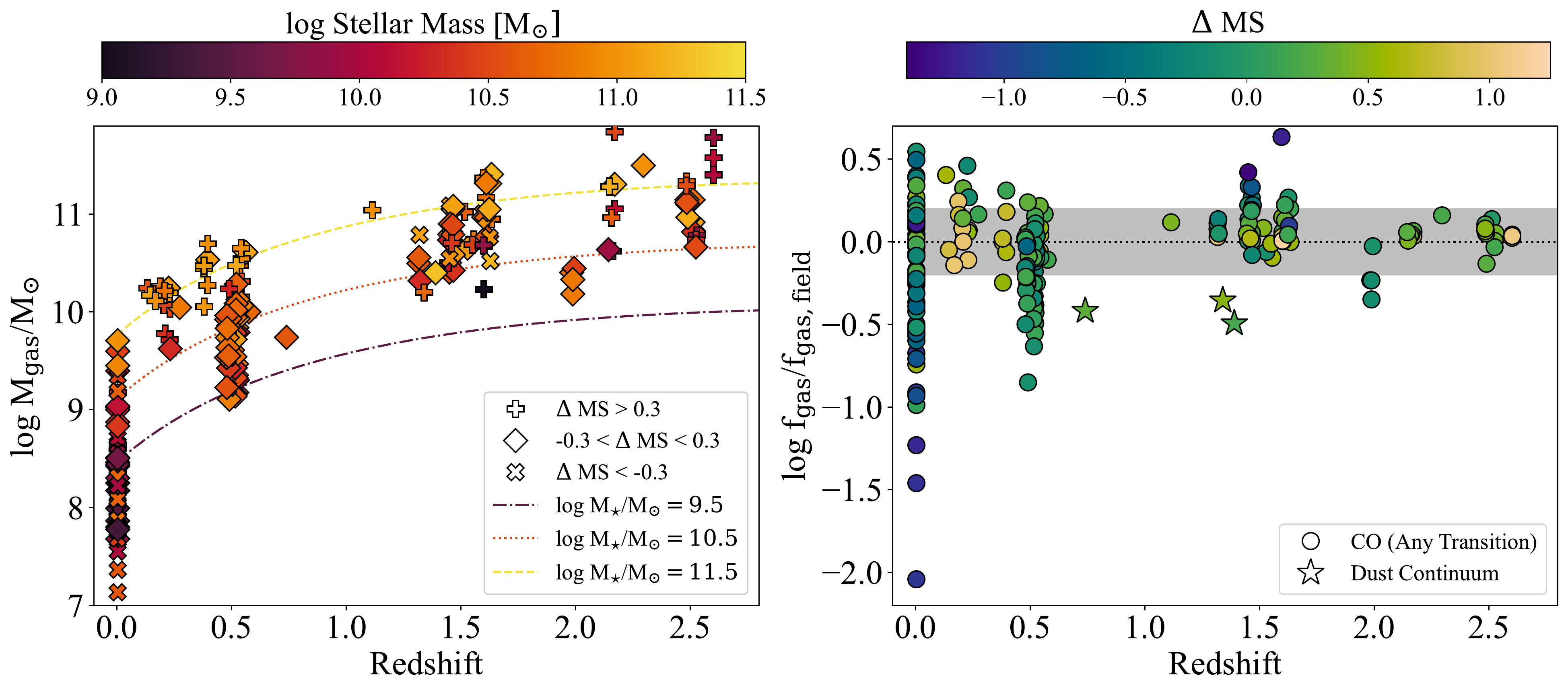}
    \caption{ $-$ {\bf (left)} Molecular gas mass as a function of redshift for Virgo galaxies\protect\footnote{Compiled from \citealt{morokuma-matsui2021} and references therein.} and other various cluster samples.\protect\footnote{Compiled from data in \citealt{castignani2020, sperone-longin2021, sperone-longin2021a, noble2017, noble2019, rudnick2017, hayashi2018, cybulski2016, alberts2022, williams2022, betti2019, coogan2018, chapman2015, wagg2012, geach2009, geach2011, aravena2012, gomez-guijarro2019, tadaki2019}.}  The lines represent the field-scaling relations at three different stellar masses, using the formulations in \citealt{tacconi2018}. There is a clear increase in gas mass with increasing redshift, stellar mass, and offset from the star-forming Main Sequence, for both field and cluster galaxies.  {\bf (right)} The molecular gas fraction offset of cluster galaxies from the field-scaling relations at a given stellar mass, star formation rate, and redshift, for the same cluster galaxies in the left panel. While low-redshift clusters display a mix of depleted and elevated gas fractions, $z>1$ cluster galaxies are preferentially elevated compared to the field with the exception of 2 galaxies at $z\sim2$ and stacked averages at $z\sim1.4$ using dust continuum as a proxy for molecular gas (stars). The gray shaded region represents the \citealt{tacconi2018} field-scaling relations with an associated $\pm0.2$ dex uncertainty.  }
    \label{fig:gas_panel}
    \end{minipage}
\end{figure}

Recently, additional studies have attempted to holistically characterize the Virgo cluster environment through phase-space diagrams, local galaxy density, and filament analyses, out to large radial extents. \citealt{morokuma-matsui2021} used a mass-limited sample (log $M_{\star}/\Msun>9$) of $\sim170$ galaxies with CO-detections (or upper limits) from the literature, with accompanying stellar mass and SFR estimates, in order to investigate molecular gas deficiencies relative to isolated galaxies as a function of a variety of environmental definitions.  They found that not only do cluster galaxies have lower SFRs (as discussed extensively in \S\,\ref{sec:fir}) and lower molecular gas fractions than field galaxies on average, but there is also a radial/accretion history dependence when normalizing for stellar mass (see Figure~\ref{fig:virgo}). Both quantities decrease and further deviate from the field quantities with decreasing clustercentric radius, increasing projected local galaxy density, and earlier accretion epoch as based on the position in the phase-space diagram, quantified as \rxv.  Moreover, they find the radial dependence transitions more sharply at $\sim1.5R_{200}$, which they interpret as a sign of possible ram-pressure stripping, given it occurs roughly where they expect the boundary to exist between the recent infall and virialized populations. The CO gas masses, SFRs, and stellar masses for Virgo galaxies that were assembled in \citealt{morokuma-matsui2021}, in addition to our own literature compilation of distant clusters, are included in Figure~\ref{fig:gas_panel}, where we plot molecular gas mass and the gas fraction offset from the scaling relations as a function of redshift.

\citealt{castignani2022a} further identified 245 galaxies within large-scale filamentary structures in Virgo, presenting an intermediate environment to contrast to the isolated field and denser core.  They detect a significant fraction of gas-deficient (\textsc{Hi}, H$_2$, or both) galaxies in filaments, and a steady decrease in SFRs and gas contents as galaxies progress from the field to filaments to the cluster core (see also the similar monotonic decrease in the fraction of SFGs in increasing density within the Coma cluster shown in Figure~\ref{fig:coma}). While the correlation is stronger for \textsc{Hi} than the molecular gas, 84\% of early-type galaxies, which tend to live in the densest group-like filamentary regions, have depleted reservoirs, possibly due to starvation and/or group preprocessing.  Conversely, late-type galaxies within Virgo, selected from the JCMT Nearby Galaxies Legacy Survey \citep[NGLS;][]{wilson2012}, have higher H$_2$ gas masses and longer depletion timescales than similarly-selected group galaxies on average \citep{mok2016}; this result holds even when accounting for CO non-detections  with upper limits using a survival analysis. Given the lower \textsc{Hi}-masses of the cluster galaxies, this could imply a more efficient conversion of \textsc{Hi} gas into H$_2$ gas in rich clusters compared to group environments (see also \S\,\ref{sec:COz1}).
It has also been suggested that disturbed and non-virialized clusters similarly represent intermediate environments.  For example, galaxies within the nearby merging Antlia Cluster contain typical molecular gas reservoirs for their mass and SFR, implying that the less-dense ICM in the disturbed cluster does not significantly effect the molecular gas component \citep{cairns2019}.

The molecular gas content of galaxies as a function of environment has now been investigated beyond the local Universe, both in the infall \citep{geach2009, cybulski2016} and virialized \citep{jablonka2013, castignani2020} regions of clusters out to $z\sim0.5$ with CO, and to $z\sim0.7$ \citep{betti2019} using dust continuum (see also \S\,\ref{sec:COz1} and \S\,\ref{sec:COproto} for examples at $z>1$). From Figure~\ref{fig:gas_panel} (left panel) it is apparent that galaxies in overdense environments follow the same general trend as field galaxies: the molecular gas mass for MS galaxies rises with increasing redshift and stellar mass. In detail, however, the picture is more complicated. Using a prescription for $\alpha_{\rm CO}$ based on SFR, \citealt{castignani2020} measured no effect on the molecular gas content in 17 FIR-selected cluster LIRGs at $z=0.2-0.6$, while \citealt{jablonka2013} concluded that three $z\sim0.4$ cluster galaxies had a lower CO luminosity at fixed stellar mass or infrared luminosity.  However, these studies primarily focused on LIRGs above the star-forming main-sequence to maximize CO detectability, and thus potentially suffer from selection effects, such as Eddington and Malmquist biases \citep{cybulski2016}. The Spatially Extended EDisCS Survey \citep[SEEDisCS;][]{sperone-longin2021, sperone-longin2021a} instead targeted 49 star-forming galaxies on the main-sequence (72\% of the sample is within 0.3 dex) in two $z\sim0.5$ clusters within 5$R_{200}$.  Matched in color, stellar mass, and redshift to PHIBSS2 field galaxies, the cluster sample comprises a unique population of galaxies with low gas-to-stellar mass ratios that is absent in the field comparison (see $z\sim0.5$ points in right panel of Figure~\ref{fig:gas_panel}); there is a slight preference for this subset to be located along the cluster infall regions, again hinting at an environmental dependence on molecular gas properties.  Supporting this, \citealt{betti2019} stacked dust continuum measurements of mostly MS galaxies at $z\sim0.7$ as a function of local environment in COSMOS, from which they derived low molecular gas content in their intermediate and high galaxy density bins. Thus, at $z\lesssim0.7$,\footnote{There is a deficit of CO cluster studies from $0.6<z<1$, which might be partially due to broad oxygen and water vapor absorption bands at 120 and 183GHz, limiting continuous coverage of CO ($3-2$) and ($2-1$) at $z\sim 0.8 - 0.9$.} environmental studies seem to be converging on the idea that the integrated molecular gas content is on average depleted compared to isolated field galaxies, though with some evidence for elevated gas masses in Virgo spirals \citep[e.g.,][]{nakanishi2006, mok2016}. 


\subsubsection{Intermediate-redshift ($1<z<2$) Cluster Trends}
\label{sec:COz1}
The picture emerging at $z>1$ is still preliminary, primarily due to large scatter from small samples, and the time-intensive nature of probing to sufficiently low CO flux limits at high redshift.  Many early studies thus focused on molecular gas in only one or two, often unique, (proto-)cluster galaxies, e.g.: an infrared luminous AGN in the outskirts of a cluster at $z=1.1$ \citep{wagg2012}; a bright AGN within a $z=1.4$ galaxy overdensity around a central radio source \citep{casasola2013}; and a potentially interacting cluster galaxy pair at $z=1.2$ \citep{castignani2018}.   Other studies have instead used high-redshift clusters to conduct pseudo-blind CO surveys, exploiting the high density of galaxies over a narrow redshift range to target numerous galaxies simultaneously due to the availability of wide receiver bandwidths and large primary beams  \citep{aravena2012, chapman2015, rudnick2017, hayashi2017, hayashi2018, noble2017,noble2019, stach2017, coogan2018, damato2020, williams2022}.  For example, \citealt{aravena2012} pointed at a $z\sim1.55$ candidate galaxy cluster, covering four spectroscopically-confirmed sources at a similar redshift. Using the JVLA, they detected CO ($1-0$) in two of the confirmed members, as well as in two serendipitous galaxies with optical counterparts.  The galaxies had SFEs consistent with high-redshift SFGs, demonstrating the utility of using overdense environments to efficiently target CO in typical distant galaxies in a reasonable integration time ($\sim20$ hours on source).  Similarly, \citealt{rudnick2017} pointed the JVLA at the peak location of MIPS cluster members in the well-known star-forming cluster ClG J0218.3-0510 at $z=1.62$ (see \S\,\ref{sec:sfr-density_relation}).  In 100-hours of integration time, only two out of 6 (9) confirmed members through spectroscopic (grism) redshifts resulted in significant CO ($1-0$) emission, both 24\um\ detected.  Despite one source having a SFR an order of magnitude below the MS, the molecular gas fractions are consistent with the field-scaling relations from \citealt{genzel2015}, and stacking on the star-forming members without CO detections did not yield any signal. 


With the increased sensitivity of ALMA, $1<z<2$ cluster surveys are now observing molecular gas reservoirs with even higher efficiency, detecting CO or dust in $\gtrsim$10 galaxies with a mere few hours of integration time.  The results have been puzzling. The majority of CO studies have measured integrated molecular gas fractions 
that are enhanced (or consistent with but systematically above) the field-scaling relations \citep{genzel2015, tacconi2018} for a given stellar mass and SFR (see solid circles at $z\sim1.5$ in right panel of Figure~\ref{fig:gas_panel}).  Using CO ($2-1$), this has been shown in a seemingly heterogeneous sample of clusters, from three rich NIR-detected SpARCS clusters (J0225, J0224, J0330) at $z\sim1.6$ \citep{noble2017, noble2019}, a mature X-ray cluster at $z=1.46$ (XMMXCS J2215.9–1738) with an overdensity of SMGs in the core \citep{hayashi2017, hayashi2018}, and a lower-mass X-ray cluster at $z=1.3$ \citep{williams2022}.  It has been postulated that these enhanced gas fractions could result from various effects, for example: (1) an environmental  pressure that increases the formation of molecular gas through compression of the ISM \citep{bahe2012}; (2) a preferential location that is conducive to efficient gas inflows \citep[e.g. ][]{dekel2006, zinger2016, zinger2018a}; (3) a mechanism that perturbs the gas such that a smaller fraction of it is available to form stars;
(4) a selection effect; or (5) an environmental dependence on the conversion factor between CO and H$_2$ ($\alpha_{\rm CO}$; see \S\,\ref{sec:caveats}). The latter two possibilities are unlikely the sole causes. Indeed, \citealt{hayashi2018} accounted for metallicity effects using a mass-dependent $\alpha_{\rm CO}$. Moreover, \citealt{noble2017} found a higher fraction of cluster galaxies offset from the scaling relations than non-detections, and \citealt{hayashi2018} stacked 12 quiescent galaxies with no measurable gas content down to $M_{\rm gas}=10^{10}\,\Msun$.
Perhaps more surprisingly, there are very few CO-detected galaxies at this redshift that lie $>1$ dex above the MS, and some of the most elevated gas fractions are in fact in galaxies $>0.5$ dex below the MS (see right panel of Figure \ref{fig:gas_panel}).  Furthermore, subsequently deeper CO observations, down to an rms of $\sim$0.1-0.2 mJy/beam over 50 km/s channels, have not produced a population of cluster galaxies with reduced gas reservoirs compared to the field-scaling relations in these clusters \citep{noble2019, williams2022}.

Conversely, some studies have found evidence of significant gas deficits at high redshift, particularly when using dust continuum as a molecular gas tracer. \citealt{alberts2022} derived the average (stacked) gas content of 126 cluster galaxies across 11 ISCS clusters at $z\sim1.4$. Performing a comparison to stacked field samples and field-scaling relations, they found that cluster SFGs (on or above the MS) have on average $2-4\times$ lower molecular gas masses and fractions than the field (shown as stars at $z\sim1.5$ in right panel of Figure~\ref{fig:gas_panel}).  This result holds out to $2R_{\rm vir}$, suggesting gas loss starting outside the cluster virial radius as the cause of the rapid quenching motivated in \S\,\ref{sec:fir} for this epoch (see discussion in \S\,\ref{sec:beyond}).  However, this work must be reconciled with CO studies before a robust interpretation can be made. 
While both CO and dust continuum are established as robust gas tracers in the field, this has not been tested in overdense environments.  As noted earlier, in the local Universe both dust and CO show signs of depletion likely due to stripping but not at the level of more extended \textsc{Hi} content, suggesting both are more tightly bound in the central disk \citep{corbelli2012}.  \citealt{cortese2016} found a tentative increase in the ratio of molecular gas to dust mass in \textsc{Hi}-deficient Virgo galaxies by $\sim1.5\times$, suggesting that dust is more easily stripped (see \S\,\ref{sec:gas_lowz}). This result, however, was \red{only} significant at the $1\sigma$ level.  Since ISM conditions at high redshift significantly differ from the local Universe, resolving this disconnect will require \red{both}\sout{simultaneous} measurements of CO and dust in a statistical sample of high-redshift cluster galaxies.

Do any $1<z<2$ CO studies find gas deficits? One study of a $z=1.99$ X-ray-detected cluster (Cl J1449+0856) has measured molecular gas fractions on or slightly below the field-scaling relations using gas masses from CO ($1-0$) in three main-sequence BzK galaxies \citep{coogan2018}.\footnote{CO ($4-3$) was detected in 8 galaxies total in the \citealt{coogan2018} study, but some lacked stellar masses and SFR estimates in order to determine how their gas fractions compare to the field-scaling relations.}  Notably, this is not due to reaching a low CO flux limit, as the ALMA observations used in the \citealt{coogan2018} study are appreciably shallower than than those of \citealt{noble2019} and \citealt{williams2022}. One important feature of Cl J1449+0856, which is NIR-selected and X-ray detected, is that it contains a significant quiescent population in the core \citep{strazzullo2018}.  In contrast, the three $z\sim1.6$ NIR-selected clusters with high gas fractions in \citealt{noble2017, noble2019} have low quenching efficiencies \citep[][see also \S\,\ref{sec:nir_highz}]{nantais2016, nantais2017}. This indicates that significant cluster-to-cluster variation is likely a culprit for some of the seemingly heterogeneous results at high redshift, where sample sizes are still small.  It is also interesting to note that using multiple CO transitions ($4-3$, $3-2$, and $1-0$), \citealt{coogan2018} found that many of the galaxies in the $z=1.99$ cluster displayed starburst-like excitation in the SLEDs, with substantial amounts of denser molecular gas\footnote{See also \citealt{daddi2015} for starburst-like excitation in CO $(5-4)$ within $z=1.5$ BzK field galaxies.}.  Similarly, an independent study of the $z=1.46$ cluster XMMXCS J2215.9–1738 found a significant fraction of dense CO ($5-4$) gas compared to  field galaxies \citep{stach2017}. This could suggest that using a larger ratio of the CO line brightness temperatures (e.g., r$_{\rm 21}$ = $L^{\prime}_{\rm CO(2-1)}/L^{\prime}_{\rm CO(1-0)}$) might be more appropriate for these high-redshift cluster galaxies, which would bring down the gas fractions when estimated from higher-order rotational transitions (see also \S\,\ref{sec:caveats}).  The current ALMA receiver bands only permit CO ($1-0$) to be observed up to $z\sim0.35$; however, this will change with the full installment of receiver Band 1 over $35-50$\,GHz \citep{huang2016} which recently had a successful first light run and will enable sensitive observations of CO ($1-0$) out to $z\sim2$ in a reasonable integration time and permit studies of environmentally dependencies on CO excitation.

\subsubsection{High redshift ($z>2$) Proto-cluster Trends}
\label{sec:COproto}
The question raised in the previous section of the distribution of CO among different transitions becomes even more salient at higher redshifts. At $z>2$, many studies rely on higher excitation transitions ($J>3$) to probe molecular gas contents with ALMA, and/or use the lower frequency receivers available with, e.g., JVLA for CO ($1-0)$ observations, albeit with less sensitivity compared to ALMA.  Again, early studies focused on a few token galaxies in proto-clusters, e.g. SMGs at $z>4$ \citep{riechers2010, daddi2009, hodge2013}; both the central radio source \citep[MRC 1138-262;][]{emonts2016} and an H$\alpha$-emitter \citep[HAE229;][]{dannerbauer2017} belonging to the Spiderweb conglomeration at $z\sim2.2$; and a single detection of an infrared-bright, optically-faint galaxy in the dense core of HS1700+64 at $z=2.3$ \citep{chapman2015}.  Proto-cluster studies of molecular gas have now expanded to a handful of systems, namely PCL1002 (and other structures associated with Hyperion), 4C23, SSA22, USS1558, GN20, SPT2349-56, and the Distant Red Core (see Table \ref{tbl:protoclusters} for details on those detected with DSFG overdensities).  Extended gas reservoirs and high gas fractions have been measured for galaxies within many of these proto-clusters \citep{riechers2010, hodge2013, chapman2015, dannerbauer2017, emonts2018, emonts2016, tadaki2019, wang2018, gomez-guijarro2019, champagne2021, polletta2022}.  For example, detecting CO ($3-2$) in 16 H$\alpha$ emitting galaxies (on or above the main-sequence) over three proto-cluster fields from $2\lesssim z\lesssim 2.5$, \citealt{tadaki2019} found a mass-dependent effect: less massive proto-cluster galaxies ($10.5<\log~M_{\star}/\Msun<11$) harbor slightly enhanced gas fractions (though consistent within errors) and longer depletion timescales relative to their isolated counterparts, while more massive members are similar to the coeval field  (albeit based on only three CO detections with log $M_{\star}/\Msun > 11$).  They propose that accelerated gas accretion along cosmic filaments might replenish the gas reservoirs and sustain star formation over long periods in lower-mass galaxies \citep[e.g.,][]{dekel2009}, while the gas could be heated by virial shocks or AGN in the more massive systems (see discussion in \S\,\ref{sec:protocluster_disc}).  


Studies of filaments associated with the Hyperion proto-supercluster at $z\sim2.5$ have yielded similar results.  Using CO ($1-0$), \citealt{champagne2021} detected a slight enhancement in the gas fractions of galaxies within PCL1002, yet also measured a integrated SFE that is comparable to the field. A nearby structure (CLJ1001) was analyzed by \citealt{wang2016} as a proto-cluster core due to its extended X-ray emission \citep[but see][which characterized this structure as a filament]{champagne2021}.  Using phase-space diagrams, they investigated the gas content of massive SFGs
as a function of their accretion history.  They found that the molecular gas reservoirs are reduced from the infall region to the center within a single radial orbit, hinting at fast quenching mechanisms like tidal or ram-pressure stripping.  \citealt{gomez-guijarro2019} utilized a variety of observatories to target multiple CO transitions in the Hyperion overdensity, as well as reporting the discovery of two new gas-rich proto-cluster cores at $z\sim2.2$ and $z\sim2.6$.  They also propose a mass-dependent trend in gas properties, finding low-mass member galaxies with high gas fractions but main-sequence SFEs.

As in the $1<z<2$ cluster studies, dust continuum measurements of proto-cluster galaxies typically estimate lower gas masses than those from CO emission lines.  For example, \citealt{aoyama2022} contend that CO $(3-2)$ fluxes overestimate the gas masses in proto-cluster USS1558-003 due to higher than expected gas excitation, while the ALMA dust continuum measurements at 1.1mm suggest the large gas reservoirs are actually consistent with the field-scaling relations.  \red{In the $z=2.49$ proto-cluster 4C23.562, CO $(3-2)$ emission also produced higher gas mass estimates compared to those from 1.1mm dust continuum and optically-thin \textsc{[Ci]} line emission by $\sim 0.15$ dex on average, but only in 2 galaxies, while the remaining 3 galaxies displayed consistent measurements from all three tracers \citep{lee2021}.} \citealt{zavala2019} similarly measured field-like gas masses from 1.3mm continuum observations, and additionally discovered a population of gas-poor members at the high-mass  end.  These studies highlight the importance of obtaining CO SLEDs in a large number of high-$z$ proto-cluster galaxies to constrain gas excitation.  A recent survey of 10 highly star-forming \textit{Herschel} sources in \textit{Planck} high-$z$ (PHz) proto-cluster candidate fields from $1.3<z<3$ has measured multiple CO transitions, finding the SLEDs to peak at quantum rotation number $J=3$, implying low excitation \citep{polletta2022}.  Increased samples sizes are needed to provide further calibration of CO gas and dust masses in high-$z$ (proto)-cluster fields.

Additionally, some surveys are placing the first constraints on the CO luminosity function and cosmic gas density in high-redshift overdense environments.  Using five CO ($3-2$)-detected sources in a $z=2.49$ proto-cluster \citep[4C23.56;][]{tanaka2011}, \citealt{lee2017} estimated a cosmic gas density that is $\sim 6-20\times$ higher than $z\sim2$ field estimates \citep[e.g.,][]{decarli2016}.  They measured a lower limit of $1-5\times10^9\,\Msun$ Mpc$^{-3}$, depending on the redshift range considered for the volume ($\Delta z_{\rm CO}$ or $\Delta z_{\rm filter}$).  A low-resolution mosaic on the Spiderweb proto-cluster with the Australia Compact Array (ATCA) has similarly exposed an enhanced CO luminosity function that is $\sim$tens of times higher than blank fields \citep{jin2021}.
Studies of gas and dust in proto-cluster members at $z>4$ have found a similarly high number of gas-rich and starburst galaxies, including: the DRC at $z=4.0$ \citep{oteo2018, long2020}; GN20 at $z=4.05$ \citep{pope2005, daddi2009, hodge2013}; SPT2349-56 at $z=4.3$ \citep{miller2018, hill2020}; and AzTEC-3 at $z=5.3$ \citep{riechers2010}.  These studies establish proto-clusters as sites of rapid stellar mass growth at early times.

In summary, a rich dataset of integrated molecular gas properties in galaxies within (proto-)cluster environments has emerged out to $z\sim3$, covering a $\sim2.5$ dex range in stellar masses and SFRs along the Main Sequence (Figure \ref{fig:gas_panel}).  Low-redshift cluster members display a wide range of gas fractions, including a clear population of Main Sequence galaxies with depleted gas reservoirs compared to coeval field galaxies.  CO-based studies at $1<z<2$ have instead discovered numerous gas-rich galaxies, with a notable absence of galaxies below the field gas scaling relations, despite probing galaxies down to low SFRs. Dust continuum tracers, however, paint a different picture, with stacked measurements estimating gas fractions that are a few times below the field.  While the enhanced gas fractions at $1<z<2$ are intriguing, we have yet to conclusively determine what drives them. Resolving the molecular gas spatially and kinematically is the first step toward answering this, as various quenching mechanisms are expected to have different effects on the gas.

\subsection{Lessons from Spatially-Resolved Studies}
\label{sec:resolved}

Spatially-resolved studies of molecular gas are now providing insight into how the gas is affected by the cluster environment, particularly through detailed studies of CO radial profiles, morphologies, and kinematics of low-redshift cluster galaxies. For example, mapping CO and far-infrared continuum in Virgo spirals from the HeViCS with the IRAM-30\,m telescope, \citealt{pappalardo2012} measured steeper molecular gas and dust profiles in \textsc{Hi}-deficient galaxies.  \citealt{fumagalli2009} similarly showed that while not all \textsc{Hi}-deficient galaxies have reduced molecular gas contents, galaxies do have depleted CO when the \textsc{Hi} is removed from within the optical disk, indicating that molecular gas is more tightly bound. Virgo studies have also found molecular gas disks to be truncated compared to field populations \citep{davis2013, zabel2022}. Moreover, the spatial extent of CO and dust is correlated with the level of \textsc{Hi} depletion, with truncated molecular gas \citep{boselli2014} and dust \citep{cortese2010} disks present in more \textsc{Hi}-deficient galaxies, as well as decreasing CO-to-optical diameter ratios (i.e., becoming more gas deficient) with decreasing \textsc{Hi}-to-optical diameters \citep{chung2017}.   On the other hand, observing a higher CO transition of ($3-2$) with the JCMT, \citealt{mok2017} found no truncation of molecular gas sizes in \textsc{Hi}-flux selected Virgo spirals, but did measure steeper CO profiles with enhanced central surface densities in the cluster sample.  These many results all suggest that the environment removes gas from the outside-in, and can potentially further increase H$_2$ formation in the galaxy center.  Simulations have shown that moderate ram-pressure stripping could result in this behavior \citep{tonnesen2009}.

The mapping of CO morphologies in low-redshift cluster galaxies similarly supports the idea that ram-pressure can affect the molecular gas, though tidal stripping features can also produce similar signatures \citep[e.g.,][]{combes1988}. Both in Coma and the merging cluster Abell 1367 (part of the Coma Supercluster), molecular gas with asymmetric and complex distributions has been revealed in member spiral and irregular galaxies \citep{vollmer2001, scott2013, scott2015}. Furthermore, in two cases, the CO was significantly offset from the optical centroid, indicative of perturbed gas from ram pressure stripping.  Even in the Fornax cluster, which is less massive and should therefore have less efficient ram-pressure stripping, many low-mass galaxies ($\lesssim  3\times10^9\,\Msun$) have disturbed molecular gas morphologies, including some with gas tails that are aligned with the cluster center \citep{zabel2019}.  \red{Molecular gas asymmetries have further been identified within lower-density group environments through peculiar CO distributions \citep{lee2022}, suggesting some amount of pre-processing might also be occurring.}

While these studies show great progress and are reaching some consensus, they have been limited by small sample sizes, marginal spatial resolution, heterogeneous comparisons, or some combination thereof. Large cluster programs on ALMA will transform this.  For example, the Virgo Environment Traced in CO (VERTICO) survey \citep{brown2021} has observed CO ($2-1$), as well as other CO isotopologues and gas tracers, at an exquisite resolution of $\sim$700\,pc in 51 Virgo galaxies over a broad range of SFRs and stellar masses, with ancillary \textsc{Hi} imaging \citep{chung2009}.  Using a homogeneously-selected and analyzed sample of spatially-resolved field galaxies from HERACLES\footnote{HERA CO-Line Extragalactic Survey \citep[HERACLES;][]{leroy2009}}, VERTICO uncovered an identical molecular gas mass-size relation in both samples, concluding that any environmental effects on the molecular gas act equally on the integrated gas mass and resolved gas distribution. They also highlighted the low scatter in the relationship when using isodensity radii sizes as opposed to 90\% flux-percentage radii.  \citealt{zabel2022} further investigated the shapes of the CO radial profiles in VERTICO, finding more compact and steeper H$_2$ profiles in cluster galaxies with larger \textsc{Hi} deficiencies, partially echoing the results in prior Virgo studies \citep{boselli2014, mok2017}.  They interpret these findings as evidence for ram-pressure stripping, and additionally identify several Virgo galaxies with clear morphological signatures of stripping events that have normal-to-enhanced amounts of molecular gas but depleted of \textsc{Hi}.  Further kinematic studies of the molecular gas in ram-pressure stripped galaxies are necessary to fully characterize the effect on this denser gas phase, and are the motivation behind the ALMA JELLY large program (PI: Jachym), which currently has observations underway.

\begin{SCfigure}[0.8][!htb]
    \parbox{0.5\columnwidth}{\centering\includegraphics[width=0.55\columnwidth]{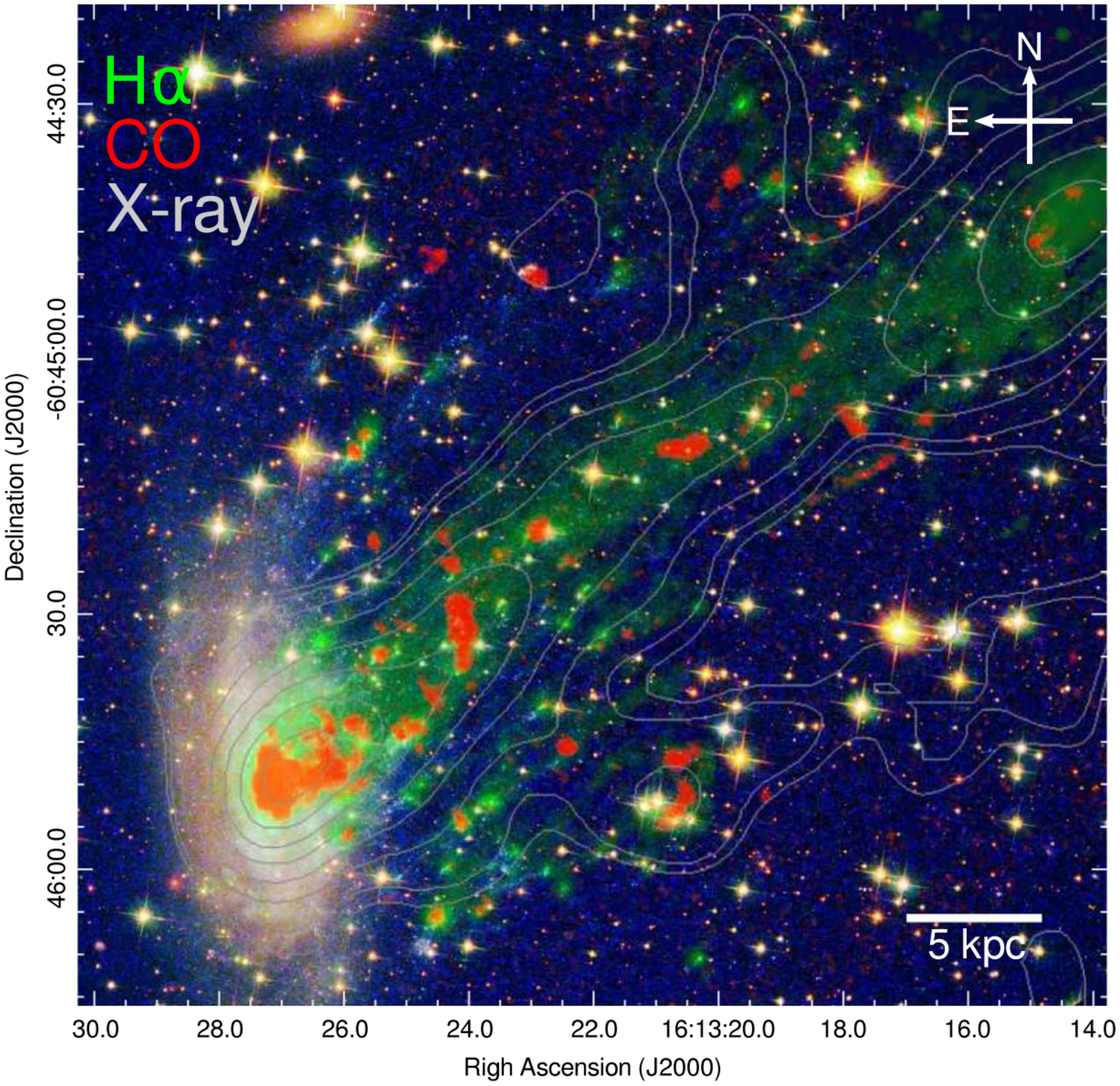}}
 \caption{ $-$ The jellyfish galaxy, ESO 137-001, with ALMA CO ($2-1$) (red), H$\alpha$ from MUSE (green), and \textit{Chandra} X-ray emission (contours), overlaid on a 3-color image from \textit{HST}.  At 350\,kpc resolution, the CO is distributed along filaments and clumps, potentially representing different phases of gas stripping.  Figure adapted from Figure 3 in \citealt{jachym2019}, reproduced by permission of the AAS.}
 \label{fig:ESO137}
\end{SCfigure}

In the meantime, dedicated studies of gas kinematics and morphologies in individual ram-pressure stripped galaxies have provided the clearest picture of environmental effects on the molecular gas component.  Large amounts of extraplanar CO \citep[including $^{13}$CO;][]{lee2018} have now been detected in tail-like features of low-redshift cluster galaxies \citep[e.g.,][]{vollmer2008, verdugo2015, moretti2018, zabel2019}, sometimes seen kiloparsecs away from the disk \citep[e.g.,][]{jachym2014, jachym2017, cramer2020, moretti2020, cramer2021}.  In particular, much focus has been on some of the most obvious ram-pressure stripped candidates, known as jellyfish galaxies due to their long tentacle-like tails of ionized gaseous debris \citep{owers2012, poggianti2016}, often with knots of star formation \citep[e.g.,][]{ebeling2014, fumagalli2014a, rawle2014}. For example, ESO 137-001 in the Norma cluster (Abell 3627) is an exquisite example of ram-pressure stripping in action; diffuse H$\alpha$ emission, discrete HII regions, and shock-excited molecular gas $-$ revealed through MIR rotational transitions of warm H$_2$ with \textit{Spitzer}/IRS\footnote{InfraRed Spectrograph \citep[IRS;][]{houck2004}} \citep{sivanandam2010, sivanandam2014} $-$ are embedded within a $\sim$80\,kpc long X-ray tail that emanates from a undisturbed spiral disk  \citep{sun2006, sun2007a}.  CO was detected in the tail out to $\sim60$\,kpc, with the amount of gas decreasing along the length of tail \citep{jachym2014}.  High-spatial resolution observations of CO ($2-1$) at 350\,pc with ALMA (Figure~\ref{fig:ESO137}) revealed both filamentary and clumpy CO, including clouds of CO at the heads of streams of young stars \citep{jachym2019}, known as ``fireballs" \citep{kenney2014}.  \citealt{jachym2019} argue that these different CO structures represent different evolutionary stages of stripping, with the filamentary structures within the tail forming in situ (potentially from stripped \textsc{Hi}), while some CO clumps might have been stripped directly from the disk.  

\begin{figure}[!htb]
    \centering
    \includegraphics[scale=0.37]{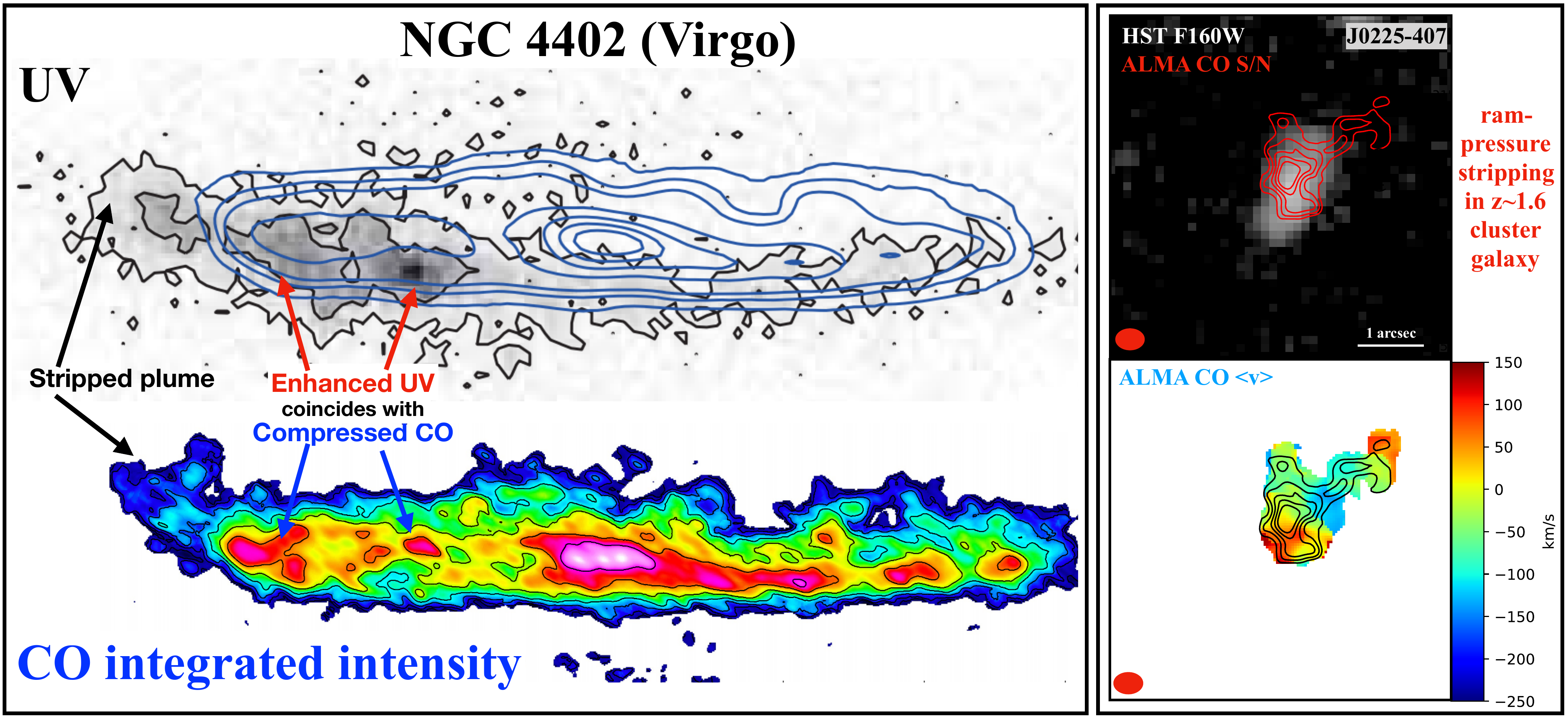}
 \caption{ $-$ {\bf (left)} Adapted from \citealt{lee2017a} and \citealt{cramer2020} in top and bottom panels, respectively.  The locations of UV peaks (gray-scale, top) correlate with sites of molecular gas compression (CO, bottom) in a ram pressure-stripped galaxy in Virgo.  {\bf (right)} Adapted from \citealt{noble2019}. ALMA CO observations of a $z\sim1.6$ cluster galaxy display similar  indications of ram-pressure stripped molecular gas tails ($2-6\sigma$ red contours on grayscale \textit{HST} F160W \red{in upper right panel}), including a strong kinematic signature--- accelerated gas toward the tip of the tail (color image in bottom right panel).  Figures reproduced by permission of the AAS.}
 \label{fig:Virgo_407}
\end{figure}

In concert with the morphological evidence for molecular gas in tails, the kinematics of CO in ram-pressure stripped galaxies has also provided insight into how this denser gas phase is affected. For example, studies have now revealed direct evidence for a marked influence on molecular gas in the Virgo cluster galaxy NGC 4402 through disturbed CO morphology with the SMA \citep{lee2017a} and kinematics with ALMA at $\sim100$\,pc resolution \citep{cramer2020}. The CO in the stripped plume, on the leading side of the galaxy, has a clear velocity offset of up to 60\,$\rm{km\,s^{-1}}$ from the galaxy's normal rotation \citep{cramer2020}.  The kinematic disturbance exhibits a radial gradient, with the velocity offset increasing with distance from the nucleus.  This is a strong indication of the influence of ram pressure, consistent with stripping acting from the outside-in, and potentially removing the more diffuse molecular gas \citep{lee2017a, cramer2021, young2022}. Moreover, compressed CO gas along the leading edge \citep{cramer2020} of NGC 4402 is coincident with enhanced star formation traced by UV \citep{lee2017a} emission (Figure~\ref{fig:Virgo_407}). Compressed gas has additionally been observed in other ram-pressure stripped galaxies \citep{lizee2021, cramer2021}, and can also result in an increase in the molecular gas fraction \citep{nehlig2016, moretti2020a}.  Therefore, ram pressure can possibly trigger star formation enhancement during the stripping phase, and prior to ultimate quenching \citep{roberts2020}.

Spatially-resolved kinematic studies of the molecular gas component in distant galaxies (within the field and clusters) are still in their infancy, with most high-redshift studies targeting single (typically star-bursting) galaxies in CO \citep[e.g.,][]{riechers2008, tacconi2010, bothwell2010, hodge2012, hodge2013, cibinel2017}. In particular, there are only a handful of higher-resolution (\red{primary} beam of a $\sim$few kiloparsecs) CO detections in main-sequence field galaxies at $z > 1$ \citep{molina2019, genzel2020}.  Instead, efficient multiplexing over dense clusters has resulted in spatially-resolved CO observations of multiple cluster galaxies at high redshift; there are currently 8 CO ($2-1$) detections  within a single ALMA field-of-view (FOV) over a $z=1.6$ cluster \citep{noble2019}, 11 CO ($4-3$) detections of H$\alpha$ emitters within a $z=2.5$ proto-cluster \citep{lee2019a}, \red{and 14 $z\sim1.6$ cluster galaxies with spatially-resolved CO kinematics \citep{cramer2022}}, all at $\sim3$\,kpc resolution.  \citealt{lee2019a} found cluster galaxies to have broader CO line widths by 50\% compared to field galaxies, despite having similar CO \red{intensities}\sout{immensities}.  They attributed this to increased merger events.  \citealt{noble2019} reported the first tentative evidence for molecular gas stripping at $z\sim1.6$ through the presence of asymmetric molecular gas tails and truncated gas disks. There is also a clear offset between the gas and stellar disk centroids in a handful of galaxies, indicative of perturbed gas. One galaxy stands out in particular (right panels in Figure~\ref{fig:Virgo_407}), due to an elongated gas tail that extends well beyond the optical stellar disk.  Moreover, the \red{gas} kinematics \red{of this cluster galaxy} display the most glaring evidence for RPS, as the tail gas is accelerated from the base to the tip. \red{Modeling the CO rotation in 13 additional galaxies within the same $z\sim1.6$ SpARCS clusters, \citealt{cramer2022} further measured a high degree of kinematic asymmetry in the molecular gas}.  These signatures, in tandem with elevated gas fractions compared to field galaxies \citep{noble2017}, are indicative of environmental processes acting on the gas component.  These results might indicate that $z\sim1.6$ cluster galaxies are undergoing modest ram pressure stripping, which could explain the apparently high gas fractions: if the stripping causes the gas to become compressed at the leading edge, it could cause an efficient conversion of neutral hydrogen into the molecular phase \citep[see also][]{moretti2020a}.

\section{Total Emission from (Proto-)Clusters: ``Total Light'' Stacking}\label{sec:totallight}

In the previous three sections, we discussed detailed studies of the galaxy populations in (proto-)clusters using near- to far-infrared (submm) observations.  These studies often involved expensive spectroscopic and photometric follow-up to confirm cluster galaxy membership and perform individual multi-wavelength characterization. This kind of follow-up is impractical on the scale of the thousands of (proto-)cluster candidates we are identifying in current and future wide-field and all-sky surveys (\S\,\ref{sec:cluster_surveys}, \S\,\ref{sec:future}).  Furthermore, we are often confined by current capabilities to studying the most luminous populations; for example, galaxies need to be bright in the optical/NIR to obtain a secure spectroscopic or photometric redshift. 
Though capabilities are growing in terms of detecting and characterizing challenging populations such as low-mass and heavily dust-obscured galaxies, the expense of confirming individual (proto-)cluster membership remains high.  
These issues combined with 1) a large intrinsic range in cluster properties (i.e. halo mass, dynamical state), 2) selection bias from different cluster selection techniques, and 3) intrinsic variations in cluster populations even in similarly selected, similar mass clusters \citep[e.g.][]{brodwin2013, alberts2016} highlight the need for complementary statistical analyses that can recover cluster properties in large cluster catalogs.  A statistical approach may in fact be the only way to take full advantage of future large (proto-)cluster surveys.  

In this section we discuss the renewed interested in using (image) stacking \citep[e.g.][]{dole2006} of full (proto-)cluster systems, in which all emission is considered without first identifying individual constituents.  Stacking boosts the collective signal over the background noise fluctuations, which often dominate over the signal of individual clusters in wide, but shallow surveys.  These ``total light'' stacks\footnote{Total light stacking is the stacking of image cutouts large enough to contain entire (proto-)cluster structures, rather than stacking on individual galaxies within the (proto)clusters.} measure the average or median flux (after background subtraction) of the (proto-)cluster sample at a given wavelength, which represents the typical integrated light from all cluster components, as well as the radial profile of that emission, given moderate spatial resolution. In \S\,\ref{sec:planck_selection}, we saw how this technique was used to characterize \textit{Planck} ``cold sources'', i.e. compact sources in the 4$^{\prime}$ \textit{Planck} beam which were resolved into extended emission in stacks at $350\mu$m (25$^{\prime\prime}$ beam; Figure~\ref{fig:planck_spire}).  We start our discussion with an overview of the history of this technique in looking for intracluster dust (ICD) and then showcase two recent examples of analyses characterizing the dust/star formation properties and concentrations of cluster populations.

\subsection{Intracluster Dust}\label{sec:icd}

An early demonstration of the power of stacking large samples of clusters was presented in \citealt{kelly1990}, who observed a strong, evolving stacked signal in the 60$\mu$m emission of 71 clusters over $z\sim0.3-0.9$ using \textit{IRAS}.  This summed MIR emission was expected to originate from two components of clusters: intracluster dust and cluster galaxies (via star formation and/or AGN).   Here we give a brief overview of ICD in the context of infrared studies. For an in-depth review of ICD, we refer the reader to \citealt{shchekinov2022}.  We discuss the second component, cluster galaxies, in the next section.

Since early confirmations of heavy elements in the hot ICM from X-ray spectroscopy \citep{sarazin1986, sarazin1988}, it has been expected that dust grains $-$ stripped or ejected from cluster galaxies $-$ are collisionally heated by X-ray-emitting gas, producing diffuse FIR emission \citep{dwek1990}.  The lifetime, and thus the observability, of this dust remains unclear, however.  Dust destruction via sputtering is expected to occur on the order of $10^6$ to $10^9$ years, depending on grain properties and gas density \citep{draine1979, dwek1992}.  Despite this potentially short lifetime, ICD may play a major role in cluster evolution, facilitating the cooling of intracluster gas \citep[e.g.][]{dwek1990, popescu2000, montier2004, weingartner2006, vogelsberger2019} and affecting scaling relations \citep{dasilva2009}.  
This diffuse component, however, has proven difficult to observe.  One common technique is to analyze the reddening of optical sources behind the clusters \citep[e.g.][]{nollenberg2003, chelouche2007, bovy2008, gutierrez2014, longobardi2020}. Another is direct observations in the infrared, which have been carried out on a few local clusters \citep[e.g.][]{stickel2002, bai2007, kitayama2009, bianchi2017}. Generally speaking, however, this diffuse, faint signal is lost in the large fluctuations of the cosmic infrared background.  As such, stacking studies provide a promising alternative, albeit with the requirement that the ICD and galaxy components need to be separated.  

Further stacking studies using \textit{IRAS} \citep{montier2005, giard2008, roncarelli2010} found that IR SEDs built from the total stacked emission were consistent with galaxy spectra, with an evolution in IR luminosity that mimicked the global cosmic evolution of star formation.  Comparing the total IR emission to X-ray indicated an extreme gas-to-dust ratio if cluster IR emission is dominated by ICD \citep{giard2008}, inconsistent with limits set by reddening studies \citep[e.g.][]{chelouche2007, bovy2008}.  Expanded coverage of the dust peak through the combination of \textit{IRAS} and \textit{Planck} 100-857 GHz (3mm-$350\mu$m) imaging of 645 low-redshift (median $z=0.26\pm0.17$) SZ-selected clusters \citep{planckcollaboration2016} enabled the modeling of the total IR emission with a modified blackbody ($\kappa_{\nu}B_{\nu}(T_{\rm dust})$ where $\kappa_{\nu}\propto \nu^{\beta}$)\footnote{A note of caution: in reality MIR/FIR emission is a summation of a series of blackbodies at different temperatures representing different dust grain sizes and compositions.  The observed dust temperature, however, is usually reported as a single, luminosity-weighted temperature or a more robust two-temperature model including ``warm'' and ``cold'' components \citep{kirkpatrick2012}.  In this review, we primarily discuss the commonly-used effective dust temperature, $T_{\rm dust,eff}$, derived from modeling a modified blackbody plus MIR power law \citep[][]{draine2007, casey2012, clemens2013}.}.  Even with no ability to remove point sources due to low resolution, the best fit model ($\beta=1.5$ and $T_{\rm dust}=24.2\pm3.0$ K) was found to be consistent with thermal emission from coeval field galaxies \citep{dunne2011, symeonidis2013}.  
Given the best-fit $T_{\rm dust}$, they derived dust masses and dust-to-gas mass ratios consistent with previous literature \citep[$\mathrm{DGR}\sim2-5\e{-4}$;][]{montier2005, giard2008}.
Using higher resolution \textit{Herschel} coverage of 327 clusters, \citealt{gutierrez2017} subtracted the contribution from (detected) point sources and set a 95\% upper limit on the surface brightness of ICD of $1.3\e{-2}$, $0.7\e{-2}$, and $0.5\e{-2}$ MJy sr$^{-1}$ at 250, 350, and 500$\,\mu$m, implying a strong deficiency in intracluster dust \citep[see also][]{bianchi2017}. Together with reddening studies and simulations, these IR studies help paint the current picture that ICD is a small component of the ICM, with on the order of $\sim0.1-3\%$ of the dust abundance in the ISM of the Milky Way.  However, this small amount of dust may still be non-negligible in the cooling of the cluster ICM \citep{longobardi2020} and there remain uncertainties in the spectral index (up to 20\%) and dust opacity \citep[up to 50\%; see \S\,4.2.1 in][]{planckcollaboration2016} used to calculate the dust mass. Given these, ICD has not been ruled out as an important contributor in cluster evolution.

\subsection{Total Emission from Cluster Galaxy Populations}\label{sec:totallight_pops}

Given the stringent upper limits placed on the ICD component of the total IR emission discussed in the previous section, we can now consider the use of ``total light'' stacking in studying the other component of the IR emission: the (proto-)cluster galaxy populations, including traditionally hard to observed populations such as low-mass and heavily obscured galaxies.  In this section, we look at recent analyses using total stacking in the M/FIR, measuring dust-obscured star formation and/or AGN activity.
Recent works have also expanded ``total light'' stacking to the near-infrared \citep[measuring the total stellar mass;][]{kubo2019, Alberts2021} and the UV \citep[measuring unobscured star formation;][]{mckinney2022}.  Given the moderate resolution (compared to the cluster radius) of all-sky or wide-field surveys from \textit{GALEX}, \textit{WISE}, \textit{Spitzer}, and \textit{Herschel}, a radial profile analysis can be done in addition to looking at the integrated light.  

\subsubsection{The Integrated Light of Cluster Galaxies: Dust Emission and the Contribution from Low Mass Galaxies}\label{sec:totalight_lowmass}

As discussed extensively in Sections~\ref{sec:nir}--\ref{sec:fir}, infrared observations of quenched populations and (obscured) SF activity in clusters have provided evidence for a transition epoch at $z\sim1.5$, above which significant star formation is found in some massive clusters, indicating the local SFR-density relation is no longer in place.
``Total light'' stacking can expand this analysis beyond the relatively luminous cluster populations to examine the typical SED of all constituent galaxies.  This includes low-mass (log $M_{\star}/\Msun<10$) cluster galaxies, which we have discussed play a unique role in identifying environmental quenching as their mass-quenching timescale can exceed the Hubble time.

Here we consider the total stacked emission in multiple bands in the M/FIR, which can be used to construct the average IR SED of cluster galaxies (neglecting the ICD component) and measure the dust temperature.  As mentioned in \S\,\ref{sec:icd}, \citealt{planckcollaboration2016} carried out this analysis on SZ-selected, massive clusters at $z\sim0.3$, finding a relatively cold (average) dust temperature of $\sim24$ K, coinciding with a lack of warm dust in the MIR.  This lack of warm dust has also been observed in the stacking of less massive (log $M_{200}/\Msun\sim13.8$) ISCS clusters from $z=0.5\rightarrow1.6$, with average effective temperatures ranging from 30 to 36 K, increasing with increasing redshift \citep[][]{Alberts2021}.
By way of comparison, massive $z\sim1$ field galaxies have a typical $T_{\rm dust, eff}$ of 42 K \citep{kirkpatrick2015}.  Modeling the IR SED instead with a two component fit, \citealt[][]{Alberts2021} reported ratios of cold to warm dust of $L_{\rm cold}$/$L_{\rm warm} =4.3\rightarrow1.7$ over $z=0.5\rightarrow1.6$, only reaching parity with the field at high redshift \citep[$L_{\rm cold}$/$L_{\rm warm}=1.36$ at $z\sim1$ in the field;][]{kirkpatrick2015}.  \red{If we assume} massive (log $M_{\star}/\Msun\geq10$) cluster galaxies have effective dust temperatures comparable to their field counterparts \citep[e.g.][]{noble2016, alberts2016} and given that $T_{\rm dust}$ increases with $L_{\rm IR}$ \citep[i.e.][]{casey2014, drew2022}, this suggests that the cluster ``total light'' IR SED has a significant contribution from low-luminosity, low-mass galaxies, which increases with decreasing redshift.  \red{We note, however, that there are studies that support both colder \citep[e.g.][]{pereira2010} and warmer \citep[e.g.][]{rawle2012, alberts2022} FIR SEDs in massive cluster galaxies, and the full diversity of the FIR SED in overdense environments in not yet well constrained.}

This lack of warm dust may also signal that (luminous and obscured) AGN activity is not a significant component in the total cluster emission, though the rise in effective temperature with redshift qualitatively parallels the rise in AGN fractions in clusters (\S\,\ref{sec:agn})\red{, as well as rising SFRs}.  The trend toward warmer stacked SEDs continues into the proto-cluster regime: \citealt{planckcollaboration2015} stacked 228 ``cold sources'' (see \S\,\ref{sec:planck_selection}, Figure~\ref{fig:planck_spire}) $-$ proto-cluster candidates probably at $z\sim2-4$ $-$ which may have warm effective temperatures. Dust temperature and redshift are degenerate in their analysis; however, in the likely redshift range of $z\sim2-3$ from \textit{Herschel} colors, they measure $T_{\rm dust,eff}\sim35-45$ K.  Similarly, \citealt[][]{kubo2019} reported high temperatures and a stacked IR SED best modeled with a very warm temperature component in their sample of 179 HSC proto-clusters at $z\sim3.8$. They conclude this warm component is only explainable by luminous AGN.  

Can we disentangle the low-mass galaxy (or AGN) component from the total emission stacks?  \citealt[][]{Alberts2021} compared the average ``total light'' stacked SSFRs of the ISCS clusters in four redshift bins over $z=0.5-1.6$ to their previous stacks on individual galaxies in mass-limited cluster member catalogs \citep[log $M_{\star}/\Msun\geq10.1$;][]{alberts2014}.  They found that the average SSFRs of both the total and high-mass  populations were consistent within the uncertainties, rising with redshift to draw even with the field by $z\sim1.4$.  
This could be interpreted two ways: 1) the SSFR is dominated by the massive galaxies, at odds with the relatively cold SEDs discussed above, or 2) the low-mass cluster galaxies are being effectively quenched at similar rates as their high-mass  neighbors.

To test this, \citealt[][]{Alberts2021} quantified the ratio of the 250$\,\mu$m flux from massive galaxies to total (stacked) emission, removing any model assumptions in deriving the SSFRs.  In the field, this ratio is $\sim60-80\%$ for log $M_{\star}/\Msun=10-11$ galaxies, derived from SPIRE observations \citep{viero2013} and simulations \citep{bethermin2017}.  This is consistent with low-mass galaxies having low obscuration \citep[$<30\%$ of SF is obscured at log $M_{\star}/\Msun<9$;][]{whitaker2017}.  Shockingly, however, the ratio for  log $M_{\star}/\Msun=10-11$ ISCS cluster galaxies to the total stacked emission is, averaged over all redshifts, $15\pm5\%$!  This suggests an improbably large contribution from low-mass galaxies to the total cluster FIR emission, given that we expect low obscuration.  Several explanations seem unlikely: the similarity between SFG SMFs in clusters and field (\S\,\ref{sec:nir}) rules out a vastly higher ratio of low- to high-mass galaxies in clusters.  Low-mass cluster galaxies appear to be on the MS \citep{old2020}, not preferentially starbursting. One might argue that massive, highly obscured cluster galaxies are simply missing, due to the difficulties in confirming membership; however, at the highest redshift, accounting for a significant amount of missing SF would require a strong reversal in the SFR-density relation \citep{Alberts2021}.  Likewise the observed total SF from massive galaxies (\S\,\ref{sec:fir}) and drop in quenching efficiency at high redshift \citep[e.g.][]{nantais2017} rule out that the massive galaxies have simply quenched.  A few studies \citep{koyama2011, hatch2016, sobral2016} have put forth tentative evidence that cluster galaxies may be more dusty then their field counterparts; however, follow-up ``total light'' stacking of the ISCS sample in the UV found the unobscured SFR was consistent with field-like obscuration in low-mass cluster galaxies \citep[][see also \citealp{tran2015}]{mckinney2022}. Further work characterizing the dust properties in cluster galaxies is needed to resolve this mystery.  

These studies, deriving averaged dust temperatures and star formation properties in large cluster samples, showcase the potential of total stacking in recovering the integrated emission from the full cluster population.  In the next section, we discuss an example that takes advantage of resolving this total emission.

\subsubsection{Radial Profiles and the $c$-$M_{\rm halo}$ relation}\label{sec:totallight_c}

When total emission stacks are more extended than the imaging beamsize (and of sufficient S/N), the (observed) radial profile can be quantified.  In current studies, the extent of the stacked stellar and dust emission in clusters seems to trace the overall Dark Matter (DM) mass distribution, as measured by stacked SZ \citep{planckcollaboration2016} or the cluster velocity dispersion \citep{Alberts2021}. Recently, the \textit{Herschel} profile of stacked \textit{Planck}-selected proto-clusters was found to extend to $\sim6-8^{\prime}$ \citep{lammers2022}, roughly comparable to the expected area covered by proto-clusters \citep{chiang2013}.   Pushing this analysis further, we can infer the intrinsic radial profile to measure the distribution of the baryons and compare to the cluster halo.  Given hierarchical structure formation, DM haloes are expected to be described by a self-similar, universal density profile over scales of 10 kpc to 10 Mpc \citep{frenk1999, gao2012}, parameterized by the magnitude of the overdensity (halo mass) and a scale radius, $r_s$. 
The Navarro-Frenk-White (NFW) profile \citep{navarro1996, navarro1997} is often assumed as a fiducial model.

\begin{figure}[!htb]
    \centering
    \includegraphics[trim=0 0 0 0, clip, width=0.8\columnwidth]{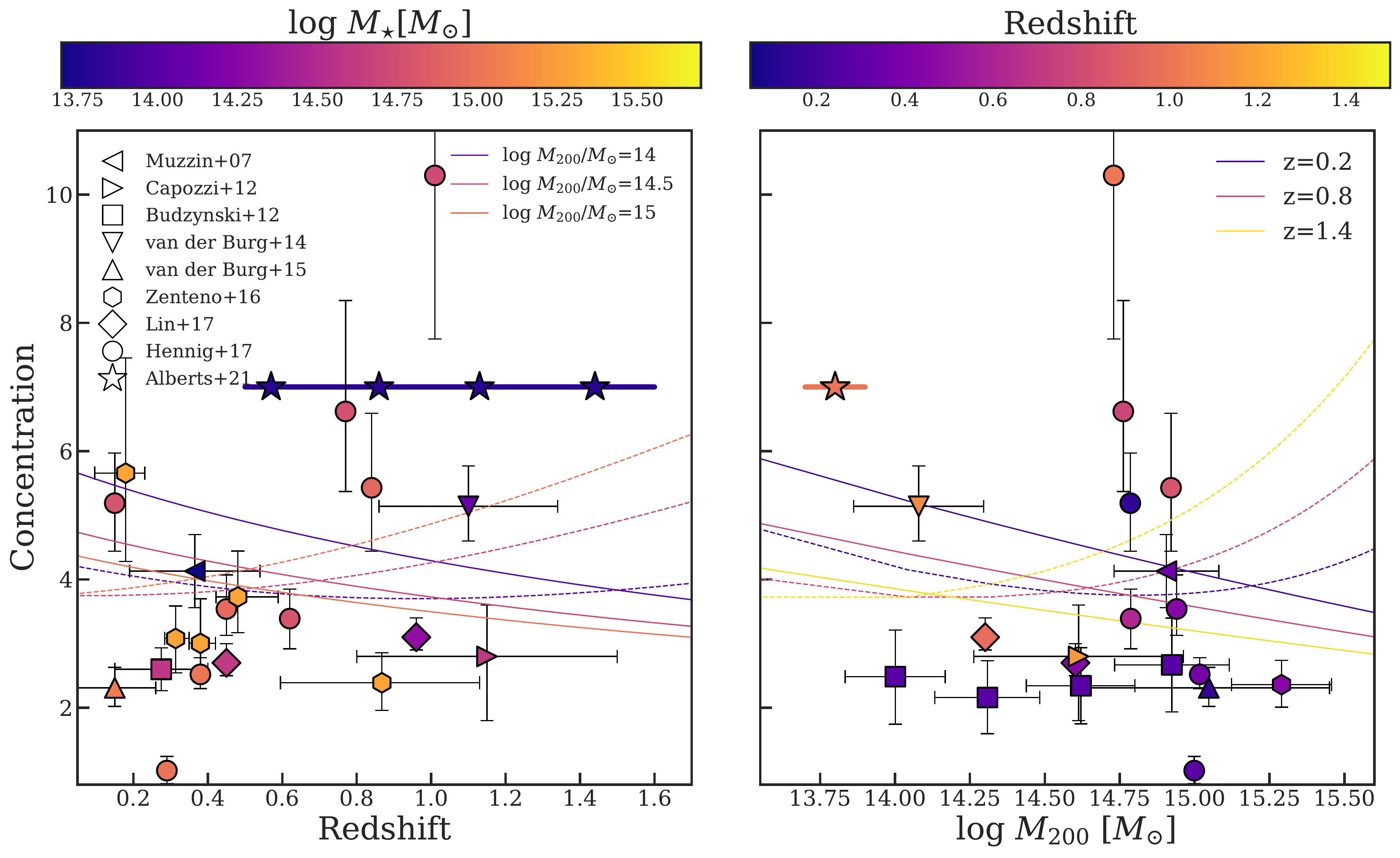}
    \caption{ $-$ Halo concentrations ($c\equiv r_{200}/r_s$) derived from galaxy density profiles \citep{muzzin2007, capozzi2012, budzynski2012, vanderburg2014, vanderburg2015, zenteno2016, hennig2017}, stellar mass density profiles \citep{lin2017}, and total light stacking \citep{Alberts2021} as a function of redshift (left) and $M_{200}$ (right).  Solid and dashed lines show the DM halo concentrations from \citealt{correa2015} and \citealt{diemer2019}, respectively. DM concentrations are relatively flat with redshift and halo mass, with the exception of an upturn at high-mass  predicted for unrelaxed clusters.  The stellar component is often observed to be less concentrated than the DM, but with exceptions at high redshift and low-mass, which may reflect complex baryonic physics in unrelaxed or merging systems.}
    \label{fig:c_Mhalo}
\end{figure}


A consequence of the supposition that the density structure of halos can be parameterized by halo mass and scale radius alone {\it and} is universal means the halo mass and scale radius must be correlated.  This is often expressed as the concentration-mass relation (where concentration $c\equiv R_{200}/r_{s}$).  The $c-M_{\rm halo}$ relation for DM has been both modeled via simulations and observed via halo mass tracers such as weak lensing, with a general lack of consensus.  At the high-mass  end (log $M_{\rm halo}/\Msun\gtrsim14$), simulations disagree on whether the $c-M_{\rm halo}$ relation is relatively flat \citep{dutton2014} or shows a sharp upturn \citep{klypin2016, diemer2015}, possibly depending on whether relaxed or unrelaxed clusters are considered \citep[][but see \citealp{klypin2016}]{ludlow2012, correa2015}.  The observed $c-M_{\rm halo}$ relation for DM has an even wider range of results \citep[see Figure 6 in][]{biviano2017}. 

In the near-infrared, the galaxy or stellar mass distribution is the observable, not the DM halo.  Though these distributions have been successfully modeled using NFW, again there is wide disagreement in $c-M_{\rm halo}$, reflecting the complicated effects of cooling, feedback, and gravitational interactions with the DM halo on the baryons \citep[e.g.][]{gnedin2004, rozo2008, deboni2013}.  Stacking provides a statistical way to measure the extent of cluster emission and the $c-M_{\rm halo}$ relation over large samples, informing our models of cluster baryonic evolution.  With stacking, information on sub-structure in individual clusters is lost; however, this provides a more robust measure of the typical concentration \citep{gao2008}.

Figure~\ref{fig:c_Mhalo} shows the concentrations derived from observations of the stellar mass density \citep{lin2017} and galaxy number density \citep{muzzin2007, budzynski2012, capozzi2012, vanderburg2014, vanderburg2015, zenteno2016, hennig2017}.  Overlaid are predictions from \citealt{correa2015} and \citealt{diemer2019} of the $c_{200}$-$M_{200}$ relation for DM halos\footnote{Simulation predictions were calculated using the Colossus python toolkit \citep{diemer2018}; \url{http://www.benediktdiemer.com/code/colossus/}.}, showing the flat relation and upturn, respectively, for Planck2015 cosmology \citep[][other cosmologies give similar trends]{planckcollaboration2016e}.  In general, galaxies seem to be less concentrated than the DM as a function of both redshift and halo mass, with several exceptions in individual clusters \citep{muzzin2007, vanderburg2014} and with \citealt{hennig2017} reporting a range of concentrations at fixed halo mass.

How do these compare to the average concentration stacked over many clusters? \citealt{Alberts2021} fit the stacked near-IR and far-IR emission from 232 clusters with NFW profiles, applying corrections for the observation beamsize and cluster centroiding to recover the intrinsic profile. 
From the corrected profiles, they found that both the near- and far-IR could be described by an NFW profile\footnote{To be accurate, an NFW profile was found to fit the far-IR at $\gtrsim0.3$ Mpc, with a relative deficit at smaller radii relative to the near-IR profile \citep{Alberts2021}.} with a high concentration ($c\approx7$) at relatively fixed halo mass \citep[log $M_{200}/\Msun = 13.8$, see][]{eisenhardt2008, brodwin2013, alberts2016} and across a wide redshift range ($z=0.5-1.5$). This result suggests the total stellar mass concentration is higher than DM, at odds with the majority of concentrations derived from galaxy populations in individual clusters. Applying this stacking technique to large, well chosen cluster samples, controlling for halo mass, dynamical state, etc., is needed to resolve this contention. 

In summary, the ``total light'' stacking technique is a promising way to perform statistical analyses on the large cluster samples enabled by current and upcoming wide-field and all-sky surveys.  Here we have overviewed just three examples, looking at ICD, the total M/FIR emission from galaxy populations in (proto-)clusters, and the $c-M_{\rm halo}$ relation, each of which has highlighted the need for further statistical analysis in these areas.

\section{Quenching in (Proto-)Cluster Galaxies (the Infrared Perspective)}\label{sec:wrap}

\begin{figure}[!htb]
    \centering
    \includegraphics[trim=0 0 0 0, clip, width=\columnwidth]{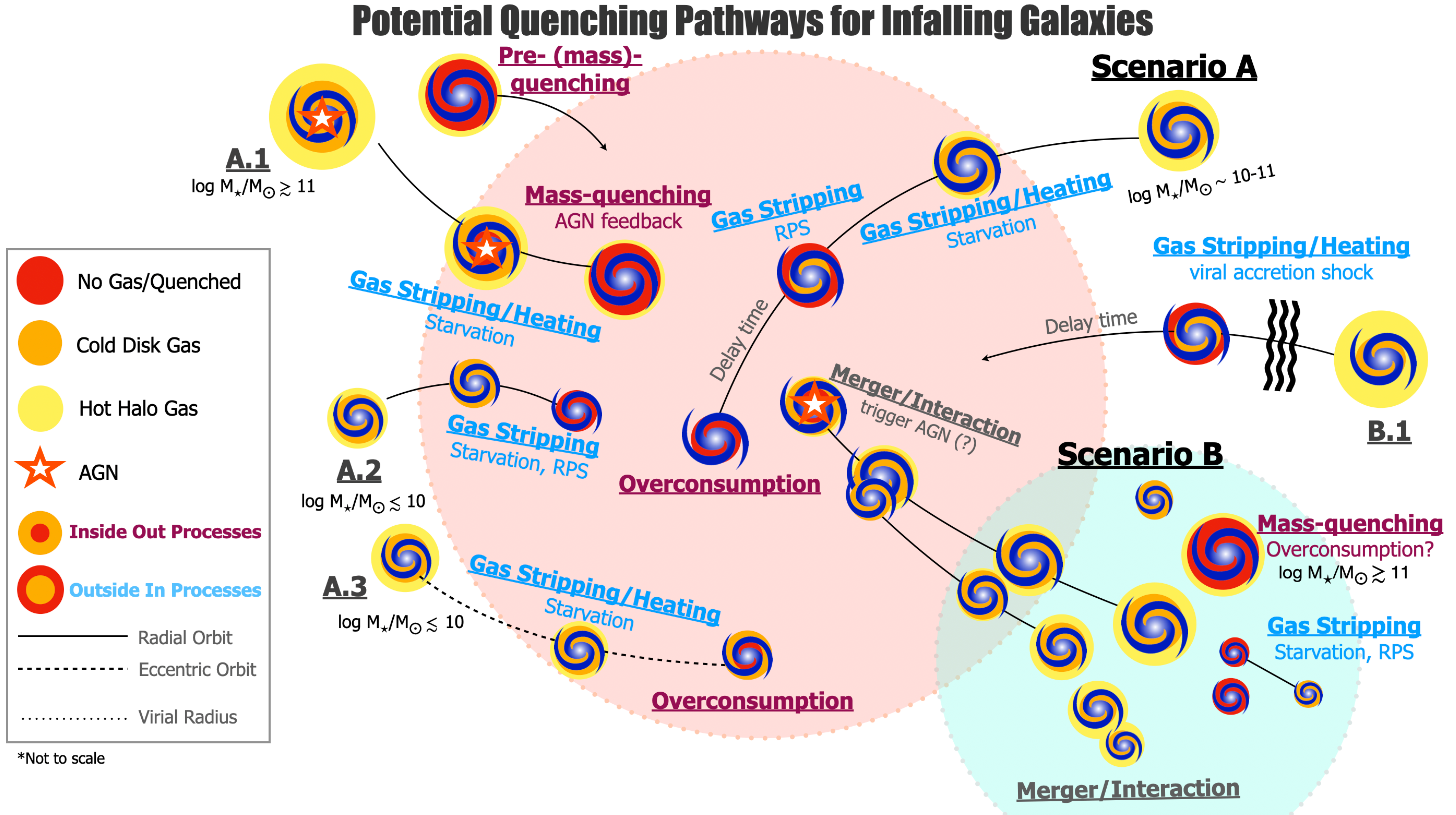}
    \caption{ $-$ A depiction of potential quenching pathways for galaxies falling into a massive (log $M_{200}/\Msun\gtrsim14$) galaxy cluster (\red{red dotted line denotes the virial radius}\sout{light red denotes the halo to $R_{200}$, dark red to $0.2R_{200}$ for reference}).  An infalling massive group (log $M_{200}/\Msun\sim13-13.5$) is depicted in light teal.  Galaxies (blue spirals) are roughly sized according to stellar mass, with solid (dashed) lines tracing their radial (eccentric) orbit around the cluster potential.  Note: morphological transformations are not shown in this cartoon for simplicity. Yellow and orange denote hot halo and cold molecular disk gas, respectively. Red denotes quenched regions with little to no gas.  A star marks AGN activity.  Scenarios A and B, along with their variations, are discussed in \S\,\ref{sec:wrap}.}
    \label{fig:quenching}
\end{figure}


In the preceding sections, we have presented the current state of the literature regarding infrared observations of galaxy populations in (proto-)cluster environments.  From this, a complex (and still developing) picture has been painted of environmentally-driven galaxy evolution and quenching.
In this section, we discuss this IR picture in the context of potential (qualitative) quenching pathways (Figure~\ref{fig:quenching}), and tabulate some of the primary quenching mechanisms with their supporting signatures and IR observations in Figure~\ref{fig:tabquenching}.
Environmental quenching mechanisms -- hydrodynamical, gravitational, and internal -- were introduced in \S\,\ref{sec:environment}.
 
\subsection{All Scenarios: Starvation}
We commence with starvation $-$ likely the most universal and potentially ubiquitous process $-$ starting in intermediate density regions ($\sim10^{-5} - 10^{-4}$ cm$^{-3}$).  The efficiency of new gas accretion via cold streams is expected to be a strong function of halo mass, decreasing as halos enter the shocked heated regime at $>10^{12}\,\Msun$ \citep[e.g.,][]{dekel2006, dekel2009}. In the group or cluster environment, the hot intragroup medium (IGrM) or ICM accelerates this process by further heating or stripping\footnote{We note that many studies refer to this stripping as ram pressure stripping.  For ease of discussion, we have defined and refer to starvation as heating and/or removal (i.e. through stripping, tidal interactions, evaporation) of the hot halo gas and RPS as the (ram pressure) stripping of cold atomic or molecular disk gas.} existing halo gas.  This process is likely slow, stripping hot halo gas on timescales longer than a Gyr \citep{kawata2008, mccarthy2008, vijayaraghavan2015} 
and its influence on star formation, fueled by cold, molecular disk gas, can occur over a few to several Gyr. In Figure~\ref{fig:quenching}, this is depicted as infalling galaxies losing \red{some or all of} their hot halo (yellow circle) on the crossing into the ICM (some, particularly low-mass, group galaxies may have also lost their hot halos).  Though the boundary is shown as sharp in the schematic, in reality the ICM is not smooth and the effects of starvation may begin at very large radii \citep[up to $\sim5R_{\rm vir}$; e.g.][]{bahe2013,gabor2015,zinger2016,morokuma-matsui2021,ayromlou2021}.

As discussed in \S\,\ref{sec:fir}, slow quenching characteristic of starvation may present as mild to moderately suppressed SSFRs.  A bimodal distribution (MS and sub-MS) has been observed in the (dusty) SFG populations of clusters at $z<1$, with sub-MS galaxies having lived in the cluster longer than the MS SFGs according to phase space diagnostics \citep{haines2013, haines2015, noble2013}. This suppression (in the SSFR of massive cluster galaxies) appears to be independent of redshift (on average) at least up to $z\sim1$ at fixed halo mass \citep{alberts2014}.  Starvation is a plausible mechanism for creating this population.  Some molecular gas studies additionally support this, as evidenced by a gradual depletion of gas compared to field galaxies in Virgo from $\sim3R_{200}$ to the core, and as a function of accretion history \citep[e.g.,][see Figure \ref{fig:virgo}]{morokuma-matsui2021}. 

Is there evidence against starvation?  Surviving hot halos have been observed in local cluster galaxies \citep[e.g.,][]{sun2007, goulding2016}, though typically in very massive galaxies with deep potential wells.  A radial analysis of Abell 1795 places galaxies with surviving halos on the outskirts of the cluster, with a clear decrease in occurrences into the cluster center \citep{wagner2018}. Still, a recent simulation suggests that not only is gas stripping ineffective for both the hot and cold gas components, but that the hot ICM can {\it feed} the hot halo \citep{quilis2017}.  This seems inconsistent with infrared observations of a sub-MS population unless that hot halo gas remains heated and unavailable for future cooling onto the disk, which would fall under our definition of starvation. On the other hand, suppressed (obscured) SSFRs are not observed at $z\gtrsim1-1.5$ (Figure~\ref{fig:sf_radial}), and molecular gas masses may be on par with or surpass field levels (Figure~\ref{fig:gas_panel}), supporting that the nature of quenching in clusters changes over cosmic time and starvation alone may be sub-dominant to more rapid mechanisms (or not have enough time to act) at higher redshifts. 

\subsection{Scenario A: Combined hydrodynamical and internal quenching} \label{sec:scenarioa}

We continue under the assumption that starvation operates on the majority of galaxies that fall into massive clusters and, given enough time, is reflected in their SFRs and (molecular) gas properties. We will now consider additional quenching mechanisms -- both hydrodynamical and internal -- first in the context of a representative intermediate-mass (log $M_{\star}/\Msun\sim10-11$) SFG on a radial orbit during its initial infall onto the cluster core (Figure~\ref{fig:quenching}, Scenario A). We will then consider how the efficacy of each mechanism may vary as a function of stellar mass and orbital trajectory, outlining possible variations/combinations of these processes in Scenarios A.1 - A.3.

\subsubsection{Ram Pressure Stripping and Overconsumption in ``Typical'' Intermediate-Mass Cluster Galaxies} \label{sec:normal}

\vspace{1mm}
\noindent\underline{\bf Ram-Pressure Stripping}
\vspace{2mm}

As previously discussed, \red{an}\sout{our} intermediate-mass galaxy eventually loses access to its hot halo upon entering the hot ICM,  with stripping and/or heating preventing the gas from cooling onto the disk. As it moves deeper into the cluster at high speeds, it encounters increasing ram pressure which may eventually \red{affect}\sout{effect} its colder, more tightly bound gas reservoir.  The magnitude of this effect is still under debate and likely depends on several parameters \citep[i.e. stellar mass, orbit, inclination, halo mass; ][]{bekki2014, merluzzi2013, merluzzi2016, quilis2017, tonnesen2019}.  Semi-analytic models and  hydrodynamic cosmological simulations generally support effective cold gas stripping on timescales much shorter\sout{($\lesssim0.5-1$ Gyr)} than the crossing time \red{\citep[$\sim2$ Gyr at $z\sim0$; see discussion in][\sout{but see also citealp{bahe2015}}]{boselli2022}}\sout{citep{quilis2000, tonnesen2007, roediger2007, singh2019}}. Detailed simulations that take into account, for example, an inhomogenous ICM with multiple gas phases support even more rapid timescales \red{\citep{tonnesen2009, tonnesen2019}}.  

In the local Universe, RPS events have been directly observed as truncated gas disks and/or stripped tails (i.e., jellyfish galaxies, see \S\,\ref{sec:submm}), strongly suggesting that gas stripping is an important quenching mechanism at late times \citep[][see also \citealt{boselli2022, cortese2021} for thorough reviews]{yagi2010, boselli2006, gavazzi2018, liu2021}.  However, the picture beyond the local Universe is much less clear.
Up to $z\sim1$, aggregate infrared studies of cluster populations may indirectly support RPS in a few ways.  Just as suppressed SSFRs may be a signature of slow quenching at $z<1$, the strong radial dependence of the star-forming fraction signals concurrent rapid quenching (\S\,\ref{sec:fir}), with galaxies preferentially quenched in the cluster cores where RPS should be most effective \citep[due to high ICM densities, $\sim10^{-3}-10^{-1}$ cm$^{-3}$, and high galaxy velocities, e.g.,][]{gunn1972, zinger2018, roberts2019}.  Along similar lines, the effectiveness of RPS increases with decreasing stellar mass and could plausibly produce the flattened low-mass slope observed in cluster QG SMFs (\S\,\ref{sec:nir_smf}). More tentatively, the halo mass dependence of the EQE suggested by the comparison of different cluster samples (Figure~\ref{fig:balogh16}, right) could be consistent with RPS.  The increased velocities and ICM densities in more massive halos should enhance RPS; however, turbulence also increases with halo mass (and perhaps redshift) and may act to weaken RPS \citep[e.g.,][]{ryu2009, schmidt2016}.

At $z>1$, the observed abrupt evolution in both the $f_{\rm SF}$ and EQE requires that rapid quenching dominate, building up a substantial portion of the quenched population 
from $z\sim1.5\rightarrow1$ (\S\,\ref{sec:nir}--\ref{sec:fir}). Cluster SFGs during this epoch are observed to live on the MS; as outside-in quenching, RPS intrinsically allows for the ``delayed, then rapid'' scenario \citep[e.g.][]{wetzel2013}, wherein star formation will continue unaffected in the central disk for some delay timescale.  Integrated gas studies at $z>1$ are decidedly mixed on the gas properties of cluster galaxies at high redshift, however.  Averaged (stacking) studies of MS cluster SFGs (using dust continuum) indicate strong molecular gas deficits, while individual CO studies find high or enhanced gas fractions, seemingly inconsistent with significant (cold gas) stripping. Resolved studies, on the other hand, show preliminary signs of perturbed gas, which may explain both the high gas fractions and point to the occurrence of stripping (see \S\,\ref{sec:submm}).

Resolved studies are likely key to solidifying the role of cold gas stripping across cosmic time. High-resolution optical studies of local \citep{schaefer2017, bluck2020} and $z\sim1$ galaxies \citep{matharu2021} have observed evidence for outside-in quenching using optical SFR tracers, a result consistent with RPS (and/or starvation). In the IR, local studies find centrally concentrated molecular gas \citep[e.g.][]{mok2017} and at high-$z$, there is tentative evidence for a jellyfish galaxy at $z\sim0.7$ \citep{vulcani2016, boselli2019} and perturbed molecular gas disks at $z\sim1.6$ \citep{noble2019} which suggest RPS at early epochs; however, spatially-resolved submm studies at high redshifts are difficult and have been limited to a few sources. As a population study, \citealt{finn2018} examined the MIPS 24$\mu$m sizes of local galaxies relative to their stellar mass distribution as a function of both global and local environment.  They found that galaxies in overdense environments tend to have more centrally concentrated star formation, a trend that correlated with increased \textsc{HI} deficiency. Centrally concentrated SF has also been observed in a $z\sim1.6$ cluster \citep{ikeda2022}. 

For \red{a}\sout{our} representative, intermediate-mass galaxy, RPS may proceed outside-in and produce differential stripping of the various gas and dust components due to their varying scale lengths \citep[e.g.][]{cortese2016}.  However, it is unlikely to be able to completely strip the disk gas given the deep potential well at log $M_{\star}/\Msun>10$ \citep{cortese2021} and may even compress the disk gas \citep{mok2017}, driving it into the nucleus \citep[e.g.][]{tonnesen2009, ramos-martinez2018} and enhancing H$_2$ formation \citep{henderson2016}. This could boost the central SFE -- as seen in jellyfish galaxies \citep[e.g.][]{poggianti2016, vulcani2018, roberts2020, durret2021} and some $z\sim1.6$ cluster galaxies \citep{ikeda2022} -- and/or feed AGN activity \citep[a contentious issue in jellyfish, see \S\,4.3 in ][for a full discussion]{boselli2022}.  Copious starbursting activity is ruled out by cluster galaxies being largely on or just below the MS\red{; however, star formation enhancements may be too small to detect easily \citep[][]{troncoso-iribarren2020}}\sout{, though the burst phase could be very short citep[$\sim10$ Myr;][]{weinberg2014, troncoso-iribarren2020}}. On the other hand, there is tentative evidence for an enhanced MIR AGN fraction in clusters at high redshift (see \S\,\ref{sec:agn}), which would require a trigger mechanism.  This leads back to the redshift dependence of RPS, which is currently an open question \citep{tecce2010, bahe2015, singh2019}. We can speculate that the increase in galaxy compactness \citep{vanderwel2014} and gas fractions \citep{geach2011, decarli2016a, lagos2011} 
at high redshift may render RPS generally less effective.  For some
galaxies at $z>1$, then, overconsumption is a sound alternative or addition. 

\vspace{1mm}
\noindent\underline{\bf Overconsumption}
\vspace{2mm}

The effectiveness of overconsumption $-$ quenching due to concurrent starvation, star formation, and modest feedack $-$ is a strong function of a galaxy's SFR and feedback strength.  In the local Universe, low SFRs mean long quenching timescales ($\sim10$ Gyr), while at higher redshift ($z>0.4$), high SFRs and strong feedback in massive galaxies can quench on timescales shorter than the dynamical time \citep{mcgee2014, balogh2016}.  From obscured SFR studies in the infrared (\S\,\ref{sec:fir}), we have seen that cluster galaxy SFRs at $z>1$ can be comparable to field galaxies prior to (rapid) quenching, with short or comparable gas depletion timescales ($\sim0.1-3$ Gyr; \S\,\ref{sec:submm}).  In MS galaxies, star formation (and feedback) scale with stellar mass, consistent with the evidence for a mass-dependent EQE (\S\,\ref{sec:nir}). Overconsumption could therefore plausibly provide the rapid quenching required at $z\sim1-2$; however, it is difficult to explain a weakening (or reversal) in the SFR-density relation in massive clusters at $z\sim1.5$ (\S\,\ref{sec:sfr-density_relation}) if overconsumption is the dominant or sole quenching mechanism.  As with RPS, resolved studies, looking for inside-out quenching, would provide strong evidence for processes such as overconsumption.  Inside-out quenching, however, appears to be more characteristic of quenching centrals, rather than satellites \citep{bluck2020}.

\vspace{1mm}
\noindent\underline{\bf Scenario A Summary}
\vspace{2mm}

Which, then, is the dominant mechanism quenching our representative galaxy?  It is apparent that no single quenching mechanism stands out as an obvious fit to the infrared observations.  Moreover, it is likely that the primary mechanism changes with cosmic time.  At $z<1$, starvation and RPS are likely both operating alongside mass-quenching.  At $z>1$, RPS and overconsumption could combine to effect rapid quenching after a delay time.  Though conventionally RPS is thought to be most effective in cluster cores, eccentric orbits \citep{ebeling2014}, inhomogenities in the ICM, and/or group membership could lead to RPS in the cluster outskirts (or beyond, see Sections~\ref{sec:deviations}, \ref{sec:interactions}).  Weaker RPS, stripping the more diffuse material of an extended \red{atomic or} molecular gas disk, could also be a key player and could combine with overconsumption by removing material heated or ejected by feedback from galactic fountains \citep{rasmussen2006,bahe2015} or AGN \citep{george2019, radovich2019}.  As such, it is unclear if a single mechanism dominates, and we conjecture that it varies from galaxy-to-galaxy which mix of quenching processes is driving the evolution in Scenario A. 
 
\subsubsection{Deviations from Scenario A in other stellar mass regimes} \label{sec:deviations}
Near-infrared studies have provided strong evidence for a stellar mass-dependence of EQE (\S\,\ref{sec:nir}, Figure~\ref{fig:balogh16}, right), which emphasizes that we should examine the fates of galaxies outside of the representative mass regime that was discussed in the previous section.  

\vspace{1mm}
\noindent\underline{\bf High-mass Cluster Galaxies (Scenario A.1)}
\vspace{2mm}

We start with high-mass (log $M_{\star}/\Msun\gtrsim10.5-11$) galaxies, for which we consider two sub-pathways: pre-quenching outside the cluster \red{(Figure~\ref{fig:quenching}, A.1, red quenched disk)} and mass- or environmental-quenching within the cluster (Figure~\ref{fig:quenching}, A.1, spiral with yellow halo\red{, orange disk gas,} and star). Mass-quenching in high-mass galaxies is largely thought to be regulated by AGN feedback \citep[e.g.,][]{dimatteo2005, hopkins2005, silk2013, somerville2015, wylezalek2016, beckmann2017}, which is likely the dominant quenching mechanism for this population at least up to group scales at $z\sim0$ \citep{peng2010}.  Via infrared studies, we have seen that the high-mass end of the SMF of cluster SFGs and QGs maintains the same shape as the field SMFs \citep[e.g.,][]{vanderburg2020}, which could indicate similar evolutionary paths. In the GCLASS and GOGREEN clusters, high-mass cluster galaxies are quenched at the same percentage as the field at $z\sim1$ \citep[Figure~\ref{fig:balogh16}, e.g.,][]{balogh2016} and, at slightly higher redshift, the cluster infall regions are found to contain an overabundance of massive halos, with nearly all high-mass galaxies quenched before crossing $R_{200}$ \citep{werner2022}. The deficit of AGN in $z<1$ clusters (\S\,\ref{sec:agn}) limits the amount of mass-quenching via AGN feedback that can be witnessed in the cluster environment.  These observations may support the pre-quenched pathway, i.e., that high-mass  galaxies quench prior to entering cluster environments \citep[as seen in simulations, e.g.,][]{donnari2021a}.  This pathway may include \red{quenching in groups (see Scenario B) or} early quenching in proto-clusters, potentially due to environmentally-driven differences in the star formation histories of massive galaxies \citep{harshan2021}. Both DSFGs, the likely progenitors of massive cluster ellipticals, and high-mass quenched galaxies have been observed in some $z>2$ systems.

On the other hand, FIR observations have identified obscured, high-mass SFGs in cluster cores at $z>1$, which are largely on the MS \cite[e.g.,][]{alberts2016}.  
Is environmental quenching (as opposed to mass-quenching) effective once these massive SFGs fall into the cluster?  Their deep potential wells suggest gas stripping may be weak or ineffective \red{\citep[e.g.][]{quilis2017, cortese2021}}\sout{citep[e.g.,][]{roediger2005,roediger2007,bruggen2008,quilis2017}}; however, even weak heating or stripping in conjunction with high SFRs and feedback would drive overconsumption in excess of pure mass-quenching \citep[e.g.,][]{bahe2015, pintos-castro2019}, as discussed in the main Scenario A. This is potentially consistent with the enhanced mass-quenching efficiency at low cluster-centric radii observed in cluster galaxies at $z\sim0.3-1$ \citep[see Figure 13 in][]{pintos-castro2019}, though we note their mass bin spans log $M_{\star}/\Msun=10-11.2$ rather than isolating very massive galaxies.  Thus, as with our representative cluster galaxy, infrared studies support a diverse range of scenarios for high-mass  (proto-)cluster galaxies, potentially involving a combination of environmental and mass-quenching across field, group, cluster, and proto-cluster environments. 

\vspace{1mm}
\noindent\underline{\bf Low-mass Cluster Galaxies: Stripping (Scenario A.2) or Overconsumption (Scenario A.3)?}
\vspace{1mm}

At the other extreme, low-mass galaxies represent a key population in testing environmental quenching, as their quenching time in isolation can exceed the Hubble time \citep[e.g.,][]{geha2012}.  Gas stripping is expected to be effective on timescales much shorter than the gas depletion timescale for log $M_{\star}/\Msun\lesssim9-10$ galaxies, given even moderate ram pressure and/or tidal interactions \citep[e.g.,][]{haines2007,boselli2008,bahe2015, benitez-llambay2013, fillingham2015, wetzel2015, roberts2019}. Locally, outside-in quenching has been tentatively observed for this population \citep[e.g.,][]{bluck2020} and NIR observations of the QG SMF low-mass slope and nonzero EQE strongly support environmental quenching at low masses (\S\,\ref{sec:nir}).  In Figure~\ref{fig:quenching} A.2, we show that low-mass galaxies can likely undergo near complete gas stripping and quenching before first passage of the cluster core
\citep[e.g.,][and discussion therein]{boselli2022}.

On the other hand, RPS also depends on both orbital parameters and halo mass.  RPS is most effective on radial orbits that bring galaxies into high-density areas at high speeds \citep[e.g.,][]{steinhauser2016}.  For cluster galaxies on eccentric orbits, there is some evidence that gradual, long timescale RPS can occur \citep{ebeling2014}, but also evidence from simulations that stripping is generally ineffective at quenching galaxies on eccentric orbits, even at low mass \citep{steinhauser2016, quilis2017}.    
In this scenario (Figure~\ref{fig:quenching}, A.3), could low-mass galaxies quench via overconsumption? Even at high redshift, field low-mass galaxies maintain SFRs of $\lesssim10\,\Msun$ yr$^{-2}$ \citep[e.g.,][]{whitaker2017}.  On the other hand, stellar feedback in dwarf galaxies has been invoked to reconcile the mismatch in dwarf galaxy abundance between simulations and observations \citep{bullock2017}.  Additionally, there is observational evidence of winds and galactic fountains in some dwarf galaxies \citep[e.g.,][]{mcquinn2019}, but the effectiveness and timescales are still unclear.  From the infrared, environmental- and mass-quenching appear inter-dependent within $R_{200}$ for $\log\,M_{\star}/\Msun<10$ galaxies up to $z\sim0.7$ \citep[\S\,\ref{sec:epsilon_env};][]{pintos-castro2019}, suggesting they should be treated simultaneously.  Given the unique role of environment in quenching low-mass galaxies in general, this population warrants a renewed focus in light of the increased capabilities of future facilities (e.g., \textit{JWST}; see \S\,\ref{sec:future}).

\begin{figure}[!htb]
    \centering
    \hspace{-5mm}
    \includegraphics[trim=0 0 0 0, clip, width=1\columnwidth]{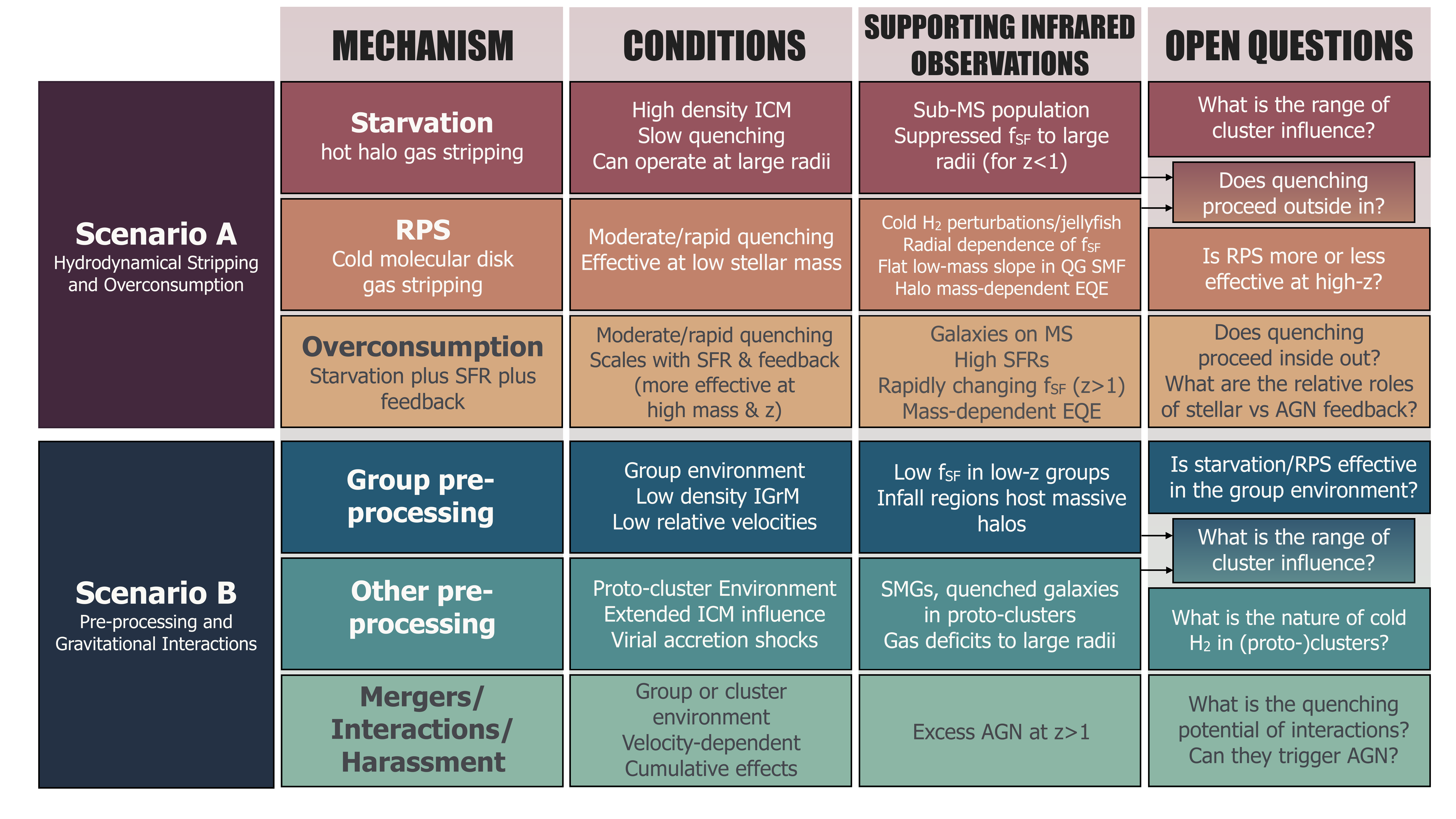}
    \caption{ $-$ A tabulated list of the primary quenching mechanisms that fall within the scenarios outlined in \S\,\ref{sec:wrap}, along with their respective signatures, some of the more robust lines of supporting evidence from infrared studies, and open questions. }
    \label{fig:tabquenching}
\end{figure}

\red{The hydrodynamical and internal quenching mechanisms and their required conditions discussed in this section are listed in Table~\ref{fig:tabquenching} alongside a summary of supporting infrared evidence and open questions.}

\subsection{Scenario B: Pre-processing and Galaxy Interactions} \label{sec:scenariob}

In the previous section, we discussed mock pathways in which hydrodynamical and internal processes operate on infalling and cluster galaxies over a range of stellar masses.  
Here we explore pre-processing and gravitational quenching mechanisms. 
These processes should not be thought of as wholly unrelated to hydrodynamical or internal quenching.  As we are discovering, the (infrared) observational data can accommodate complex quenching pathways and, as such, we will continue to refer to mechanisms like stripping in this section's discussion.

\subsubsection{The environment beyond the virial radius} \label{sec:beyond}

\red{Twenty to forty per cent}\sout{Half of all galaxies live in groups and $20-40\%$} of the galaxies in  $\log\,M_{200}/\Msun\sim14$ clusters at $z=0$ were likely accreted as a group or low-mass cluster \citep{mcgee2009,boselli2022}.  In infrared observations, the star-forming fraction remains well below the field to large \red{cluster-centric} radii \citep{haines2015, pintos-castro2019} and a low $f_{\rm SF}$ has been directly observed in groups falling into the LoCuSS clusters at $z\sim0.3$ \citep{bianconi2018}. Similarly, at $z\sim1$, galaxies in the infall regions of clusters are found to live in comparatively massive halos with a overabundance of satellites and a high quenched fraction \citep{werner2022}. 

In Scenario B (Figure~\ref{fig:quenching}), we consider the conditions in a group halo that will eventually merge with a massive cluster.  In the group environment, galaxies experience lower relative velocities, which could facilitate mergers and interactions.  Group halos are within the shock heated regime ($>10^{12}\,\Msun$), though the IGrM presents less extreme conditions (density, temperature) than in the ICM. Nevertheless, starvation is likely effective \citep[][]{kawata2008, peng2015, trussler2020}, particularly for low-mass galaxies with shallow potential wells.  RPS may also occur, though observations are currently mixed: some works find gas stripping \red{in groups} \citep{rasmussen2006, vulcani2018, roberts2022, lee2022} while others find none, even of the more extended \textsc{HI} reservoir \citep{yoon2015, odekon2016}.  The effects of gas stripping would likely take longer to realize in groups \citep[$\sim3$ Gyr for stripping, up to $\gtrsim5$ Gyr for the SSFR;][]{delucia2012,oman2021}, but could combine with overconsumption to speed along the process in high-mass  galaxies.  

\vspace{1mm}
\noindent\underline{\bf Group pre-processing or extended cluster influence? (Scenario B.1)}
\vspace{1mm}

That environmental quenching occurs in groups seems well supported by the observations \citep[e.g.,][]{vulcani2018, bianconi2018}, even if the mechanism remains unclear. However, as discussed in \S\,\ref{sec:nir_preprocessing}, we should be careful about ascribing all environmental effects outside the cluster to pre-processing in the group environment.  It is common to define the virial radius as the ``edge'' of cluster influence; however, this assumption is not physically well motivated \citep{diemer2017}.  Recently, the steep dropoff in cluster matter density profiles has been termed the splashback radius \citep[$R_{\rm sp}$;][]{adhikari2014, diemer2014, more2015, mansfield2017}, a physically-motivated radius at which bound particles reach the apocenter\footnote{The apocenter is the point of an orbit farthest from the center of attraction.} of their orbit after first infall \citep{shin2021}. $R_{\rm sp}$ depends on the matter accretion rate onto the halo as well as redshift, such that slowly (rapidly) accreting halos at low-$z$ (high-$z$) have a splashback radius up to 2x (1x) $R_{\rm vir}$ \citep{more2015}. 

Is the splashback radius the boundary we should consider when separating cluster quenching from pre-processing?  At low redshift, this is complicated by the backsplash population, which may be found up to $2-2.5R_{\rm vir}$ \citep{haggar2020}.  An additional complication is the unknown range of influence of the ICM.  Gas infalling onto massive halos is expected to compress and form shocks \citep[termed virial accretion shocks, e.g.,][]{birnboim2003, dekel2006}, which we consider in Figure~\ref{fig:quenching}, Scenario B.1.  Such shocks are now supported by observational evidence in the form of high-energy gamma-ray rings around local clusters  \citep{keshet2017, keshet2018, keshet2020}.  Using high-resolution, zoom-in simulations of 16 clusters, \citealt{zinger2018} found that at $z>0.6$ virial accretion shocks were present out to $2-3R_{\rm vir}$ \citep[see also][]{voit2003, molnar2009, book2010, lau2015, hurier2019,anderson2021,baxter2021}, and were effective at starting the process of starvation \citep[see also ][]{tonnesen2007,bahe2013,gabor2015,lotz2019}, with the effect of quenching some galaxies before crossing $R_{\rm vir}$.  They concluded that early starvation could dominate over pre-processing as a quenching mechanism in the infall regions \citep{bahe2013, cen2014, jaffe2015}.  Subsequent simulations suggest an even more dramatic effect, that RPS associated with virial accretion shocks can deplete molecular gas directly \citep{mostoghiu2021}.  In the infrared, there is tentative evidence for molecular gas depletion out to $2R_{\rm vir}$ at high redshift \citep{alberts2022}, which would lend itself to this scenario.  However, the gas properties of cluster galaxies in that epoch are still poorly understood (see \S\,\ref{sec:submm}).  Further complicating this picture are potential shocks from cluster mergers \citep{zhang2020}, and gas streams and cold fronts permeating the ICM \citep{zinger2016, zinger2018a}.  As such, we must move forward cautiously in disentangling the effects of group pre-processing from cluster influence beyond the virial radius.

\subsubsection{Gravitational quenching in group and cluster environments} \label{sec:interactions}

Our last consideration is the role of gravitational quenching, facilitated by tidal interactions, fly-bys (harassment) and strong interactions, or major and minor mergers. Of these, the most well-studied is (major) mergers, and enhanced merger fractions have been reported across several cluster studies out to high redshift \citep[][but see \citealp{delahaye2017}]{lotz2013, hine2016, coogan2018, deger2018, watson2019, sazonova2020}.  The causal link between mergers and quenching has been difficult to establish, however. In the field, local (gas-rich) mergers are thought to trigger starbursts \citep[e.g.][]{alonso-herrero2000, barnes2004, evans2008} and AGN \citep[e.g.][]{ellison2011, ellison2013, ellison2015, ellison2019, weston2017} by driving gas inflows into the nucleus, though the ability of a merger to drive inflows depends strongly on the details of the encounter \citep[][see \citealp{u2022} in this Special Issue for a review on the local merger-AGN relationship]{blumenthal2018}. At high redshift, a link between mergers and AGN activity has not been cleanly established \citep{kocevski2012, mechtley2016, shah2020}, which could be due to changing ISM conditions and enhanced gas fractions \citep[e.g.][]{scudder2015,fensch2017}. It could also be due to the difficulties in merger identification, which is sensitive to the technique used and often biased against late-stage mergers, particularly at high redshift \citep{abruzzo2018}.  As AGN activity is expected to peak in the late stages of a merger \citep[e.g.][]{springel2005, capelo2015} and is likely to be heavily obscured \citep{urrutia2008, glikman2015, donley2018, blecha2018, koss2016, koss2018} or even Compton thick \citep{kocevski2015, ricci2017}, we may be missing the connection with current capabilities. Harassment and interactions are even more difficult to establish, requiring high-resolution imaging to establish the presence of features such as asymmetries, bars, warped disks, which are then not unique signatures of harassment but point only to some perturbing force (i.e. RPS, tidal interactions).

In Figure~\ref{fig:quenching}, Scenario B, we consider a group, with some pre-processing, on infall into a cluster. Compared to gas stripping, the conditions for interactions/mergers are favorable in groups due to low relative velocities \citep[e.g.,][]{knebe2006, vijayaraghavan2013, bahe2019}, and we postulate that some of the group members may merge before infall.  By contrast, the high velocities in clusters may suppress merger activity while facilitating high speed fly-bys that accumulate into harassment.  This is not entirely clear-cut, however.  Recent analysis of the hydrodynamical simulation {\tt Illustris} showed that, compared to isolated infalling galaxies, infalling group members retain lower relative velocities \citep{benavides2020}, which may facilitate mergers in cluster outskirts \citep[e.g.][]{deger2018}.  As such, our Scenario B includes two group galaxies destined to merge after passing the cluster $R_{200}$.  

Will merging quench these galaxies? As stated earlier, the end result of merger activity is unclear; however, we can gain some context based on infrared observations. In terms of triggering star formation, as in local mergers, copious starbursting activity is disallowed as we typically observe cluster galaxies at all redshifts to be on or below the MS (\S\,\ref{sec:fir}). On the other hand, there is tentative evidence of an excess in the MIR AGN fraction in clusters at high redshift \citep{alberts2016}, for which a triggering mechanism is needed (\S\,\ref{sec:agn}).  Enhanced interaction/merger fractions would be a natural explanation for this triggering and potential subsequent quenching by AGN feedback \citep[e.g.,][]{brodwin2013}.  Similarly, the centrally concentrated star formation and molecular gas observed in resolved studies of both local and high-redshift clusters \citep{finn2018, ikeda2022} could be driven by gravitational mechanisms; however, \citealt{ikeda2022} found no correlation between enhanced central SFE and close galaxy pairs.

Progress in understanding quenching by gravitational mechanisms will require more high-resolution studies and improved infrared capabilities, such as with \textit{JWST}, in identifying merger and AGN activity in more obscured phases.  Gravitational mechanisms in overdense environments likely also need to be invoked to explain morphological changes in cluster galaxies; however, that is a topic for another review. \red{As in the previous section, pre-processing and gravitational quenching mechanisms and their required conditions are listed in Table~\ref{fig:tabquenching} with a summary of supporting infrared evidence and open questions.}

\section{Conclusions and Future Prospects}\label{sec:future}

This review has presented the state of infrared studies investigating environmental influences on galaxy evolution in the most extreme regions, clusters and proto-clusters, out to high redshift.  While the resounding conclusion is that environment plays an important role in driving quenching in excess of secular processes, it is also clear that current infrared evidence can accommodate a variety of quenching scenarios, with multiple mechanisms potentially acting concurrently, and with the dominant mechanism(s) likely changing over cosmic time.  In Figure~\ref{fig:tabquenching}, we attempt to summarize the possible scenarios (discussed in some detail in \S\,\ref{sec:wrap}) in terms of the more robust signatures from IR observations, along with listing some of the many open questions.  As surveys and capabilities continue to expand and selection biases are mitigated, we will refine our understanding of these quenching pathways.  While there is still much work to be done, the near-, mid-, and far-infrared regimes have provided unique perspectives, including quantifying quenching efficiencies, measuring radially-dependent obscured star formation rates, and probing spatially-resolved molecular gas reservoirs.  Expanding on this work and continuing to unite it with powerful observables in other wavelength regimes is necessary to address the pressing questions that remain.  

Thankfully, the future prospects for infrared studies of environment are promising, from cluster detection to cluster population analyses.  In \S\,\ref{sec:nir_selection} and \S\,\ref{sec:sz_selection}, we briefly mentioned some of the upcoming surveys and facilities that will expand the number of (proto-)cluster candidates selected in the NIR and via SZ and submm colors by orders of magnitude.  These surveys will greatly broaden the dynamic range of our group and (proto-)cluster surveys and allow considerable binning by cluster properties, breaking the degeneracies between halo properties and galaxy properties discussed throughout this review.  Overlap in (optical/)NIR and SZ selections will produce large samples with known redshifts and halo masses.

The sheer number of (proto-)cluster candidates will necessitate confirmation from wide-field spectroscopic surveys and analyses using statistical techniques such as the total light stacking, discussed in \S\,\ref{sec:totallight}.  Upcoming spectroscopic facilities with multiplexing and/or slitless capabilities (e.g. MOONS\footnote{Multi-Object Optical and Near-infrared Spectrograph; \url{https://vltmoons.org/}} and ERIS\footnote{Enhanced Resolution Imager and Spectrograph; \url{https://www.eso.org/sci/facilities/develop/instruments/eris.html}} on the Very Large Telescope (VLT), \textit{JWST}, \textit{Euclid}, \textit{Roman}, and future Extreme Large Telescopes) will provide confirmation (and detailed characterization, see below) of populations in a moderate sample of clusters.  On the more distant horizon, two proposed space missions would provide wide-field NIR spectroscopic follow-up: SPHEREx\footnote{SPHEREx; \url{https://spherex.caltech.edu/}} and  ISCEA\footnote{Infrared Satellite for Cluster Evolution Astrophysics; \url{https://iscea.ipac.caltech.edu/}}.  SPHEREx will conduct an all-sky spectral survey at $\sim1-5\mu$m, providing spectroscopic redshifts of known clusters up to $z\sim0.9$ \citep{dore2016}.  ISCEA, funded for a NASA mission concept study, has the science goal of mapping large scale structure at cosmic noon and would observe 50 proto-cluster fields on $>10$ Mpc scales, gathering spectra for $\sim1,000$ galaxies per field over 1-2$\mu$m and mapping the 3D cosmic web with better than 50 km s$^{-1}$ resolution \citep{wang2021a}. In the redshift range $\sim1-2$, ISCEA would measure the H$\alpha$ and H$\beta$ emission lines to a few solar masses per year in SFR, mitigating the effects of dust extinction discussed throughout this review.

Targeted follow-up will continue to play an outsized role in understanding cluster populations. Of particular importance in disentangling environmental processes is understanding the behavior of gas, particularly the cold molecular gas that fuels star formation.  Interferometers such as ALMA and NOEMA will continue to be invaluable on this front, particularly via resolved studies; however, it is imperative that we also utilize the more wide-field capabilities of single-dish submm observatories \red{\citep{dannerbauer2019}}.  For example, the 50-meter Large Millimeter Telescope (LMT) will offer 5$^{\prime\prime}$ spatial resolution at 1-2 mm via its new bolometer camera TolTEC\footnote{Commissioning in 2022, \url{http://toltec.astro.umass.edu/}.}, with fast mapping speeds (see also NIKA2\footnote{New IRAM KID Arrays 2 \citep[NIKA2;][]{adam2018}} on IRAM, MUSTANG2 on GBT\footnote{MUSTANG2 \citep[][]{dicker2014}, Green Bank Telescope (GBT)}, \red{and the proposed Atacama Large Aperture Submillimeter Telescope \citep[AtLAST;][]{klaassen2020}}).  This will enable us to quantify the dust in cluster galaxies to $z\sim2$ and obscured star formation at higher redshift.  Resolving the contention in the gas-to-dust ratio in overdense environments and the gas and dust scale lengths relevant for stripping processes (\S\,\ref{sec:submm}) is then paramount to (re-)calibrate dust as a molecular gas proxy and/or establish cold dust content as a unique signature of stripping.

There are other areas where we are poised to make great strides.  Slit and slitless spectroscopy (and narrow/medium band imaging) via \textit{JWST} and sensitive, large FOV ground-based NIR spectrographs will return highly complete cluster catalogs; for example, \textit{JWST} can detect and spatially resolve the Pa$\alpha$ or Pa$\beta$ emission lines $-$ robust SFR tracers with minimal dust extinction $-$ to a few solar masses per year in just a few hours at $z\sim1-2$\footnote{For example, see \textit{JWST} Cycle 1 GO Program 1572 \url{https://www.stsci.edu/jwst/science-execution/program-information.html?id=1572}}. This will account for previously missed obscured and low-mass cluster members. Indeed, we have emphasized in several contexts that low-mass galaxies (and populations like AGN) may be key to understanding environmental quenching.  With the successful launch and commissioning of \textit{JWST}, we now have a unique opportunity to efficiently carry out global and local environmental studies to $\log\,M_{\star}/\Msun<8$.  In a similar vein, \textit{JWST}'s mid-infrared capabilities will revolutionize our understanding of AGN populations by revealing and characterizing obscured supermassive black holes to cosmic noon and beyond. \textit{JWST} further provides the high-resolution (0.06-0.1$^{\prime\prime}$), sensitive rest-frame NIR imaging needed to robustly quantify merger and interaction rates and tie these processes to the triggering of star formation and AGN across different environments and spanning cosmic time.  Given our current and upcoming capabilities, the infrared will continue to deliver a wealth of information regarding the role of environment in shaping the evolution of galaxies.



\vspace{6pt} 




\funding{This research received no external funding.
}

\acknowledgments{The authors first thank Anna Sajina and Asantha Cooray for organizing this Special Issue, as well as Yi-Kuan Chiang, Anthony Gonzalez, Kyoung-Soo Lee, Adam Muzzin, Irene Pintos-Castro, Alex Pope, George Rieke, Alex Van Engelen, Tracy Webb, and Jorge Zavala for valuable content editing, discussions, and suggestions.  We further thank Jianwei Lyu, Kana Morokuma-Matui, Alex Pigarelli, Irene Pintos-Castro, and Damien Sp\'{e}rone-Longin for assistance with data catalogs and figures.  We are also grateful to the original authors
of many of the figures shown here for giving us permission to reproduce their work for this review. S.A. acknowledges support from the James Webb
Space Telescope (\textit{JWST}) Mid-Infrared Instrument (MIRI)
Science Team Lead, grant 80NSSC18K0555, from NASA
Goddard Space Flight Center to the University of Arizona. A.N. gratefully acknowledges support from the Beus Center for Cosmic Foundations at Arizona State University, from the NSF through award SOSPA7-025 from the NRAO, and from \textit{HST} program number GO-16300. Support for program number GO-16300 was provided by NASA through grants from the Space Telescope Science Institute, which is operated by the Association of Universities for Research in Astronomy, Incorporated, under NASA contract NAS5-26555.  This review made use of the following software: NumPy \citep{harris2020}, Matplotlib
\citep{hunter2007}, Astropy \citep{astropycollaboration2018},
pandas \citep{reback2021}, seaborn \citep{waskom2021}, CMasher \citep{vandervelden2020}, a3cosmos-gas-evolution \citep{liu2019}}

\conflictsofinterest{The authors declare no conflict of interest.}



\begin{minipage}[l]{0.45\textwidth}
\small
\vspace{5mm}
The following abbreviations are used in this manuscript:
\renewcommand*{\arraystretch}{1}
\begin{tabular}{@{}ll}
\\
2MASS & Two Micron All Sky Survey\\
ACT & Atacama Cosmology Telescope\\
AGES & Spectroscopy from the AGN and\\ 
& Galaxy Evolution Survey\\
AGN & Active Galactic Nuclei\\
AlFoCS &  ALMA Fornax Cluster Survey\\
ALMA &  Atacama Large Millimeter Array\\
ATCA & Australia Compact Array\\
AtLAST & Atacama Large Aperture \\
& Submillimeter Telescope\\
BCG & Brightest Cluster Galaxy\\
BGG & Brightest Group Galaxy\\
BIMA & Berkeley Illinois Maryland Association\\
\textsc{[Ci]} & Neutral Atomic Carbon\\
CO & Carbon Monoxide\\
COLD GASS & CO Legacy Database for GASS\\
COSMOS & Cosmic Evolution Survey\\
CT & Compton-thick (AGN)\\
DECaLS &  Dark Energy Camera Legacy Survey\\
DES & Dark Energy Survey\\
DGR & Dust-to-Gas Ratio\\
DM & Dark Matter\\
(Hot) DOGs & (Hot) Dust-obscured Galaxies\\
DRC & Distant Red Core\\
EDisCs &  ESO Distant Cluster Survey\\
EQE & Environmental Quenching Efficiency\\
ERCSC & \textit{Planck} Early Release Compact\\ 
& Source Catalog\\
ERIS & Enhanced Resolution Imager and\\ 
& Spectrograph\\
ETG & Early Type Galaxy\\
FOV & Field-of-view\\
GASP &  GAs Stripping Phenomena in \\
& galaxies survey\\
GBT & Green Bank Telescope\\
GCLASS & Gemini Cluster Astrophysics \\
& Spectroscopic Survey\\
GOGREEN & Gemini Observations of Galaxies\\
& in Rich Early Environments\\
H$_2$ & Molecular Hydrogen \\
HERACLES & HERA CO-Line Extragalactic Survey\\
HerMES & \textit{Herschel} Multi-tiered Extragalactic Survey\\
HeViCS & \textit{Herschel} Virgo Cluster Survey \\
\textsc{[Hi]} & Neutral Atomic Hydrogen \\
HRS & \textit{Herschel} Reference Survey\\
HSC-SSP & Hyper-Surprime Cam-Subaru \\
& Strategic Program\\
ICBS & IMACS Cluster Building Survey\\
ICD & Intra-cluster Dust\\
ICM & Intra-cluster Medium\\
IGrM & Intra-group Medium\\
IMF & Initial Mass Function\\
\end{tabular}
\end{minipage}
\hfil
\begin{minipage}[l]{0.45\textwidth}
\small
\renewcommand*{\arraystretch}{1}
\begin{tabular}{@{}ll}
\\
(N/M/F)IR & (Near/Mid/Far-)Infrared\\
IRAM & Institut de Radioastronomie Millim\`{e}trique\\
\textit{IRAS} & \textit{InfraRed Astronomy Satellite}\\
IRS & InfraRed Spectrograph\\
ISCS/IDCS &  IRAC Shallow and Distant Cluster Surveys\\
\textit{ISO} & \textit{Infrared Space Observatory}\\
ISS &  IRAC Shallow Survey\\
JCMT & James Clerk Maxwell Telescope\\
JVLA & Karl G. Jansky Very Large Array\\
\textit{\textit{JWST}} & \textit{James Webb Space Telescope}\\
LAE & Lyman-$\alpha$ Emitter\\
LABOCA & Large Apex BOlometer CAmera\\
LF & Luminosity Function\\
LMT & Large Millimeter Telescope/\\
& Gran Telescopio Milimétrico Alfonso Serrano\\
LoCuSS & Local Cluster Substructure Survey\\ 
MaDCoWS & Massive and Distant Clusters \\
& of \textit{WISE} Survey\\
MAGAZ3NE & Massive Ancient Galaxies At \\
& z > 3 NEar-infrared\\
MIPS & Multi-Band Imaging Photometer\\
MOONS & Multi-Object Optical and \\
& Near-infrared Spectrograph\\
& for \textit{Spitzer}\\
MS & Main Sequence\\
MUSE & Multi-Unit Spectroscopic Explorer\\
NDWFS &  NOAO Deep Wide-Field Survey\\
NFW & Navarro-Frenk-White (profile)\\
NGLS & Nearby Galaxies Legacy Survey\\
NIKA2 & New IRAM KID Arrays 2\\
NOEMA & NOrthern Extended Millimeter Array\\
NMBS &  NEWFIRM Medium-Band Survey\\
ORELSE & Observations of Redshift Evolution\\
& in Large-Scale Environments\\
PACS & Photodetector Array Camera \& Spectrometer\\
Pan-STARRS & Panoramic Survey Telescope and\\
& Rapid Response System\\
PCCS(2) & (Second) \textit{Planck} Catalogue\\
& of Compact Sources\\
PdBI & Plateau de Bure Interferometer\\
PHIBSS & Plateau de Bure High-$z$
Blue Sequence Survey\\
PICO & Probe of Inflation and Cosmic Origins\\
Photo-$z$ & Photometric Redshift\\
QG & Quiescent Galaxy\\
RPS & Ram Pressure Stripping\\
RS & Red Sequence\\
SCUBA & Submillimeter Common-User\\
& Bolometer Array\\
SDSS & Sloan Digital Sky Survey\\
SDWFS & \textit{Spitzer} Deep Wide-field Survey\\
SED & Spectral Energy Distribution\\
SEEDisCS & Spatially Extended EDisCS Survey\\
\end{tabular}
\end{minipage}

\begin{minipage}[l]{0.45\textwidth}
\small
\renewcommand*{\arraystretch}{1}
\begin{tabular}{@{}ll}
\\
SF & Star Formation\\
SFE & Star Formation Efficiency\\
(D)SFG & (Dusty) Star-Forming Galaxy\\
SFR & Star Formation Rate\\
SFRD & Star Formation Rate Density\\
SHELA & \textit{Spitzer}-HETDEX Exploratory Large\\
& Area survey\\
SLED & Spectral Line Energy Distribution \\
SMG & Sub-Millimeter Galaxy\\
SMF & Stellar Mass Function\\
Spec-$z$ & Spectroscopic Redshift\\
SpARCS & \textit{Spitzer} Adaptation of the Red-sequence\\
& Cluster Survey\\
SPIRE & Spectral and Photometric Imaging REceiver\\
SPT & South Pole Telescope\\
SSDF & \textit{Spitzer} South Pole Telescope Deep Field\\
SSFR & Specific-Star Formation Rate\\
Submm & Submillimeter\\
SWIRE & \textit{Spitzer} Wide-area InfraRed\\
& Extragalactic survey\\
SZ & Sunyaev-Zel'dovich (Effect)\\
(U)LIRG & (Ultra-)Luminous Infrared Galaxy\\
UMG & Ultra-Massive Galaxy\\
UV & Ultraviolet\\
VERTICO & Virgo Environment Traced in CO survey\\
VLT & Very Large Telescope\\
WINGS & WIde-Field Nearby Galaxy-cluster Survey\\
\textit{WISE} & \textit{Wide-field Infrared Survey Explorer}\\
\end{tabular}
\end{minipage}
\footnotesize
\externalbibliography{yes}
\bibliographystyle{apj}
\bibliography{main}

\end{document}